\documentclass[11pt,a4paper,oneside]{book}

\usepackage[english]{babel}
\usepackage[latin1]{inputenc} 
\usepackage[T1]{fontenc}
\usepackage{graphicx}
\usepackage{microtype} 
\usepackage{indentfirst} 
\usepackage{bm} 

\usepackage{epigraph}
\usepackage{geometry}
\usepackage{cite}
\usepackage{yhmath}
\usepackage{amsmath}
\usepackage{mathtools}
\usepackage{amssymb}
\usepackage{listings}
\usepackage{csquotes}
\usepackage{fancyvrb}
\usepackage{xcolor}
\usepackage{doi}
\usepackage{float}
\usepackage{caption}
\usepackage{subcaption}
\usepackage{multirow}
\usepackage{setspace}
\usepackage[ruled,vlined]{algorithm2e}

\newcommand{\mymatrix}[1]{\hat{\bm{#1}}}

\begin{document}

\title{Optical spin glasses: a new method to simulate glassy systems with scattering disorder}
\author{Erik Hörmann}
\date{\today}

\newgeometry{left=3cm,bottom=3cm,top=3cm,right=3cm}
\begin{titlepage} 
\begin{center}
\begin{figure}[h!]
\begin{center}
\includegraphics[width=10cm]{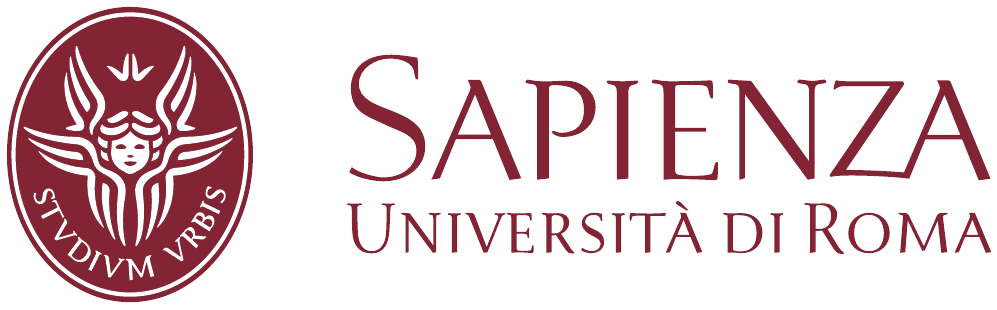}
\end{center}
\end{figure}
\vspace{1cm}
{\normalsize{\bf FACOLT\`A DI SCIENZE MATEMATICHE, FISICHE E NATURALI\\ \vspace{5mm}
Corso di Laurea Magistrale in Fisica}}
\end{center}
\vspace{\stretch{4}}\null
\begin{center}
\huge{\bf{Optical spin glasses:\\ a new model for glassy systems}}\\
\end{center}
\vspace{\stretch{8}}\null
\par
\noindent
\begin{minipage}[t]{0.49\textwidth}
{\large{\bf Relatore:\\
Chiar.mo Prof.\\
Giancarlo Ruocco\\ \\
Relatore esterno:\\
Dott. Marco Leonetti}
}
\end{minipage}
\hfill
\begin{minipage}[t]{0.49\textwidth}\raggedleft
{\large{\bf Candidato:\\
Erik H\"{o}rmann}\\
matricola 1617761}
\end{minipage}
\vspace{\stretch{8}}\null
\begin{center}
{\large{ Sessione Autunnale \\Anno Accademico 2017-2018}\\
Dipartimento di Fisica }
\vspace{\stretch{1}}\null
\end{center}
\end{titlepage}
\restoregeometry
\clearpage

\frontmatter
{\setstretch{1.2}
\tableofcontents
}

\mainmatter
{\setstretch{1.4}
\chapter*{Abstract}
\addcontentsline{toc}{chapter}{Abstract}

The aim of this work is to prove that it is possible to realise an optical system which produces as output a light intensity that can be expressed in the same mathematical form of the spin glass Hamiltonian. The optical system under study is controlled through an adaptive optics device composed by millions of switchable mirrors in an ON and OFF position. These user controlled mirrors are playing the role of the spin states. Furthermore, such optical system can be used to run simulations and, applying the same Metropolis Monte Carlo algorithm used in computer simulations, extract the quantities of interest for Spin Glass Physics.\\

The proposed system has a great advantage over existing computer-based spin glass simulations: the simulation step has no scaling dependence on the total number of spins $N$ and the whole simulation has a linear dependence on $N$, due to the fact that we want to keep the number of moves \emph{per spin} constant. This should be compared to the $N^2$ dependence of a single Monte Caro move in computer simulation, which yields a global $N^3$ dependence. Some experimental problems limit such advantage, but are partially addressed at the end of the presentation.\\

Firstly, in chapter \ref{Ch:spin_glass_theory}, an overview of spin glass theory is presented. The defining properties of spin glass systems are analysed to prepare for the comparison with the experimental results: strong emphasis is given to the two main measured quantities: the \emph{spin autocorrelation function} and the \emph{Edward-Anderson parameter} $q_{EA}$. Finally, the scaling of the problem with respect to the number of spins in the system is given both for the optical approach and the standard computational simulations.\\

In chapter \ref{Ch:theory_optics} the optical theory for the proposed system is presented. The optical phenomenon involved in the realizatdion of the experiment - the \emph{speckle pattern} - is presented both experimentally and theoretically. The complete calculation of the output intensity is performed, in terms of the \emph{multipole expansion} and the \emph{dipole propagator}. The mathematical equivalence between the intensity function and the spin glass Hamiltonian is therefore proven.\\

The description of the experimental setup follows in chapter \ref{Ch:experimental_setup} , with an emphasis on the correspondence between each piece of the experimental setup and the constitutive part of the spin glass model. A detailed explanation of the experimental hardware and software is given to ensure the reproducibility fo the experiment. Finally, some experimental limitations are taken into account and addressed.\\

In chapter \ref{Ch:preliminary} the calibration of the experimental setup is presented. This includes the stability test and the DMD control test. Also the measure of the couplings is included in this section because its results are used to tune the main simulation, in which the overlap function is calculated at different effective temperatures.\\

Chapter \ref{Ch:data_analysis} is the central part of the present work, in which the experimental results are analysed and compared with the ones from both theory and computer simulations. The two quantities analysed, namely the \emph{coupling probability distribution} and the \emph{spin autocorrelation function} are used to prove that the optical systems indeed behaves like a spin glass system. The data analysis relies on the fit of the overlap function, from which the Edward-Anderson parameter is extracted and, in particular, its temperature dependence is studied and compared with the reference in the literature.\\

Finally, in chapter \ref{Ch:conclusions}, a future development of the optical system is proposed. The aim to optically simulate systems made by very high number of spins is addressed and the steps to improve the experimental setup to obtain the necessary stability are included. Other systems of interest, obtained by the generalisation of the one presented, are briefly discussed.
\chapter{Some topics in spin glass theory}\label{Ch:spin_glass_theory}

This thesis wants to investigate the relation between spin glasses and optical systems. In particular, we want to show that it is possible to simulate a spin glass though an appropriate optical system. Therefore, a short summary of spin glass theory is given. This will be a highly non-exhaustive treatment and only the concepts which are useful to the aim of the present work will be considered.

\section{A spin glass introduction: the Ising model}

We may start from the very basic question of \emph{what is a spin glass ?} Due to the extreme variety of the spin glass models and phenomenology, we do not want to address this question directly, but rather build up a prototype model of spin glass, suitable for the purpose of the present work. We begin with the \emph{Ising model}: this path is a sort of historical introduction, in the sense that the Ising model is older and simpler that a fully featured spin glass model, but the latter can be obtained quite straightforwardly from the previous one. However, we do not aim for a complete historical description of the spin glass model: this can be found, for example, in \cite{BaityJesi2016}.\\

The Ising model is a lattice system of atoms. This means that the constitutive parts of the model are located on a discreet set of points in the real space. For simplicity, we consider a hyper-cubic, $N$-dimensional lattice. Mathematically, this is defined as the $\mathbb{Z}^N-span$. Our lattice $\mathcal L$ will therefore be defined as:
\begin{equation}
	\mathcal{L} = \left\{\bm{x}\in \mathbb{R}^N|\bm{x}=\lambda \bm{n}, \bm{n}\in\mathbb{Z}^N\right\}
\end{equation}
in which $\lambda$ is the \emph{spacing} of the lattice, but it will not play any significant role in the theory.\\

On each site of the lattice, namely for each $\bm{x}\in\mathcal{L}$, we define a variable $\sigma(\bm{x})$, which represent the magnetic moment  of the atom sitting in the corresponding site of the lattice. However, due to historical reasons, the variable $\sigma$ is called \emph{spin}. Various representations for the spins are possible, but the Ising model constrains such variables to assume only the discrete values $\{+1,-1\}$. The other common choice is the so-called \emph{Heisenberg model}, in which the spins are unitary vectors which can point in any direction.\\

Physically, the choice between the two models is dictated by the geometry of the system we want to represent. Since the spin variables represent the magnetic moment of an atom, for isotropic materials the Heisenberg model is normally chosen, as the atomic magnetic moment is approximately constant in magnitude but can vary in all directions. On the other hand, a strong anisotropy in the structure of the material could constrain the direction of the magnetic moment along an axes, giving rise to the representation typical of the Ising model.  It is clear now that the two models represent quite different physical systems. For example, the  magnetic susceptibility at high temperature is different \cite{PhysRev.128.168}, and the Ising model has a phase transition in two dimension, while the Heisenberg model has none \cite{PhysRevLett.17.1133}. Several other studies have proven different results for many common observables in spin glass theory such as free-energy, magnetic susceptibility at all temperatures and correlation functions \cite{doi:10.1119/1.1970340,YAMAGUCHI1993201,steiner1972three}\\

We will always refer to the Ising model; the next step is to introduce the Hamiltonian describing the system. Its fundamental characteristic is to contain only pairwise interaction among the spins: three-body terms and higher orders of many-bodies expansion are neglected. This approximation is assumed not only in the simple Ising and Heisenberg model, but will be retained in every generalisation we will perform on these models to define a full spin glass model. Mathematically, this constrain gives rise to a Hamiltonian which can be written as:
\begin{equation}\label{eq:general_Hamiltonian}
	H = -\sum_{(i,j)\in\,l} J_{ij} \sigma_i \sigma_j
\end{equation}
in which $l$ is an appropriate subset of $\mathcal L\times \mathcal L$ (potentially $l \equiv \mathcal L\times \mathcal L$, if each pair of spins has an associated interaction) and $J$ is either a tensor when the spins are vector-valued (like in the Heisenberg model) or a matrix if the spins are scalar (like the Ising model). We note that different choices of the subset $l$ and the form of the coupling $J$ leads to different interaction models for the same system of spins.\\

The Ising model specialises equation (\ref{eq:general_Hamiltonian}) by assuming \emph{constant, nearest neighbour interaction}. The interaction is said to be \emph{constant} because the coupling constant $J_{ij}$ is the same for each pair of interacting spin: namely $J_{ij}=J$. Note that, of course, it is still possible to have $J_{ij}=0$ for some spin pairs: these are said to be \emph{non-interacting pairs}. In addition, the sign of the constant $J$ can be either positive or negative. In particular:
\begin{itemize}
	\item if $J_{ij}>0$, the spin pair $\left(\sigma_i,\sigma_j\right)$ will tend to stay aligned and the two spins to point towards the same direction. The coupling is said to be \emph{ferromagnetic} 
	\item conversely, if $J_{ij}<0$, the spin pair $\left(\sigma_i,\sigma_j\right)$ will tend to stay anti-aligned and the two spins to point in opposite directions. The coupling is said to be \emph{anti-ferromagnetic} 
\end{itemize}

Furthermore, the interaction is said to be \emph{at nearest neighbours} because only pairs of atoms which are distant one lattice spacing are interacting. Having chosen an hyper-cubic lattice, this means that the only interacting pairs are the one whose atoms are neighbours in exactly one dimension. A graphical illustration of the nearest neighbour pairs is given in figure \ref{nearestneigh}.\\

\begin{figure}[ht]
\begin{center}
	\includegraphics[width=0.5\textwidth]{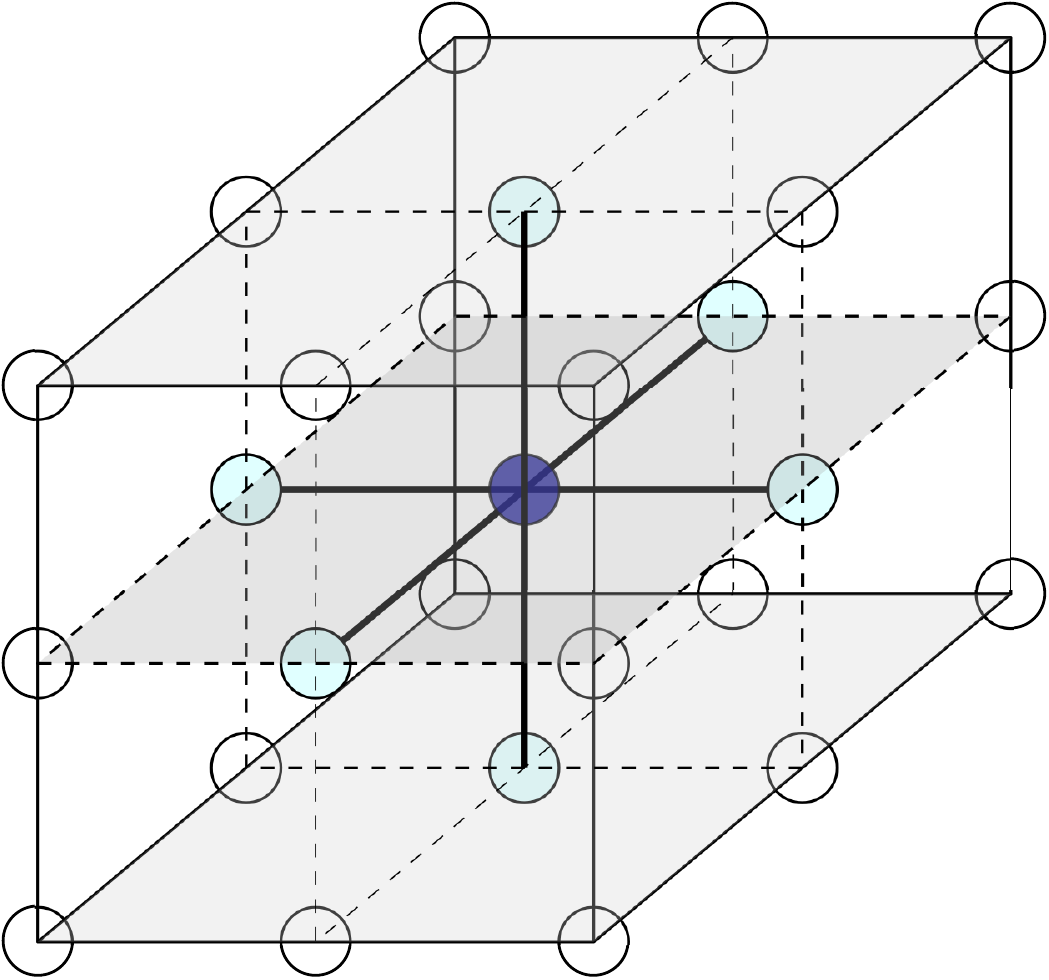}
	\caption{A representation of the concept of nearest neighbours in three dimensions. The blue atom at the center has six nearest neighbours, the ones in light blue, connected to it by the fatted links.}\label{nearestneigh}
\end{center}	
\end{figure}
 
The common mathematical expression used to indicate that the sum in equation \ref{eq:general_Hamiltonian}) runs only over nearest neighbours pairs of spins is $\left<\cdot,\cdot\right>$. The Hamiltonian of the Ising model will therefore be:
\begin{equation}\label{eq:Ising_nofield}
	H = -\frac{1}{2}\sum_{\left<i,j\right>}J\,\sigma_i\sigma_j =-\frac{J}{2}\sum_{\left<i,j\right>}\sigma_i\sigma_j
\end{equation}
in which the $\frac{1}{2}$ factor is included to prevent double counting each pair. Note that, because we have chosen the spins to assume the values $\{+1,-1\}$ only, the sum on the right-hand side is just a sum of values $+1$ when the two considered spins are aligned (namely, when they have the same sign) or $-1$ when are anti-aligned (they have opposite signs).\\

The Ising model is trivial if the lattice is one-dimensional, and has been solved by Onsager in 1944 for a two dimensional lattice \cite{onsager1944crystal} by using the \emph{transfer matrix formalism}. At higher dimensions, it produces a variety of phenomena \cite{barber1985finite,ferrenberg1991critical}, which we will not treat as we want to straightforwardly proceed to the spin glass case.\\

However, an intermediate step is indeed needed. The Ising model described by the Hamiltonian (\ref{eq:Ising_nofield}) correspond to a perfectly isolated system of atoms. However, it is possible to generalise the system by enforcing a \emph{constant magnetic field} from the external environment. Because the tendency of the magnetic moment to align with an existing magnetic field, we will get a second piece in the Hamiltonian describing the system:
\begin{equation}\label{eq:isingh_wf}
	H = -\frac{J}{2}\sum_{\left<i,j\right>}\sigma_i\sigma_j+h\sum_{i=1}^N\sigma_i
\end{equation}
in which $h$ is the intensity of the applied magnetic field. Note that this simple generalisation completely changes the behaviour of the system: the Onsager solution for 2D systems is not valid except in the case of vanishing external field. Furthermore, the external field can be chosen both static and time-dependent: the phenomenology of phase transition in higher dimensions becomes much more varied and complex \cite{lo1990ising,berg1993simulation,delfino1995spin}.

\subsection{Two equivalent Ising models} \label{sec:twoising}
We have defined the Ising model in such a way that the spins can assume only two values, namely $\{-1,+1\}$. This is by far the most common choice in literature: for this reason, we will refer to it as the \emph{Ising representation}. It is also a very convenient one because it specifies the direction of the angular momentum along the axis. If the $z$ axis is taken as reference in a 3D model, this means that if $\sigma=+1$ the spin point upwards, conversely it points downward when the associated variable is $\sigma=-1$. \\

However, the fundamental assumption of the Ising model is that the spin is a scalar number that can only assume two values: there is no physical meaning in the actual numbers we use to represent the state of the spin. Of course, the Hamiltonian as written equations (\ref{eq:Ising_nofield}) and (\ref{eq:isingh_wf}) are only valid for the $\{-1,+1\}$ choice: if we change the variables for the spins, the coupling constants and - possibly - the magnetic field must be rescaled accordingly.\\

We want now to derive the equations to change the spin representation from the standard $\{-1,+1\}$ to the representation in which each spin can assume the values $\{0,+1\}$. This latter representation is also well known in literature, mainly for the neural network application of the Ising and its derived spin glasses \cite{amit1992modeling}, and will be useful for the purposes of the present work: we will call it the \emph{Amit representation} in honour of the pioneering work of Daniel J. Amit in applying the spin glass theory to the world of neural networks.\\

We will call $\left\{\sigma_1,\sigma_2,...,\sigma_N\right\}$ a set of spins in Ising representation and $\left\{\rho,\rho,...,\rho\right\}$ the very same set of spins in Amit representation. We want to map the two representations in such a way that if a spin has an associated value of $\sigma_i=+1$ in the Ising representation, it will maintain the same number in the Amit representation: $\rho_i=+1$. Of course this means that if a spin was in the $\sigma_j=-1$ state in the first case, we want it to have an associate value of $\sigma_j=0$ in the latter.\\

The equations to pass from one representation to the other are trivial and can be found by direct check:
\begin{equation}\label{eq:isingtransf}
	\left\{
	\begin{array}{l}
		\rho_i = \frac{1}{2}\sigma_i + \frac{1}{2}	\\
		\sigma_i = 2\rho_i -1	
	\end{array}
	\right.	
\end{equation}

In which the first relation allows us to pass from the Ising to the Amit representation, and the second one is used to to the opposite transformation.\\

We now need to write down the correct Hamiltonian in the Amit representation. This is done simply by plugging into the normalised version of (\ref{eq:Ising_nofield}) the second transformation formula, thus expressing the intensive Hamiltonian in terms of the Amit spin representation:

\begin{align}
	\frac{H}{N} =& -\frac{J}{2N}\sum_{\left<i,j\right>}\sigma_i\sigma_j +\frac{h}{N}\sum_{i=1}^N\sigma_i= -\frac{J}{2}\sum_{\left<i,j\right>}  \left(2\rho_i-1\right)\left(2\rho_j-1\right) +\frac{h}{N}\sum_{i=1}^N\left(2\rho_i-1\right)=\\
\label{eq:dropconst}	=& -\frac{J}{2N}\sum_{\left<i,j\right>} 4\rho_i\rho_j + \frac{J}{2N}\sum_{\left<i,j\right>} 2\rho_i + \frac{J}{2N}\sum_{\left<i,j\right>} 2\rho_j  -\frac{J}{2N}\sum_{\left<i,j\right>} 1 +\frac{2h}{N} \sum_{i=1}^N \rho_i - \frac{h}{N} \sum_{i=1}^N 1=\\
\notag	=& -2\frac{J}{N}\sum_{\left<i,j\right>} \rho_i\rho_j + 2\frac{J}{N}\sum_{\left<i,j\right>} \rho_i + 2\frac{h}{N} \sum_{i=1}^N\rho_i=\\
\label{eq:nnexpansion}	=& -2\frac{J}{N}\sum_{\left<i,j\right>} \rho_i\rho_j + 2\frac{J}{N}\,2d \sum_{i=1}^N\rho_i  +2\frac{h}{N} \sum_{i=1}^N\rho_i=\\
\notag	=& -2\frac{J}{N}\sum_{\left<i,j\right>} \rho_i\rho_j + 2\left(2d\,\frac{J}{N}+h\right)\sum_{i=1}^N\rho_i=\\
\label{eq:Amtiham}	=& -\frac{\tilde{J}}{2N}\sum_{\left<i,j\right>} \rho_i\rho_j + \frac{\tilde{h}}{N}\sum_{i=1}^N\rho_i
\end{align}
In (\ref{eq:dropconst}) we have dropped all the terms which represent constant contributes, from the very general assumption that only energy differences has physical meaning. In (\ref{eq:nnexpansion}) we have expanded the sum over the nearest neighbour couples: since on a hyper-cubic lattice in $d$ dimensions each spin has $2d$ nearest neighbours\footnote{If we are studying high-dimensionality Ising models, there is a \emph{caveat} on this point: we cannot of course count more nearest neighbours than the number of spins in the system. We therefore need to substitute $2d$ with the following quantity: $$ \mbox{min } \left(2d,N\right)$$ This avoids the problem, and it's commonly used in infinite dimensional Ising models, in which all couples of spins are nearest neighbours and, of course, $N$ is always the upper limit for the sum.}, counting a single spin over all possible nearest neighbour pairs is equivalent to count each spin exactly $2d$ times. We reached  an expression for the Hamiltonian in the Amit representation (\ref{eq:Amtiham}), identical to the one in Ising representation (\ref{eq:isingh_wf}). As we forecast, we had to redefine the coupling constant and the external magnetic field as:
\begin{equation}
	\left\{
	\begin{array}{l}
		\tilde{J} = 4J	\\
		\tilde{h} = 2\left(J\,\frac{2d}{N}+h\right)
	\end{array}
	\right.	
\end{equation}

We note that in many dimensions, when the $2d$ term is limited by the number of available spins in the system, we simply  get:
\begin{equation}
	\left\{
	\begin{array}{l}
		\tilde{J} = 4J	\\
		\tilde{h} = 2\left(N\frac{J}{N}+h\right)= 2(J+h)
	\end{array}
	\right.	
\end{equation}

In both cases, the equation for the coupling constant is just a rescaling, but the equation for the magnetic field entails an important physical consequence: passing from the Ising representation to the Amit representation causes a shift in the magnetic field. In particular, \emph{an Ising model with no magnetic field acquires an effective magnetic field in passing from one representation to the other, due to the shift in the mean value of the spin variables}. This is not only conceptually important, but will actually determine some of the results obtained in this thesis.\\

Keeping in mind this correspondence, we can proceed and concentrate on the Ising representation only: we will use equations (\ref{eq:isingtransf}) when we will need to pass form the Ising representation to the Amit representation and vice-versa.

\section{Towards spin glasses}
We have introduced most of the components of a complete spin glass model. Nevertheless, the systems described up until now bears no resemblance to the final model we want to obtain. Indeed, two key ingredients are missing: \emph{randomness} and \emph{frustration}. To introduce these two concepts, we need to generalise the coupling constants $J_{ij}$. Up until now we have used only a single value for the coupling constant $J$ in the Hamiltonian: to move on to spin glasses, we need to drop this simplification. There are two ways to do so: we can let the interaction vary in time, be different for different couples or a combination of both.\\

If the coupling constants are fixed in time like they are in the Ising model, we say that the interaction is \emph{quenched}; if instead the interaction coefficients changes in time we talk about \emph{annealed} interaction. For our purpose, we will only use spin glasses with quenched interactions, so we do not need to further generalise the coupling constants as functions of time. Nevertheless, annealing in spin glass-like systems is frequently used in optimisation problems and has given rise to a number of applications \cite{kirkpatrick1983optimization,kirkpatrick1984optimization}.\\

We are left free to change the value of the interaction coefficients from one pair of spin to another. This can be done in a number of ways. Firstly, we recap the meaning of the sign of the coupling constant for different spin pairs in the Ising model in table \ref{tab:Js}.

\begin{table}[h]
	\begin{center}
		\begin{tabular}{|c|c|c|}
		\hline
		Value of $J_{ij}$ 	&	Reciprocal position of $\sigma_i$ and $\sigma_j$	 &	Interaction\\
		\hline
			$J_{ij} > 0$	&	$\sigma_i$ and $\sigma_j$ are nearest neighbour & Ferromagentic interaction\\
		\hline
			$J_{ij} < 0$	&	$\sigma_i$ and $\sigma_j$ are nearest neighbour & Anti-ferromagentic interaction\\
		\hline
			$J_{ij} < 0$	&	$\sigma_i$ and $\sigma_j$ are NOT nearest neighbour & No interaction\\
		\hline
		\end{tabular}
		\caption{Possible values of the coupling constant for different spin pairs}\label{tab:Js}
	\end{center}
\end{table}

To go further we want to generalise the magnitude of the coupling constant, too. We do so by assigning a probability distribution for the coupling constants rather than their specific values. In other words, we say that each $J_{ij}$ is a \emph{random variable} extracted with probability $P(J)$. It is important to notice that, while the specific value of the coupling constant differs from one pair to the other, they are all extracted from the same probability distribution.

\subsection{Frustration and complex energy landscape}

The sign of the coupling constant between a pair of spins is fundamental, as it determines the reciprocal position of the two spins. In particular, two spins that have a ferromagnetic link between them are more likely to have the same reciprocal orientation, because this reduces the total energy of the system as given in the equation (\ref{eq:general_Hamiltonian}).\footnote{The underlying assumption is that the probability of each state is distributed as the Boltzmann probability, namely $Prob \propto e^{-\beta H}$. What is presented here is just a way to ensure that the couple of spins under consideration is giving a negative contribution to the total energy.} Note that in absence of external field there is no difference between the two cases in which both are $+1$ and both are $-1$. On the other hand, two anti-ferromagnetic links spins will tend to have opposite values, for the very same reason.\\

If the $J_{ij}$ can assume both positive and negative values, \emph{frustration} can happen. A system is said to be frustrated if there is no spin configuration in which each couple of interacting spins gives a negative contribution to the total energy. To better understand this phenomenon, a simple example is given in figure \ref{fig:frustration1} and \ref{fig:frustration2}.\\

\begin{figure}[H]
	\begin{center}
		\includegraphics[height=3.8cm]{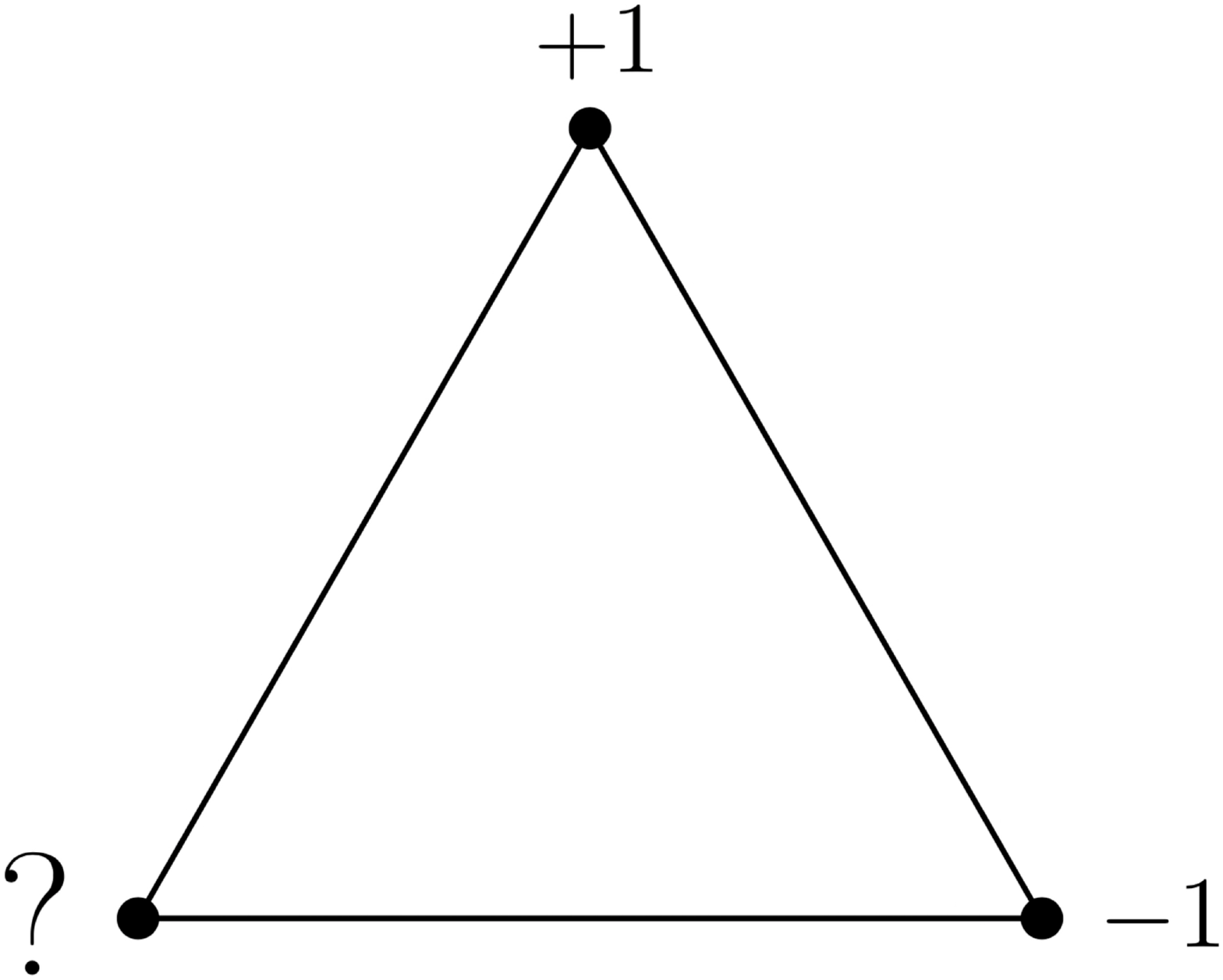}
		\caption{An example of a lattice that presents frustration. All couplings are assumed to be anti-ferromagnetic.}\label{fig:frustration1}
	\end{center}
\end{figure}

We want to minimise the energy in this spin glass: this is obtained by satisfying the greatest number of couplings as we have seen before. We start by choosing for the first spin on top a value $\sigma=+1$. This is perfectly legitimate: the Hamiltonian - in absence of external magnetic filed - is invariant under the contemporary inversion of all spins and therefore we are not precluding any possible solution with this choice. Then we fix the spin in the lower, right hand-side corner at a value $\sigma=-1$, so to satisfy the antiferromagnetic link between it and the already fixed upper spin.

\begin{figure}[ht]
	\begin{center}
		\includegraphics[height=3.8cm]{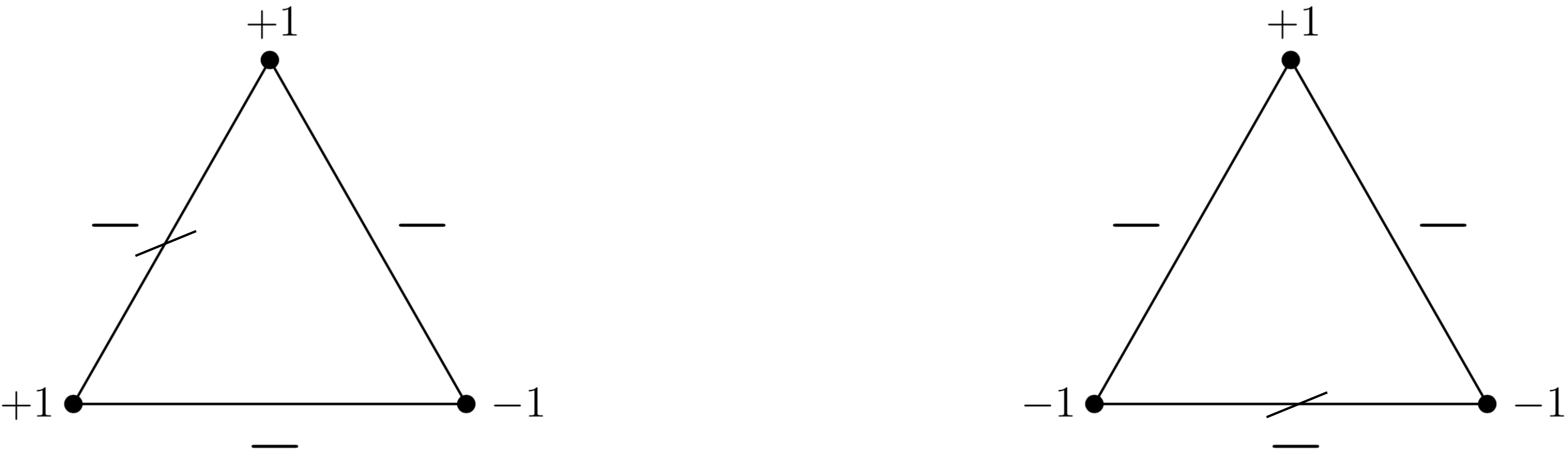}
		\caption{Frustration on a triangular lattice: there is no possible way of selecting the spins without breaking a coupling. All couplings are anti-ferromagnetic. Broken couplings are indicated with a slash.}\label{fig:frustration2}
	\end{center}
\end{figure}

We now have to choose the state of the third spin: we are using Ising representations and therefore we have two choices, namely $\{-1,+1\}$. It is easy to see that by selecting $\sigma=+1$ we break the coupling with the top spin, whereas the choice $\sigma=-1$ breaks the coupling with the lower, right hand-side spin.\\

An immediate immediate consequence of frustration is that the system has no unique ground state. Conversely, there exist many states with the same energy and therefore the energy function defined on the phase space has many minima: this representation of the energy function is usually called \emph{energy landscape}. The complexity of the energy landscape is one of the most appalling traits of the spin glass systems and can sometimes become really exotic \cite{garstecki1999energy} (see also section \ref{sec:NP}).

\subsection{Sherrington-Kirkpatrick model}\label{sub:SKmodel}
In 1975 David Sherrington and Scott Kirkpatrick introduced a new model for of spin glass that they were able to solve analytically in an appropriate temperature range \cite{PhysRevLett.35.1792}.\\

We may ask in which sense they solved the spin glass model: by \emph{solving} we mean \emph{finding the order parameter}. The order parameter of a spin glass was defined by Edwards and Anderson \cite{edwards1975theory} as:
\begin{equation}\label{eq:qeadef}
	q_{EA} = \lim_{t\rightarrow\infty} \overline{\left<\sigma_i(0)\sigma_i(t)\right>}
\end{equation}
in which the angle brackets denotes the thermal or dynamical average and the bar denotes the average over the randomly chose coupling constant. Indeed there are other definitions of the Edward-Anderson parameter \cite{sommers1982static}, but we will refer to the one in equation (\ref{eq:qeadef}) because it's the most common and the most convenient to calculate through  Monte Carlo simulations \cite{PhysRevB.17.4384}.\\

Sherrington and Kirkpatrick defined their system to be a set of $N$ Ising spins with an infinite-range interaction\footnote{We note that the requirement of infinite range is equivalent to assume a infinite-dimensional hyper-cubic lattice} in the form:
\begin{equation}\label{eq:SKham}
	H = -\sum_{i\neq j} J_{ij}\sigma_i\sigma_j
\end{equation} 
with the coupling constants extracted from a gaussian distribution:
\begin{equation}
	P\left(J_{ij}\right) = \frac{1}{\sqrt{2\pi}J}\,e^{-\frac{{\left(J_{ij}-J_0\right)}^2}{2J^2} }
\end{equation}

This will be used from now on as a reference model for the optical model that will be proposed in Chapter \ref{Ch:theory_optics}.\\

We now finish this general introduction with some examples of physical situations in which the spin glass model is useful to study the system under consideration. The model was originally developed to describe magnetic alloys and glass forming liquids \cite{binder1986spin}. The framework has seen substantial development since then and today it is successfully applied in the most diversified fields and topics, such as the study of colliods \cite{zaccarelli}, quantum glasses \cite{mueller2004glass}, random lasers \cite{angelani2006glassy}. Also outside statistical mechanics the spin glass mathematical framework is used to model - for example - neural networks \cite{amit1992modeling} and economical models \cite{minniti2004entrepreneurial}.

\section{Spin glass simulations}
The phase space $\Omega$ of an Ising spin glass composed of $N$ spins is composed by $2^N$ possible configuration. Thus a direct solution of the problem by the calculation of the partition function is out of question, even for relatively small system size.

\subsection{Metropolis-Hastings algorithms}\label{sec:metrhast}
The aim to the Metropolis-Hasting algorithm is to produce a Markov chain whose states are the possible configuration of the phase space $\Omega$ and its asymptotic probability $\Pi(x)$ is the Standard Boltzmann distribution, namely:
\begin{equation}
	\Pi(x) \propto e^{-\beta H(x)}
\end{equation}

We want the Markov chain to satisfy the \emph{detail balance principle}, namely to be represented by a transition matrix $\mymatrix{P}_{xy}$, satisfying:
\begin{equation}
	{\Pi}_x\,{P}_{xy} = {\Pi}_y\,{P}_{yx}\quad \forall\, x,y \in \Omega
\end{equation}

The Metropolis algorithm is composed of two steps, the \emph{proposal} and the \emph{acceptance}. The proposal defines the efficiency of the algorithm, while the acceptance is defined by the algorithm itself. The change in configuration from the state $x$ to the state $y$ happens with a total probability:

\begin{equation}
	P_{xy} =\left\{
     \begin{array}{ll}
      P^{(0)}_{xy}\,A_{xy} & \mbox{if } x \neq y \\
      1-\sum_y P^{(0)}_{xy}\,A_{xy} & \mbox{if } x = y
     \end{array}\right.	
\end{equation}
in which $P^{(0)}_{xy}$ is the proposal probability, the probability that the final state $y$ is selected being in the current state $x$, and $A_{xy}$ is the acceptance probability, the probability that the suggested configuration change actually takes place.\\

We now want to enforce the detail balance condition. Namely, we obtain:
\begin{equation}
	\Pi_x\,P^{(0)}_{xy}\,A_{xy} = \Pi_y\,P^{(0)}_{yx}\,A_{yx}
\end{equation}
or, equivalently:
\begin{equation}\label{condition_fordb}
	\frac{A_{xy}}{A_{yx}}=\frac{\Pi_y\,P^{(0)}_{yx}}{\Pi_x\,P^{(0)}_{xy}}
\end{equation}
if the proposal is \emph{symmetric} (that is: $P^{(0)}_{yx}=P^{(0)}_{xy}$ $\forall x,y \in \Omega$), we can simplify it in equation (\ref{condition_fordb}). We instead stick to the general case and define:
\begin{equation}\label{Metroplis_eq}
	R_{xy}\coloneqq\frac{\Pi_y\,P^{(0)}_{yx}}{\Pi_x\,P^{(0)}_{xy}}=\frac{A_{xy}}{A_{yx}}
\end{equation}
in which the first part is a \emph{defintion} of the matrix $\mymatrix{R}$ which defines the ratios between each pair of states in the Markov chain. The second part is instead a matrix equation to be solved for the matrix $\mymatrix{A}$. It is easy to see that the matrix $\mymatrix{R}$ has the following property:
\begin{equation}
	R_{xy}=\frac{\Pi_y\,P^{(0)}_{yx}}{\Pi_x\,P^{(0)}_{xy}}={\left[\frac{\Pi_x\,P^{(0)}_{xy}}{\Pi_y\,P^{(0)}_{yx}}\right]}^{-1}=\frac{1}{R_{yx}}
\end{equation}

There are several choices of $\mymatrix{A}$ which solve equation (\ref{Metroplis_eq}). The \emph{Metropolis choice} is obtained by choosing:
\begin{equation}
	A_{xy} = \mbox{min}\left(1,R_{xy}\right)
\end{equation}

Firstly, we want to see that the Metropolis choice satisfy equation (\ref{Metroplis_eq}). It is possible to show that this is also the best choice \cite{metropolis1953equation}, in the sense that it gives the maximum acceptance ratio and therefore wastes less time in rejecting moves. To prove the correctness of the Metropolis choice, we need to distinguish two cases:
\begin{itemize}
	\item if $R_{xy}\leq1$, by using the property for $\mymatrix{R}$, we get $R_{xy}=\frac{1}{R_{yx}}>1$. In such case we have:
	\begin{align}
		A_{xy} &= \mbox{min}\left(1,R_{xy}\right)=R_{xy}\\
		A_{yx} &= \mbox{min}\left(1,R_{xy}\right)=1\\
	\end{align}
	and equation (\ref{Metroplis_eq}) becomes:
	\begin{equation}
		\frac{A_{xy}}{A_{yx}}=R_{xy} \quad\Rightarrow\quad \frac{R_{xy}}{1}=R_{xy}
	\end{equation}
	\item if $R_{xy}>1$, by using the property for $\mymatrix{R}$, we get $R_{xy}=\frac{1}{R_{yx}}>leq1$. In such case we have:
	\begin{align}
		A_{xy} &= \mbox{min}\left(1,R_{xy}\right)=1\\
		A_{yx} &= \mbox{min}\left(1,R_{xy}\right)=R_{yx}\\
	\end{align}
	and equation (\ref{Metroplis_eq}) becomes:
	\begin{equation}
		\frac{A_{xy}}{A_{yx}}=R_{xy} \quad\Rightarrow\quad \frac{1}{R_{xy}}=\frac{1}{\frac{1}{R_{xy}}}=R_{xy}
	\end{equation}
\end{itemize}
We immediately note that for the asymptotic probability distribution to be the Boltzmann distribution:
\begin{equation}
	\Pi_x \propto e^{\beta H(x)s}
\end{equation}
we get
\begin{equation}
	P_{xy}=\frac{\Pi_y\,P^{(0)}_{yx}}{\Pi_x\,P^{(0)}_{xy}}=\frac{P^{(0)}_{yx}}{P^{(0)}_{xy}} e^{-\beta(H(y)-H(x))}
\end{equation}
By choosing the proposal probability to be symmetric, we can simplify it in all the equations, obtaining:
\begin{equation}
	P_{xy}=e^{-\beta(H(y)-H(x))}
\end{equation}
We then can write the Metropolis choice for the acceptance matrix $\mymatrix{A}$ as:
\begin{equation}\label{eq:Metropchoice}
	A_{xy} = \mbox{min}\left(1,e^{\beta(H(x)-H(y))}\right)
\end{equation}

It is important for the aim of the present work to notice that the only required condition to ensure the Metropolis algorithm produces the correct Boltzmann distribution is that \emph{the proposal probability is symmetric}.\\

The Metropolis algorithm for Ising spin glass is usually implement in such a way that the proposal move is nothing but a spin flip. Namely, a spin is randomly chosen and the state of the spin is reversed.
Two configurations in phase space $x,y \in \Omega$ will have a proposal probability:
\begin{equation}
	P^{(0)}_{xy} =\left\{
     \begin{array}{ll}
      0 & \mbox{if } x \mbox{ and } y \mbox{ differ by no spin}\\
      0 & \mbox{if } x \mbox{ and } y \mbox{ differ by two or more spins}\\
      \frac{1}{N} & \mbox{if } x \mbox{ and } y \mbox{ differ by exactly one spin}\\
     \end{array}\right.	
\end{equation}
Hence the proposal probability is intrinsically symmetric.

\subsection{Multi spin flips and cluster algorithms}\label{sec:msf}
As it was remarked in the previous section, the fundamental property for easily obtain the stationary Boltzmann distribution is to perform a symmetric proposal.

This requirement does not involve the number of spins in the proposed configuration change. If we flip $n$ random spins among the $N$ spins that constitute the spin glass at each proposal, we connect the configuration in phase space as follow:
\begin{equation}
	P^{(0)}_{xy} =\left\{
     \begin{array}{ll}
      \frac{1}{\binom{N}{n}} & \mbox{if } x \mbox{ and } y \mbox{ differ by exactly $n$ spin}\\
      0 & \mbox{otherwise}
     \end{array}\right.	
\end{equation}
hence the Metropolis algorithm is correct no matter how many spins are involved in the proposed move.

We should note, however, that changing more spins in the same step does not relate to a faster algorithm. Two states separated by one accepted move will be \emph{more different} (more uncorrelated); on the other hand, the suggested move will be rejected more often, as it will imply a greater change in the overall energy difference.\\

It is possible, however to choose a \emph{cluster} of $N$ spins to be flipped together during the trial flip, in such a way that the probability of accepting the flip is higher than the probability of flipping $N$ spins chosen randomly and independently.\\

The \emph{Swendsen-Wang algorithm} \cite{swendsen1987nonuniversal,wang1990cluster,barbu2005generalizing} construct the cluster to be flipped by assigning to each pair of nearest neighbour spins $\sigma_n,\sigma_m$ a random variable $b_{n,m}$ in such a way that $b_{n,m}=0$ means that the two spins will be connected in the resulting cluster, while  $b_{n,m}=1$ means there is no connection. The $b$ variables are then assigned to each nearest neighbour pair in the following way:
\begin{equation}
	\left\{
     \begin{array}{l}
      P\left[b_{n,m}=0|\sigma_n\neq\sigma_m\right] = 1\\
      P\left[b_{n,m}=1|\sigma_n\neq\sigma_m\right] = 0\\
      P\left[b_{n,m}=0|\sigma_n=\sigma_m\right] = e^{2\beta J_{nm}}\\
      P\left[b_{n,m}=1|\sigma_n=\sigma_m\right] = 1-e^{2\beta J_{nm}}\\
     \end{array}\right.
\end{equation}
in which $J_{nm}$ is the coupling constant for the given pair.\\

Once this variables have been set, the clusters are established and the proposal step consist in the flipping of all variables in the cluster with probability $1/2$. We note that after the assignment of the variables $b_{n,m}$ has to be repeated at every step, thus generating different clusters every time. The advantage of this approach is near the critical point: while the standard Metropolis algorithm slows down because of the  correlation length divergence, the Swedensen-Wang algorithm is robust. In fact, the divergence is related to the creation of percolation clusters, which in this algorithm are flipped as a single spin, greatly reducing the relaxation time.\\

The \emph{Wolf algorithm} choses the cluster as the set of neighbouring spins sharing the same value of the spin and this allows a larger probability of flipping bigger clusters in a move \cite{wolff1989collective}, compared to the Swendsen-Wang algorithm. This is considered the best cluster algorithm to beat the dynamical slow down near the critical point \cite{ferrenberg1992monte}.

\section{The spin autocorrelation function}
Given an Sherrington Kirkpatrick model, we perform a Monte Carlo simulation on the system. Hence we will have the spin set $\{\sigma_1^{(0)},\sigma_2^{(0)},...,\sigma_N^{(0)}\}$ at the beginning of the simulation. We then can save the state of the system at each Monte Carlo step and we will have:
\begin{align}
\notag	&\{\sigma_1^{(1)},\sigma_2^{(1)},...,\sigma_N^{(1)}\}\\
\notag	&\{\sigma_1^{(2)},\sigma_2^{(2)},...,\sigma_N^{(1)}\}\\
\notag	& \quad \vdots\\
\notag	&\{\sigma_1^{(N)},\sigma_2^{(N)},...,\sigma_N^{(N)}\}
\end{align}

We then use all the saved configurations to calculate the following quantity:
\begin{equation}
	C(t) = \frac{1}{N}\,\frac{1}{T-t}\,\sum_{i=1}^N \,\sum_{\tau=1}^{T-t}\sigma_i(\tau)\,\sigma_i(t+\tau)
\end{equation}

Since this last equation correspond to average the correlation of a single spin over all spins of the system and over all configurations reached during the Monte Carlo dynamics, the average encompass a great number of spin values and it becomes a correct estimator for the spin correlation function:
\begin{equation}\label{eq:spin_acorrf}
	C(t)  = \frac{1}{N}\sum_{i=1}^N\left<\sigma_i(0)\sigma_i(t)\right>
\end{equation}

Hence it is possible to \emph{estimate the Edward Anderson order parameter throught the self autocorrelation function}. In fact, in the limit of many measurements over different coupling parameters $J_{ij}$ (extracted from the same probability distribution), we average over the disorder and, recalling equation (\ref{eq:qeadef}), we obtain:
\begin{equation}
	C(t)  \underset{t \to \infty}{=} q_{EA}
\end{equation} 


\begin{figure}[!ht]
  \centering
  \begin{subfigure}[b]{\linewidth}
    \centering\includegraphics[width=0.7\linewidth]{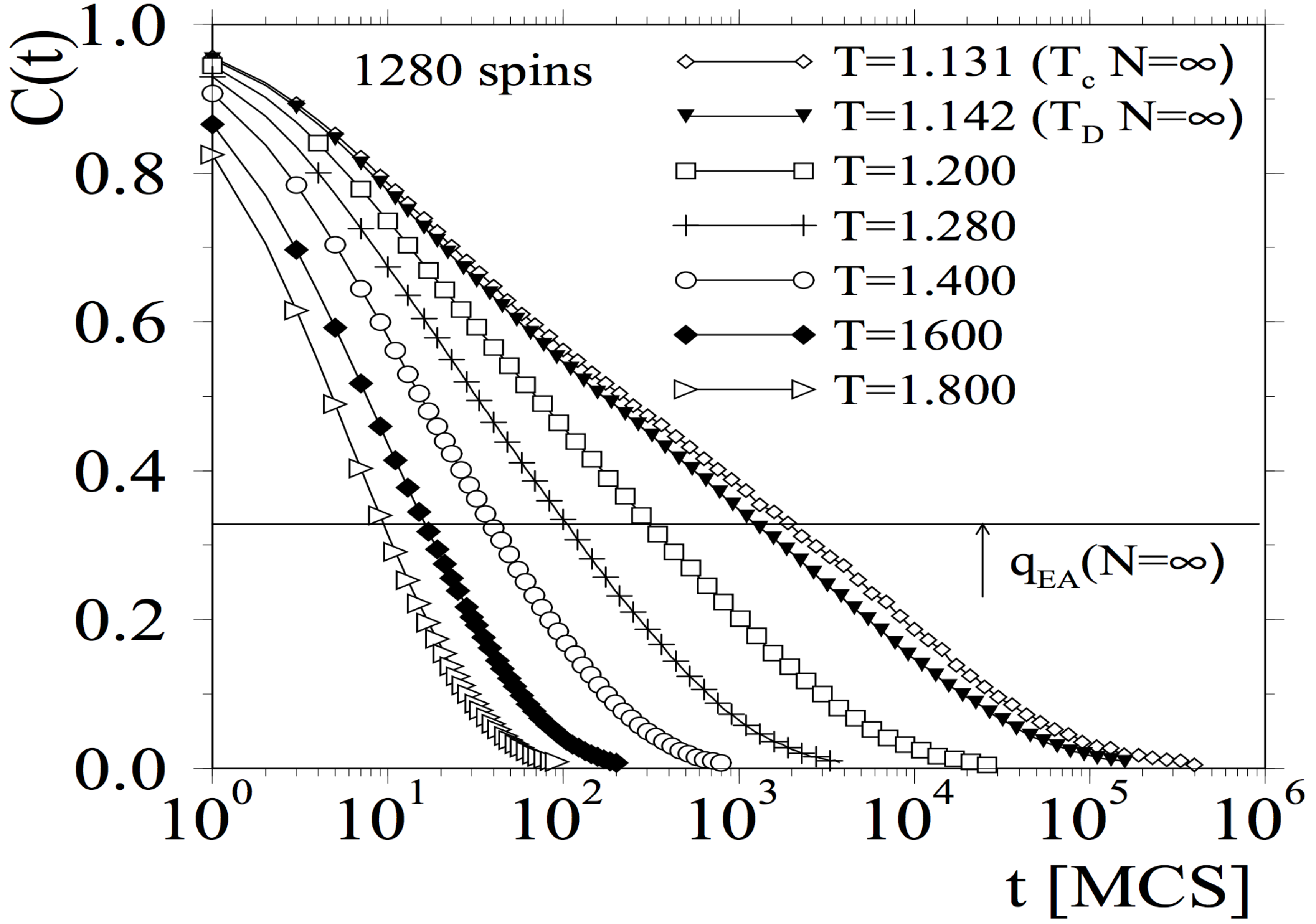}
    \caption{\label{fig:acorr1}Spin autocorrelation function $C(t)$ versus $t$ (measured in units of Monte Carlo steps per spin) in a mean field Potts spin glass. Figure from reference \cite{brangian2002high}}
  \end{subfigure}\\\vspace{0.2cm}
  \begin{subfigure}[b]{\linewidth}
    \centering\includegraphics[width=0.7\linewidth]{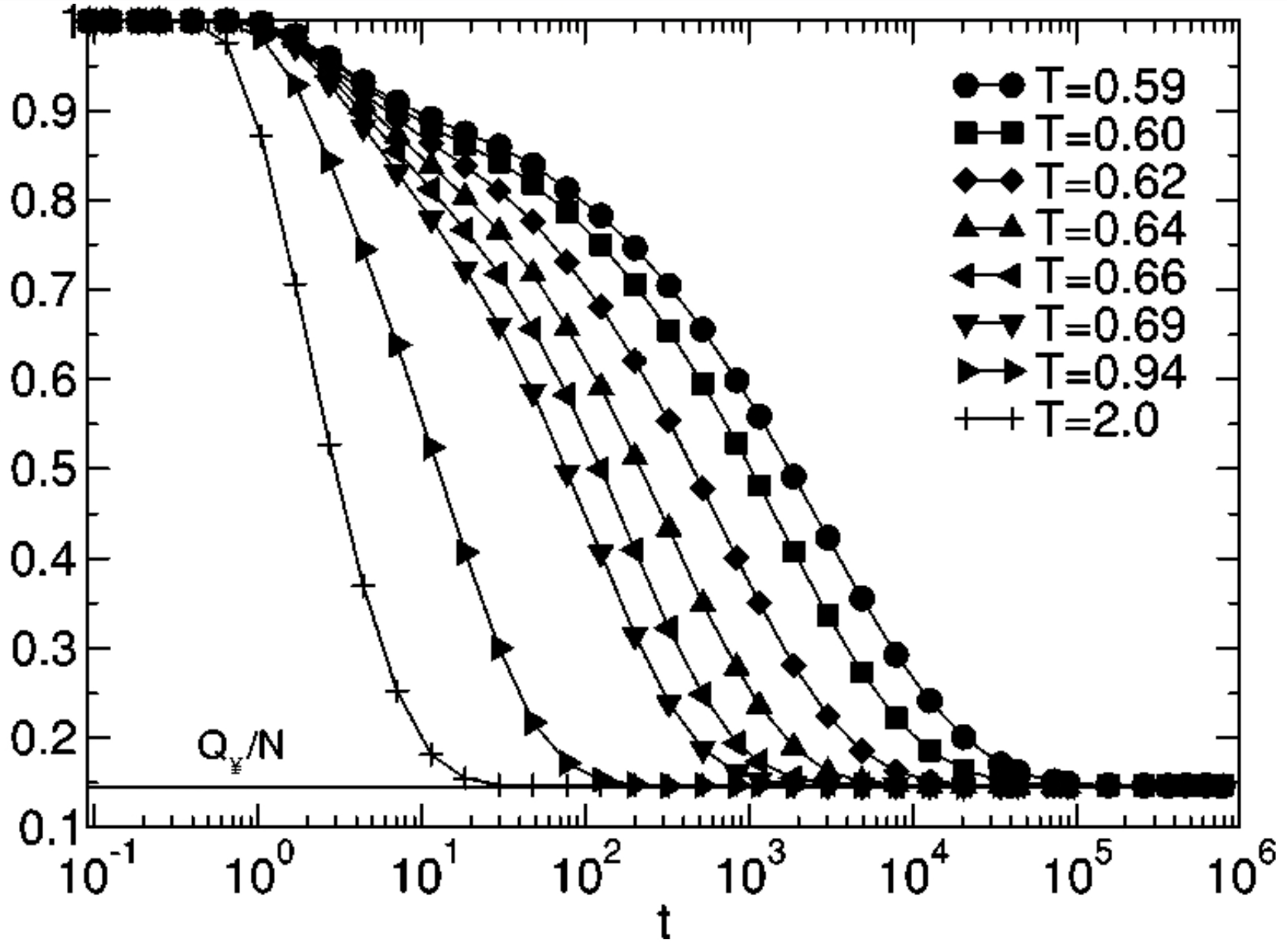}
    \caption{\label{fig:acorr2}Intermediate scattering function obtained in the study of the relaxation time in supercooled liquids through a time-dependent four-point density correlation function. Figure from reference \cite{lavcevic2003spatially}}
  \end{subfigure}\hspace{0.05\linewidth}
  \caption{Comparison between a spin autocorrelation function for a spin glass system (\subref{fig:acorr1}) and an intermediate scattering function for a supercooled liquid (\subref{fig:acorr2}).}
  \label{fig:acorr}
\end{figure}

This approach is widely used, and produces an interesting similarity that can explain why the spin systems we are dealing with are usual called spin \emph{glasses}. In glass theory a standard quantity is the \emph{intermediate scattering function}; for a glass made by $N$ particle it is defined as:
\begin{equation}
	F\left(\bm{k},t\right) = \frac{1}{N} \sum_{i=1}^N \sum_{j=1}^N \left<e^{i\bm{k}\cdot\left(\bm{r}_i(t)-\bm{r}_j(0)\right)}\right>
\end{equation}
which can be rewritten in terms of the the Fourier transform of the microscopic density \cite{hansen1990theory} as follows:
\begin{equation}\label{eq:hansen}
	F\left(\bm{k},t\right) = \frac{1}{N}\left<\rho_{\bm{k}}(t)\rho_{-\bm{k}}(0)\right>
\end{equation}

In this form, it's quite obvious the similarity between equations (\ref{eq:spin_acorrf}) and (\ref{eq:hansen}). Indeed, the two function has the same meaning in the two theories and behaves very similarly, as shown in figure \ref{fig:acorr}. The two equation are exactly the same only in the case of the \emph{p-spin model} \cite{kirkpatrick1987p,crisanti1993sphericalp,castellani2005spin}, but this is one example of glassy behaviour that justify the very sam name of \emph{spin glasses}.

\section{Brief concepts on scaling in spin glass problems}\label{sec:NP}

The direct approach of finding the partition function of an Ising spin glass is out of question, since the number $N_C$ of possible configurations grows exponentially as:
\begin{equation}
	N_C=2^N
\end{equation}

A simpler problem might be the search for the ground state, but also this has been proven to be a NP-hard problem if the graph on which the system is defined is non planar \cite{barahona1982computational,Yucesoy_2007,cipra2000ising}.\\

In the context of Monte Carlo simulations, if we take the Sherrington-Kirkpatrick model as reference, we immediately see that the energy calculation requires to solve the sum in equation (\ref{eq:SKham}), which has exactly $N(N-1)$ terms. From this property, we immediately obtain that the scaling of a Monte Carlo simulation for the Sherrington-Kirkpatrick model is at best of order $O(N^2)$. The energy calculation, needed at every Monte Carlo step for the acceptance probability, is the most computationally expensive part of the problem. Indeed, all of the other parts of the algorithm scale linearly in the number of spins.
\chapter{The theory of the ``optical spin glass''}\label{Ch:theory_optics}

\epigraph{Although it is true that it is the goal of science to discover rules which permit the association and foretelling of facts, this is not its only aim. \\It also seeks to reduce the connections discovered to the smallest possible number of mutually independent conceptual elements.}{\textit{Albert Einstein, Science and Religion (1941)}}

We consider the spin glass Hamiltonian in equation (\ref{eq:general_Hamiltonian}). Suppose we can write the coupling matrix $\mymatrix{J}$ as a complex product of two tensor:
\begin{equation}
	\mymatrix{J}=\left(\mymatrix{M},\mymatrix{M}\right)=\mymatrix{M}^*\,\mymatrix{M}
\end{equation}
then the Hamiltonian is expressible as a product among the vectors obtained by applying the tensor $\mymatrix{M}$ to the vector $\bm\sigma$, encoding the entire state of the system:
\begin{equation}\label{Hamiltonian_dyadic}
	\left. H = \left<\bm{\sigma}\,\right|\mymatrix{M}^*\mymatrix{M}\left|\bm{\sigma}\right>=\left<\mymatrix{M}\bm{\sigma}\right|\mymatrix{M}\bm{\sigma}\right>
\end{equation}

If we now suppose that the spin variables of the system are real, which is always the case in standard spin glass models, equation (\ref{Hamiltonian_dyadic}) is a scalar product of complex vectors. This is the very product used in optics to calculate the intensity of an electric field. The main idea behind this work is to transfer the relation between the spins of the system and the Hamiltonian to a suitable optical configuration. To do so, we need a system acting like the matrix $\mymatrix{M}$, which scramble the light and make the light beams from various sources to interact in a dyadic way as in equation (\ref{Hamiltonian_dyadic}).\\

The rest of this chapter is devoted to find a suitable optical system to simulate a spin glass and the theoretical proofs that the chosen model actually work.

\section{Speckle patterns}
When the first HeNe laser was created in 1960, scientists had a coherent light source to work with. Almost immediately, it was noted \cite{speckle_discovery1,speckle_discovery2} that most object appeared granulated when illuminated with coherent light. The phenomenon was known since the time of Newton, but the laser made it much simpler to observe. This  optical pattern, made of fine-resolved light spots which have a very high contrast with respect to the dark areas outside the spots, was called \emph{speckle pattern}. Also, the phenomenon is not limited to the visible light, but happens at any wavelength. See some examples in Figure \ref{specklepatterns}.\\

Such a pattern is due to light interference: the problem is the complex nature of the speckle, whose intensity map is linked with the phase difference of each light path that creates the speckle, relative to all the others. By far the most common situation in which it is observed is the case in which the interference is due to the scattering of the light caused by the reflection or transmission by a turbid medium. Two ingredients are therefore  necessary: firstly, the incident beam must be composed by \emph{coherent light}, (otherwise no interference is possible); furthermore, the reflective surface must be rough, otherwise the reflection happens geometrically, following the well-known \emph{law of reflection} $\theta_{inc}=\theta_{refl}$, and no scrambling happens to the incident beam.\\

\begin{figure}[ht]
\begin{center}
	\includegraphics[height=8cm]{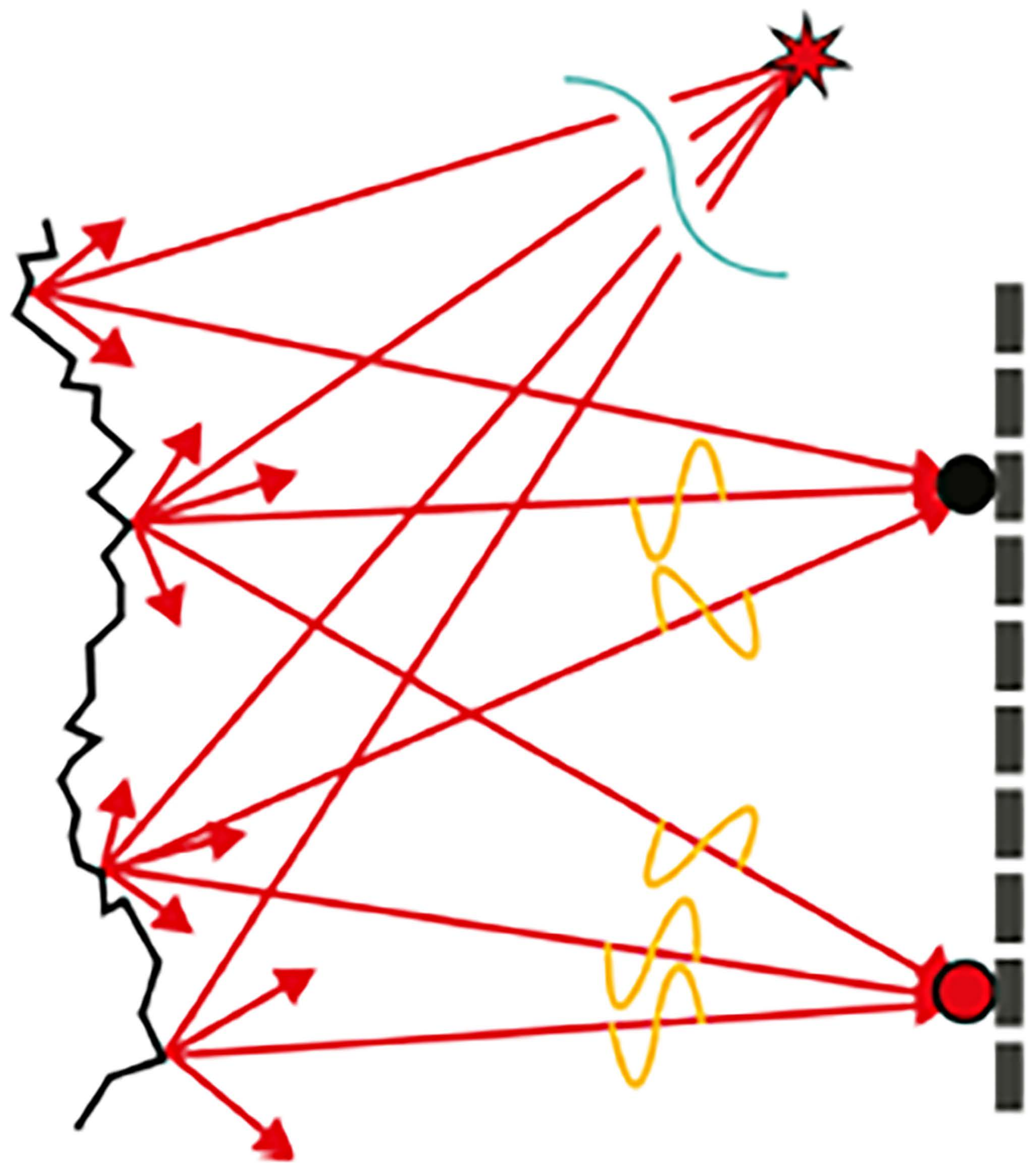}
	\caption{Graphical representation of the creation of the speckle pattern by a rough surface. The upper point of the screen receives light beams out of phase, resulting in little or no light intensity. The lower point receives coherent light beams and the measured intensity is at a maximum.}\label{speckle_draw}
\end{center}
\end{figure}

Another important aspect of speckle patterns is the distinction between a \emph{subjective} speckle pattern and an \emph{objective} speckle pattern. The first term is used to refer to a speckle pattern which is formed in the image plane of a lens. On the other hand, when a rough object is illuminated by a coherent wave, we have the \emph{objective speckle pattern} and this will always be the case from now on. Such speckle pattern will depend on the interference between the waves from various scattering points, and its size will increase linearly with the distance $d$ between the observation plane and an object which creates it.\\

It is important to notice, however, that while the size will only depend on the distance $d$ mentioned above, the specific shape of the pattern will depend also on:
\begin{itemize}
\item the wavelength $\lambda$ of the light
\item the size of the laser beam which illuminates the scattering surface: it is common to assume it perfectly cylindrical with radius $R$ and characterised by an intensity which only depends on the radial distance from the centre of the beam
\item the aforementioned distance $d$ between the observation plane and the scattering surface
\end{itemize}

The dependence on the wavelength $\lambda$ is due to the fact that the particular structure of the speckle pattern depends on the detailed travel path of the photons inside - or rather along the surface of - the scattering target. Changing $\lambda$ will change the size $L$ of the speckle according to the formula (see \cite{BookDainty}):
\begin{equation}
	L = \frac{\lambda\,R}{d}
\end{equation}
and therefore also the specific photon path changes.
\clearpage

\begin{figure}
  \centering
  \begin{subfigure}[b]{0.45\linewidth}
    \centering\includegraphics[width=0.99\linewidth]{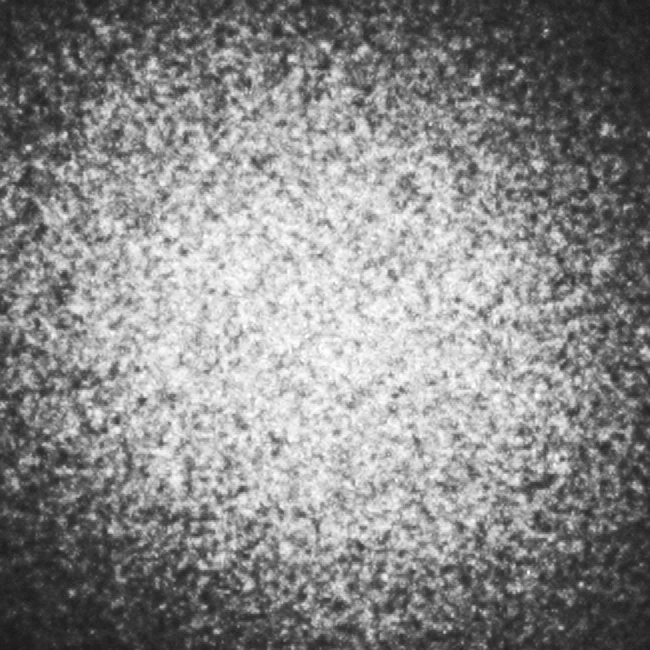}
    \caption{\label{fig:fig1}Speckle generated by a laser beam reflected by a rough sample\\}
  \end{subfigure}\hspace{0.05\linewidth}
  \begin{subfigure}[b]{0.45\linewidth}
    \centering\includegraphics[width=0.99\linewidth]{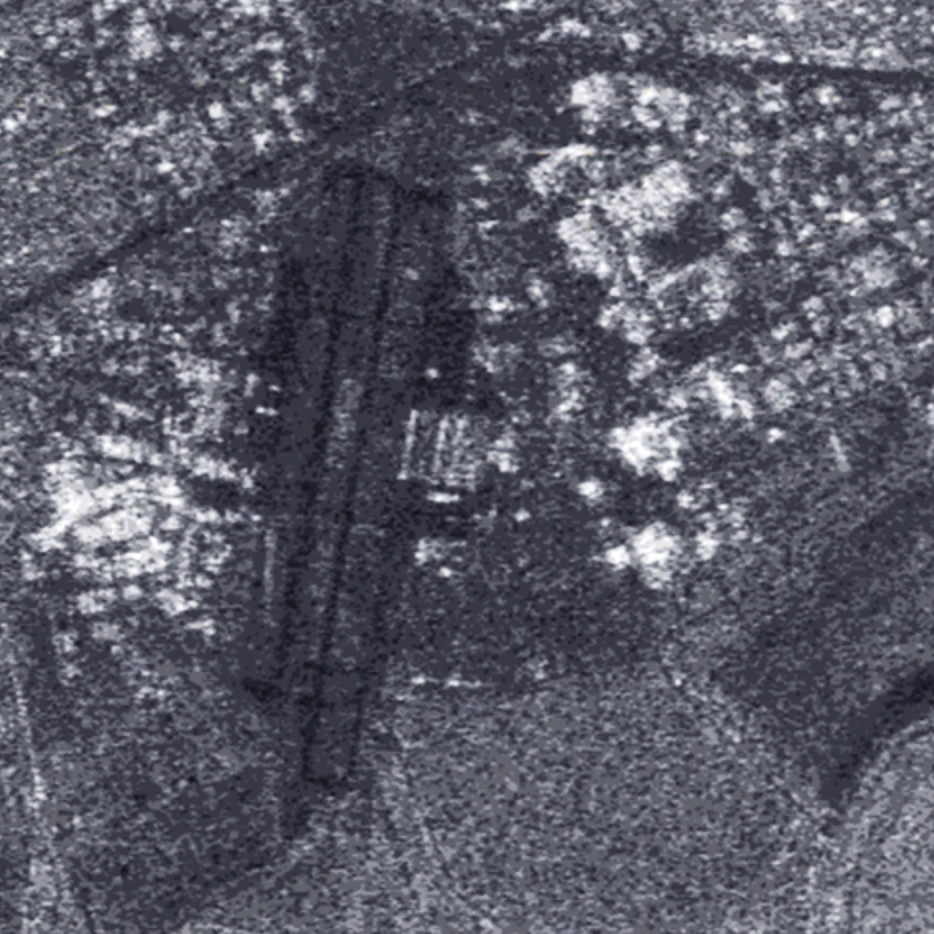}
    \caption{\label{fig:fig2}A speckled image of Moffet field in California, taken by a radar with radar wavelength of 5.76 cm.}
  \end{subfigure}\\\vspace{0.2cm}
  \begin{subfigure}[b]{0.45\linewidth}
    \centering\includegraphics[width=0.99\linewidth]{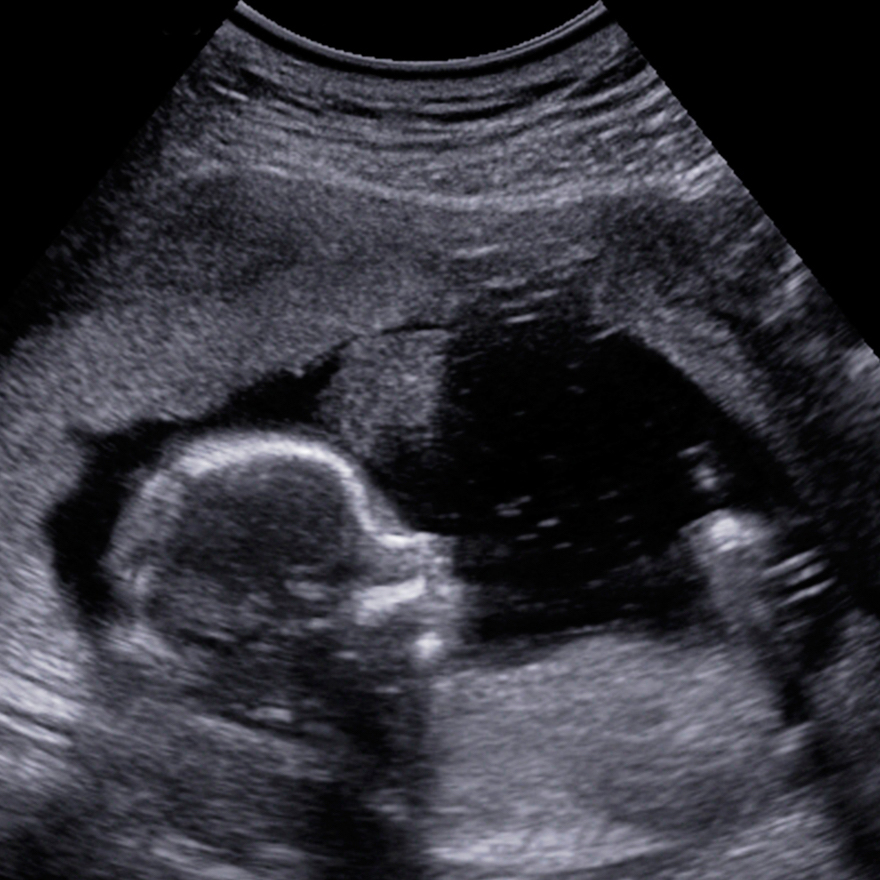}
    \caption{\label{fig:fig3}Ultrasound image of a 5 weeks human fetus. Speckles are clearly visible in the entire image}
  \end{subfigure}\hspace{0.05\linewidth}
  \begin{subfigure}[b]{0.45\linewidth}
    \centering\includegraphics[width=0.99\linewidth]{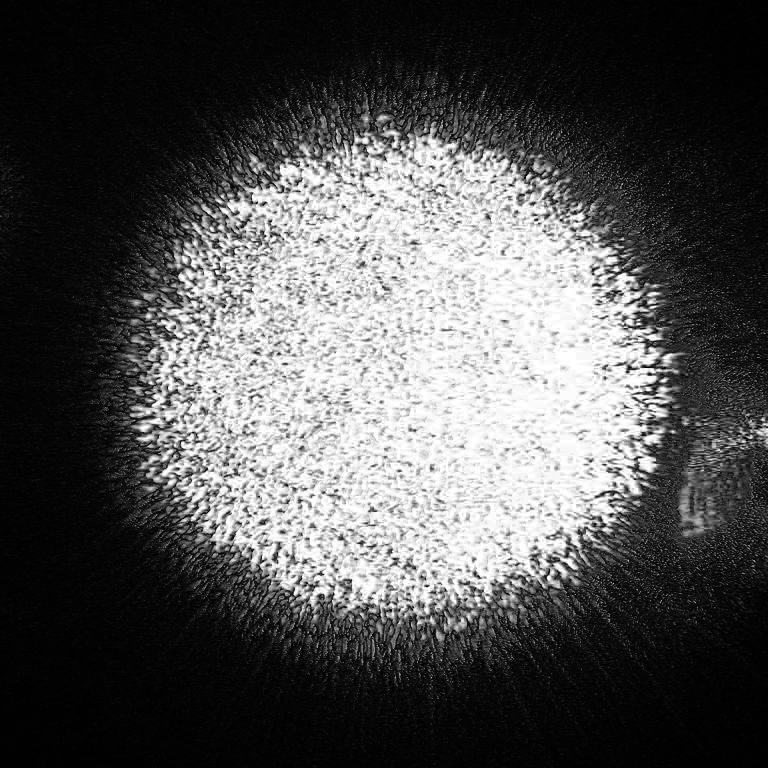}
    \caption{\label{fig:fig4}Speckle pattern formed by a laser beam directly pointed towards the camera that took the picture}
  \end{subfigure}
  \caption{Different examples of speckle patterns appearing in different situations. Images (\subref{fig:fig1}) (\subref{fig:fig2}) and (\subref{fig:fig3}) are examples of \emph{objective speckle pattern}, while (\subref{fig:fig4}) is a \emph{subjective speckle pattern}}
  \label{specklepatterns}
\end{figure}
\clearpage

\subsection{Mathematical description of speckle pattern}
To represent more conveniently an electromagnetic wave, in this section \emph{phasors} are used. This is noting but a complex representation of the wave, which is obtained by suppressing the positive components of the frequency spectrum of the wave and doubling the negative ones \cite{BookGoodman}. To obtain the frequency spectrum, the Fourier transform is used:
\begin{equation}
	\mathcal{F}[f](\nu) = \int_{-\infty}^\infty f(x)e^{-i\nu x} dx
\end{equation}
this allows us to write, for example, a cosinusoidal wave with a rotating complex vector:
\begin{equation}
	\cos(2\pi\nu_0t) \quad \rightarrow \quad e^{-i2\pi\nu_0t}
\end{equation}

Since the speckle pattern is due to the interference of scattered waves and the waves are essentially independent from one another, we will be looking at the statistical properties of the sums of many random phasors, each with its own amplitude and phase. Furthermore, we will be interested in the local properties: the specific intensity (which is related to the magnitude) and phase in a given point of space. Mathematically, we can write a phasor sum as follows:
\begin{equation}\label{phasor_sum}
	\bm{A} = Ae^{i\theta} = \frac{1}{\sqrt{N}}\sum_{n=1}^N\bm{a}_n= \frac{1}{\sqrt{N}}\sum_{n=1}^N a_n e^{i\phi_n}
\end{equation}
in which $\bm{A}$ is the resulting complex phasor, $A$ its associated magnitude and $\theta$ its associated phase, while $\bm{a}_n$ are the independent complex phasors, $a_n$ their associated magnitude and $\phi_n$ their associated phases. The rescaling factor $1/\sqrt{N}$ is needed to keep the intensity finite also when the number of phasors to sum approaches infinity.

We then separate the real and imaginary part, and write:
\begin{align}
	\mathcal{R} &= \mbox{Re}(\bm{A})= \frac{1}{\sqrt{N}}\sum_{i=1}^N\,a_n \cos\phi_n\\
	\mathcal{I} &= \mbox{Im}(\bm{A})= \frac{1}{\sqrt{N}}\sum_{i=1}^N\,a_n \sin\phi_n
\end{align}

We now want to assume that the phasors in the sum are independent. This assumption, called the hypothesis of \emph{statistical independence}, can be expressed as follows:
\begin{itemize}
	\item the amplitude $a_n$ and the phase $\phi_n$ of each phasor is uncorrelated with the ones of other phasors. This means that it is not possible to obtain information of one phasor by the knowledge of another phasor
	\item for a given phasor, its amplitude $a_{\tilde{n}}$ and its phase $\phi_{\tilde{n}}$ are uncorrelated, hence it is not possible to obtain information on the latter by having information on the previous, nor vice-versa
	\item each phase $\phi_n$ is uniformly distributed in the interval $(-\pi,\pi)$
\end{itemize}
these are quite general assumptions and satisfactory in all cases of the present work. For some example of generalisations, see \cite{exotic_phasors_sum,BookGoodman}. This is a mathematical model representing the speckle pattern, the random phasors being the light waves scattered from the surface scrambling the light. The uncorrelation hypothesis are due to the fact that different waves producing the speckle comes form different portions of the scattering medium. 

We therefore want to extract some properties out of the model: we start from the average values. We can write:
\begin{align}
	\left<\mathcal{R}\right> &=\left<\frac{1}{\sqrt{N}}\sum_{i=1}^N\,a_n \cos\phi_n\right> = \frac{1}{\sqrt{N}}\sum_{i=1}^N\left<a_n \cos\phi_n\right>=\\
\notag	&= 	\frac{1}{\sqrt{N}}\sum_{i=1}^N\left<a_n\right>	\left<\cos\phi_n\right>=0\\
	\left<\mathcal{I}\right> &=\left<\frac{1}{\sqrt{N}}\sum_{i=1}^N\,a_n \sin\phi_n\right> =\frac{1}{\sqrt{N}}\sum_{i=1}^N\left<a_n \sin\phi_n\right>=\\
\notag	&= 	\frac{1}{\sqrt{N}}\sum_{i=1}^N\left<a_n\right>	\left<\sin\phi_n\right>=0
\end{align}
in which we use the uncorrelation between the phase and the amplitude to separate he average of the product into the product of the averages, and the uniform distribution of the phase $\phi$ to obtain a zero average of $\sin\phi$ and $\cos\phi$. Hence, the electric field which constitute the speckle pattern has a zero average. This result is not unexpected, as the pattern we see is an intensity pattern. Therefore, we need to calculate the average of the \emph{amplitude} of the electric field, to which the intensity is proportional. 

To proceed further and give some information about the intensity, we need to assume that the number of phasor we are summing is very high. This way, we can apply the \emph{central limit theorem} to equation (\ref{phasor_sum}): the sum of these $N$ independent random variables becomes a gaussian in the limit of $N \rightarrow\infty$. In such limit, we can focus only on the variance of the amplitude distribution, and neglect all higher-order momenta.

\begin{align}
\notag	\sigma_{\mathcal{R}}^2&=\left<\mathcal{R}^2\right>-{\left<\mathcal{R}\right>}^2=\left<\mathcal{R}^2\right>= \frac{1}{N}\sum_{i=1}^N\sum_{j=1}^N\left<a_ia_j\cos\phi_i\cos\phi_j\right>=\\
	&=\frac{1}{N}\sum_{i=1}^N\sum_{j=1}^N\left<a_ia_j\right>\left<\cos\phi_i\cos\phi_j\right>=\frac{1}{N}\sum_{i=1}^N\left<a_i^2\right>\left<\cos\phi_i^2\right>=\\
\notag	&=\frac{1}{N}\sum_{i=1}^N\left<a_i^2\right>\left<\frac{1}{2}+\frac{1}{2}\cos\left(2\phi_i\right)\right>=\frac{1}{N}\sum_{i=1}^N\frac{\left<a_i^2\right>}{2}\\\notag
\end{align}
\begin{align}
\notag \sigma_{\mathcal{I}}^2&=\left<\mathcal{I}^2\right>-{\left<\mathcal{I}\right>}^2=\left<\mathcal{I}^2\right>= \frac{1}{N}\sum_{i=1}^N\sum_{j=1}^N\left<a_ia_j\sin\phi_i\sin\phi_j\right>=\\
&=\frac{1}{N}\sum_{i=1}^N\sum_{j=1}^N\left<a_ia_j\right>\left<\sin\phi_i\sin\phi_j\right>=\frac{1}{N}\sum_{i=1}^N\left<a_i^2\right>\left<\sin\phi_i^2\right>=\\
\notag	&=\frac{1}{N}\sum_{i=1}^N\left<a_i^2\right>\left<\frac{1}{2}+\frac{1}{2}\sin\left(2\phi_i\right)\right>=\frac{1}{N}\sum_{i=1}^N\frac{\left<a_i^2\right>}{2}
\end{align}
in performing these latter calculations, we have neglected the terms in which $i\neq j$ because of the uncorrelation hypothesis. In fact, we have:
\begin{equation}
	\left<\cos\phi_i\cos\phi_j\right>=\left<\cos\phi_i\right>\left<\cos\phi_j\right>=0
\end{equation}
For the very same hypothesis, it is immediate to show that there is no correlation between the real and the imaginary part:
\begin{align}
\notag	\Gamma_{\mathcal{R},\mathcal{I}}&=\left<\mathcal{R}\mathcal{I}\right>-\left<\mathcal{R}\right>\left<\mathcal{I}\right>= \left<\mathcal{R}\mathcal{I}\right>=\frac{1}{N}\sum_{i=1}^N\sum_{j=1}^N\left<a_ia_j\cos\phi_i\sin\phi_j\right>=\\
&=\frac{1}{N}\sum_{i=1}^N\sum_{j=1}^N\left<a_ia_j\right>\left<\cos\phi_i\sin\phi_j\right>=\\
\notag &=\frac{1}{N}\sum_{i=1}^N\sum_{j=1}^N\left<a_i\right>\left<a_j\right>\left<\cos\phi_i\right>\left<\sin\phi_j\right>=0
\end{align}

By applying the central limit theorem to equation (\ref{phasor_sum}) and using the calculation just performed, we can calculate the joint probability for the real and imaginary parts of the resulting phasor:
\begin{equation}
	p_{\mathcal{R},\mathcal{I}}\left(\mathcal{R},\mathcal{I}\right)= \frac{1}{2\pi\left(\sigma_{\mathcal{R}}^2+\sigma_{\mathcal{I}}^2\right)}\;\mbox{exp}\left\{-\frac{\mathcal{R}^2+\mathcal{I}^2}{2\left(\sigma_{\mathcal{R}}^2+\sigma_{\mathcal{I}}^2\right)}\right\}
\end{equation}
hence the joint distribution of the real and imaginary part of the phasor created by the superposition of a great number of independent phasors is a multivariate gaussian.\\

As it has been mentioned, we are interested in the mean intensity $I$ of the speckle pattern. This is defined as the squared amplitude of the resulting phasor. We do our calculation in two steps, firstly calculating the probability distribution for the amplitude $A$ and then - through a second change of variables - the probability distribution for the intensity $I$.

To pass from $(\mathcal{R},\mathcal{I})$ to the amplitude and the phase $(A,\theta)$ we need to perform the following change of variables:
\begin{equation}
     \left\{
     \begin{array}{rl}
      A &= \sqrt{\mathcal{R}^2+\mathcal{I}^2}\\
     \theta &= \arctan\left(\frac{\mathcal{I}}{\mathcal{R}}\right)
     \end{array}
     \right.
\end{equation}
The corresponding relation between the two probability distribution is:
\begin{equation}
	p_{A,\theta}\left(A,\theta\right)= p_{\mathcal{R},\mathcal{I}}\left(A\cos\theta,A\sin\theta\right)\left\lVert J\right\lVert
\end{equation}
in which $\left\lVert J\right\lVert$ is the determinant of the Jacobian of the transformation, namely:
\begin{equation}\label{eq:changeofvariable}
	\left\lVert J\right\lVert = \left\lVert 
	\begin{array}{cc}
		\frac{\partial\mathcal{R}}{\partial A}	&	\frac{\partial\mathcal{R}}{\partial\theta}\\
		\frac{\partial\mathcal{I}}{\partial A}	&	\frac{\partial\mathcal{I}}{\partial\theta}
	\end{array}
	\right\lVert =A
\end{equation}
This allows us to explicitly calculate the joint probability function for the amplitude and the phase of the resulting phasor:
\begin{equation}\label{eq:jointAt}
	p_{A,\theta}\left(A,\theta\right)=\frac{A}{2\pi\sigma^2}\;\mbox{exp}\left\{-\frac{A^2}{2\sigma^2}\right\}
\end{equation}
To obtain the probability distribution for the amplitude $A$ alone we have to marginalise the joint probability distribution: that is, we integrate equation (\ref{eq:jointAt}) over $\theta$:
\begin{equation}\label{eq:prob_amplitude}
	p_A\left(A\right)=\int_{-\pi}^{+\pi}p_{A,\theta}\left(A,\theta\right)\,d\theta = \frac{A}{\sigma^2}\;\mbox{exp}\left\{-\frac{A^2}{2\sigma^2}\right\}
\end{equation}

Now we are ready to calculate the probability distribution of the intensity. To do so, we need to use a change of variable like in equation (\ref{eq:changeofvariable}). However, this time we need to pass from the  variable $A$ to its square value $I=A^2$. Therefore this time the transformation involves scalar quantities only, and equation (\ref{eq:changeofvariable}) can be simplified as follows. In general, given the scalar random variable $\chi$ with a probability distribution $p_{\chi}(\chi)$, we perform a scalar change of variable by defining a new scalar random variable $\xi$ through the relation:
\begin{equation}
	\xi = f(\chi)
\end{equation}
and we want to calculate the probability distribution $p_{\xi}(\xi)$ of the newly defined variable in terms of the probability distribution of the old variable. The general relation becomes:
\begin{equation}
	p_{\xi}(\xi) = p_{\chi}(f^{-1}(\xi))\left|\frac{d\chi}{d\xi}\right|
\end{equation}
In our specific case we set $\xi=I$, $\chi=A$ and $I=A^2$ and we obtain:
\begin{equation}
	p_I(I) = p_A\left(\sqrt{I}\right)\left|\frac{dA}{dI}\right|
\end{equation}
By plugging in the functional form in equation (\ref{eq:prob_amplitude}) we obtain the probability density for the intensity of the speckle pattern:
\begin{equation}
	p_I(I)=\frac{\sqrt{I}}{\sigma^2}\;\mbox{exp}\left\{-\frac{I}{2\sigma^2}\right\}
\end{equation}

Having the probability distribution, we can easily calculate the average value of the intensity of the speckle pattern:
\begin{equation}
	\left<I\right>= 2\sigma^2
\end{equation}
which expresses the average intensity of the speckle pattern as a function of the variance of the single scattering element that constitutes the speckle pattern.\\

Being the probability distribution an exponential decay, it's actually easy to calculate also higher order momenta:
\begin{equation}
	\left<I^q\right>= \left(2\sigma^2\right)^qq!= \left<I\right>^qq!
\end{equation}
We can rewrite the intensity probability distribution as:
\begin{equation}\label{eq:fspeckleint}
	P_I(I)=\frac{1}{\left<I\right>}\,\mbox{exp}\left\{-\frac{I}{\left<I\right>}\right\}
\end{equation}
a speckle pattern which satisfy (\ref{eq:fspeckleint}) is often called \emph{fully developed speckle pattern}.\\

Two other important parameters of the speckle pattern - especially in the experimental framework - are the \emph{contrast} $C$, defined as:
\begin{equation}
	C=\frac{\sigma_I}{\left<I\right>}
\end{equation}
and its inverse, often called \emph{signal-to-noise ratio}:
\begin{equation}
	\mbox{SNR} = \frac{\left<I\right>}{\sigma_I}
\end{equation}
In short, the \emph{contrast} is a comparative measure between the fluctuations of the intensity in the speckle pattern and its average intensity; the signal-to-noise ratio is the inverse concept. We note that for fully developed speckle patterns we get:
\begin{equation}
	\mbox{SNR} = 1 = C
\end{equation}

\section{The dipole propagator}
After having described - both experimentally and theoretically - the speckle pattern, we want to extract a method to explicitly calculate the intensity of the scattered field in a given point of the space: this is obtained through the theory of the \emph{dipole propagator}\cite{DipolePropagatorBook1,DipolePropagatorBook2,bornwolf,Ruocco_dipole}.\\

Rayleigh firstly gave the theoretical explanation of the light scattering for a simple particle in 1881, proving that the the intensity of the light scattered by non-interacting particles is proportional tho the inverse fourth power of the wavelength. This result was originally meant to explain the reason for the sky to appear blue, and was therefore derived under the assumption - totally justified for the Earth atmosphere - that the scattering molecules have a mean free path much larger than the wavelength of the light involved in the scattering process.\\

This is not true in the case of a scattering dense medium. In such case, indeed, the mean free path of the molecules is actually smaller than the light wavelength and therefore it is not possible to treat them independently: the intensities of the \emph{microfields} (the elementary fields scattered by the single particles) no longer combine and a different description is needed. The fundamental difference is that the microfields themselves combine, rather than their intensity. Mathematically speaking, the difference between the two case comes from the order in which we apply the square absolute value and the summation over the elementary fields (see Table \ref{scattering_conpound}).

\begin{table}[ht]
\begin{center}
	{\renewcommand{\arraystretch}{1.5}
	\begin{tabular}{c|cc}
	Scattering & Intermolecular distance $d$ & Scattered intensity\\
	\hline
	\hline
	Independent molecules  & $d>>\lambda$ & ${I}_{{sc}}=\sum_i\,I_i = \sum \left|E_i\right|^2 $\\
	Dense medium  & $d	 \leq \lambda$ & I = $\left|\bm{E}^{(sc)}\right|^2$ = $\left|\sum E_i \right|^2$ 
	\end{tabular}}
\caption{Differences between the two cases of multi-molecular light scattering}\label{scattering_conpound}
\end{center}
\end{table}

\subsection{Multipole expansion in the simple case of one molecule}
We are interested in the case of a dense scattering medium, for which we have showed that the total resulting scattered field $\bm E_{sc}$ is the superposition of the \emph{microfields} generated by the single particles. We want to start from the calculation of the single particle contribution. \\\null\\We begin by writing Maxwell equations:
\begin{equation}\label{maxwelleqns}
	\left\{
     \begin{array}{l}
     \bm\nabla\times\bm E = -\frac{1}{c} \frac{\partial \bm H}{\partial t}  \\
     \bm\nabla\times\bm H = \frac{4\Phi}{c}\bm{j} + \frac{1}{c} \frac{\partial \bm E}{\partial t} \\
     \bm\nabla\cdot\bm E = 4\Phi \rho  \\
     \bm\nabla\cdot\bm H = 0
     \end{array}
     \right.
\end{equation}

By standard notation, the equations are given in Gaussian units and $\bm E$ is the electric field, $\bm H$ is the magnetic field, $c$ is the velocity of light in vacuum, $\rho$ and $\bm{j}$ are the density of charge and the current density respectively.

The two latter quantities are related by the \emph{continuity equation} for the conservation of the charge; namely, the following equation holds:
\begin{equation}
	\frac{\partial\rho}{\partial t} +\bm\nabla\cdot\bm j=0
\end{equation}
It is possible to express both $\rho$ and $\bm j$ in terms on just one vector $\bm P$, called the \emph{polarizability vector}, as:
\begin{equation}
	\rho=-\bm\nabla\cdot\bm P \qquad \bm j = \frac{\partial\bm P}{\partial t}
\end{equation}

\noindent Substituting this last result in Maxwell equations, we get:
\begin{equation}
	\left\{
     \begin{array}{l}
     \bm\nabla\times\bm E = -\frac{1}{c} \frac{\partial \bm H}{\partial t}  \\
     \bm\nabla\times\bm H = \frac{4\Phi}{c}\frac{\partial\bm P}{\partial t} + \frac{1}{c} \frac{\partial \bm E}{\partial t} \\
     \bm\nabla\cdot\bm E = 4\Phi\, \bm\nabla\cdot\bm P \\
     \bm\nabla\cdot\bm H = 0
     \end{array}
     \right.
\end{equation}
which automatically satisfies the continuity equation.\\

When written in this form, Maxwell equations has an explicit solution in terms of the \emph{Hertz vector} $\bm\Phi$. Details of this theory can be found -for example - in \cite{essex1977hertz,sein1989solutions}; for our purpose we only need the explicit expressions of $\bm E(\bm\Phi)$ and $\bm H(\bm\Phi)$ and the consistency equation for $\bm\Phi$, which are:
\begin{equation}\label{maxwellHertzsolution}
	\left\{
     \begin{array}{l}
     \bm E = \bm\nabla\left(\bm\nabla\cdot\bm\Phi\right)-\frac{1}{c^2}\frac{\partial^2\bm\Phi}{\partial t^2}\\
     \bm H = \frac{1}{c}\bm\nabla \times \frac{\partial\Phi}{\partial t}
     \end{array}
     \right.
\end{equation}
\begin{equation}\label{Hertzconsistency}
	\nabla^2\bm\Phi-\frac{1}{c^2}\frac{\partial^2\bm\Phi}{\partial t^2}= -4\Phi \bm P
\end{equation}
The explicit expression for $\bm\Phi$ can be found through the \emph{method of Green function}. This is a widely used method and the details can be found in \cite{PhysRevLett.57.2168}.\\

However, we do not want to obtain an exact solution: in fact, this can be done in this simple case of just one molecule, but is doomed to fail in the more general case of a scattering medium composed by a great number of molecules, the one we are actually interested in. We want, instead, perform an \emph{expansion in series of multipoles}. For such expansion to be valid, we need be in the limit of large wavelength with respect to the linear mean distance $l$ among the charges:
\begin{equation}\label{multipolehp}
	\frac{\omega l}{c}=\frac{2\pi l}{\lambda} \ll 1
\end{equation}
which must hold for all significant frequencies in the spectrum of the scattered light. Note that this is indeed the case of visible light and small molecules. Therefore, we can expand the Fourier transform of the Hertz vector:
\begin{equation}
	\tilde{\bm{\Phi}}_{\bm{\omega}}(\bm r) = \int_V P_{\bm{\omega}} (\bm{r'}) \frac{e^{i\bm k \cdot \bm R}}{R}d\bm{r'}
\end{equation}
in which we have defined $\bm R = \bm r -\bm{r'}$ and $\bm{\omega}=\frac{\bm k}{c}$.

If we now retain only the first two term of the expansion, we get:
\begin{equation}\label{Hertzexpansion}
	\tilde{\bm{\Phi}}_{\bm{\omega}}(\bm r)= \tilde{\bm{\Phi}}_{\bm{\omega}}^{(0)}(\bm r)+\tilde{\bm{\Phi}}_{\bm{\omega}}^{(1)}(\bm r)+...
\end{equation}

We want to investigate the dependence of these two terms with respect to the ones coming from the expansion of $\bm P$. Such dependence has the form:
\begin{equation}\label{multipole_ord1}
\bm{\Phi}_{\bm\omega}^{(0)}=\,\bm{P}_{\bm\omega}^{(1)}\frac{e^{i\bm{k}\cdot\bm{r}}}{r}
\end{equation}
\begin{equation}
\bm{\Phi}_{\bm\omega}^{(1)}=\left(\bm{M}_{\bm\omega}^{(1)} \times \bm{e}_r -\frac{i\omega}{2c} \bm{P}_{\bm\omega}^{(2)} \cdot \bm{e}_r\right)\left(\frac{1}{r}+\frac{i}{kr^2}\right)e^{i\bm{k}\cdot\bm{r}}\label{multipole_ord2}
\end{equation}
In which $\bm{e}_r$ is the versor relative to the vector $\bm{r}$ and the explicit formula of the three multipoles is:
\begin{align}
	\bm{P}_{\bm\omega}^{(1)}&=\int_V\bm{r'}\,\rho_{\bm{\omega}}\,d\bm{r'}\label{multipole_p1}\\
	\bm{M}_{\bm\omega}^{(1)}&=\frac{1}{2c}\int_V\bm{r'}\times\bm{j}\left(\bm{r'}\right)\,d\bm{r'}\label{multipole_m1}\\
	\bm{P}_{\bm\omega}^{(2)}&=\int_V \bm{r'}\cdot\bm{r'}\,\rho_{\bm{\omega}}\, d\bm{r'}\label{multipole_p2}
\end{align}

\subsection{The dipole approximation}
We have calculated the multipole expansion for a single particle. In the limit of equation (\ref{multipolehp}), the expansion reconstruct the exact solution as the number of retained terms increases. If we let this happen, we would encounter the very same problem we came across when trying to solve equations (\ref{maxwellHertzsolution}) and (\ref{Hertzconsistency}) with the method of Green function: the solution, however exact, would be too difficult to generalise to the case of interest, namely when there is a great number of scattering particles.\\

We want to retain only the quantities $\bm{P}_{\bm\omega}^{(1)}$, $\bm{P}_{\bm\omega}^{(2)}$ and $\bm{M}_{\bm\omega}^{(1)}$. Even further, we show in this section that \emph{only the dipole term can be retained} under some reasonable assumptions, making the solution easily applicable to the case of the scattering dense medium. Therefore, we now want to compare the relative magnitude of the three terms above.\\

To do so, we firstly need to consider some approximations. Firstly, the field arriving on a single particle will consist by two contributes: one will be the wave arriving directly by the source, the other will be the contribution from all the microfields scattered by the other molecules. It is important to notice that, despite the fact that we are considering the fields scattered from other particles, this theory still describes only the filed produced by a single particle. That being said, we want to enforce some approximation on both contributions:
\begin{itemize}
	\item the incoming wave is assumed to be generated by a perfectly monochromatic and time-coherent source, that is: $\bm{E}^{(in)} = \bm{E}_0\,e^{i\left(\bm{k}\cdot\bm{r}-\omega_0t\right)}$ 
	\item the microfields, on the other hand, could in principle undergo a doppler effect if the molecules from which are generated are in motion: this would introduce a non trivial term, different from one molecule to the other. We therefore make the assumption - well satisfied by standard molecules at room temperature - that the scattering medium is \emph{nonrelativistic}. This way, the frequency shift of the microfields is negligible and they can be assumed to have the same pulsation $\omega_0$ of the incoming wave.
\end{itemize}

Furthermore, we want to distinguish between \emph{permanent} dipole (quadrupole) momentum and \emph{induced} dipole (quadrupole) momentum. In general, both contributions determine the elements in the multipole expansion. Instead, it is assumed from now on that the \emph{scattered radiation depends only on the induced part}. In fact, the scattered field is generated by the molecule in response to a perturbation in the form of a monochromatic wave with pulsation $\omega_0$. For the permanent multipoles to influence the produced field, a physical rotation of the molecules is needed. However, for $\omega_0$ big enough, the molecule can not follow the rapid variation of the perturbation. This is indeed the case in presence of the visible light, when the pulsation is in of order of some terahertz: $\omega_0=O(10^{14}\,\mbox{Hz})$.\\

The direct consequence of the aforementioned property is that the scattered light depends only on the deviation of the multipoles from their permanent value. This deviation is considered to be \emph{linear}. This assumption induces multipoles having the same pulsation as the incoming wave.\\

Under these assumption, it is relatively simple to use \emph{Drude-Lorentz model} to calculate the quantities $\bm{P}_{\bm\omega}^{(1)}$, $\bm{P}_{\bm\omega}^{(2)}$ and $\bm{M}_{\bm\omega}^{(1)}$. We consider the electron to be bounded by an elastic force in a position $\bm{r}_0$ with respect to center of the positive charge. The force acting on the electron is therefore:
\begin{equation}
	\bm{F}_e=-F_0\,\left(\bm{r}-\bm{r}_0\right)
\end{equation}
the electron is also subject to the incoming wave $\bm{E}^{(in)} = \bm{E}_0\,e^{i\left(\bm{k}\cdot\bm{r}-\omega_0t\right)}$ and the resulting motion will satisfy the standard equation for Drude-Lorentz model:
\begin{equation}
	m_e\frac{d^2\bm{r}}{dt^2}+\gamma\frac{d\bm{r}}{dt}+s\left(\bm{r}-\bm{r}_0\right)= e\,\bm{E}_0\,e^{-i\omega_0t}
\end{equation}
in which $\omega_s=\sqrt{s/m_e}$ is the pulsation associated to the spring that binds the electron to the postion $\bm{r}_0$.
The solution can be explicitly expressed as:
\begin{equation}
	\bm{r}(t)=\bm{r}_0+\frac{e}{m_e}\frac{\bm{E}_0\,e^{-i\omega_0t}}{\left(\omega_s^2-\omega_0^2\right)+i\omega_0\gamma}
\end{equation}

This will be the value for $\bm{r}$ to be plugged in into equations (\ref{multipole_p1}) (\ref{multipole_m1}) and (\ref{multipole_p2}). By performing those integrals we obtain:
\begin{align}
	\bm{P}_0^{(1)}=&\,\frac{e^2}{m_e}\frac{1}{(\omega_s^2-\omega_0^2)+i\omega_0\gamma}\bm{E}_0\\
	\bm{M}_0=&\frac{i\omega_0}{2c}\bm{P}_0^{(1)}\times\bm{r}_0\\
	\bm{P}_0^{(2)}=&\bm{P}_0^{(1)}\bm{r}_0+\bm{r}_0\bm{P}_0^{(1)}
\end{align}

in which we are using the standard notation for tensors:
\begin{equation}
	\left[\bm{P}_0^{(1)}\bm{r}_0\right]_{ij} = \left(\bm{P}_0^{(1)}\right)_i\,\left(\bm{r}_0\right)_j
\end{equation}

Now that we have the explicit expression of these three quantities, we can plug them back in equations (\ref{multipole_p1}) and (\ref{multipole_p2}), which are hereby copied for the sake of commodity:
\begin{equation}
\bm{\Phi}_0^{(0)}=\,\bm{P}_{0}^{(1)}\frac{e^{i\bm{k}\cdot\bm{r}}}{r} \tag{\ref{multipole_p1}}
\end{equation}
\begin{equation}
\bm{\Phi}_{0}^{(1)}=\left(\bm{M}_{0}^{(1)} \times \bm{e}_r -\frac{i\omega}{2c} \bm{P}_{0}^{(2)} \cdot \bm{e}_r\right)\left(\frac{1}{r}+\frac{i}{kr^2}\right)e^{i\bm{k}\cdot\bm{r}} \tag{\ref{multipole_p2}}
\end{equation}

From these latter equations, we can extract the relative magnitude of each term of $\bm{\Phi}_{0}^{(1)}$ with respect to $\bm{\Phi}_{0}^{(1)}$. By a direct check, we find:
\begin{align}
	&\left(\bm{M}_{0}^{(1)} \times \bm{e}_r -\frac{i\omega}{2c} \bm{P}_{0}^{(2)} \cdot \bm{e}_r\right)	\frac{1}{r}\propto\frac{r_0}{\lambda_0} \bm{\Phi}_{0}^{(1)}\\
	&\left(\bm{M}_{0}^{(1)} \times \bm{e}_r -\frac{i\omega}{2c} \bm{P}_{0}^{(2)} \cdot \bm{e}_r\right)\frac{i}{kr^2}\propto\frac{r_0}{r}\bm{\Phi}_{0}^{(1)}
\end{align}
both of the terms on the left are negligible with respect to $\bm{\Phi}_{0}^{(1)}$. In fact, for the first one have the relation:
\begin{equation}
	\frac{r_0}{\lambda_0} \ll \frac{r_0}{l} \ll 1
\end{equation}
in which the inequalities follows from the fact that we have assumed the wavelength much larger than the linear dimension $l$ of the molecules to perform the multipole expansion: see equation (\ref{multipolehp}). Furthermore, the  typical distance of the electron from the center of the positive charge $r_0$ is reasonably be much smaller than the average intermolecular distance $l$, hence the result.

For the second term, on the other hand, we simply notice that
\begin{equation}
\frac{r_0}{r}\leq\frac{r_0}{l_0}
\end{equation}
and the thesis follows straightforwardly for this.


\subsection{Explicit expression of the dipole propagator}\label{sec_dipoleporp}
We have proved that the only relevant term to the multipole expansion of the Hertz vector $\bm{\Phi}$ is the dipole contribution. If we write explicitly such expansion retaining only the relevant term and recall equation (\ref{Hertzexpansion}), we get:
\begin{align}
\notag	\bm{\Phi}\left(\bm{r},t\right) \approx&\,  \bm{\Phi}_0^{(0)}= \bm{P}_{0}^{(1)}\frac{e^{i\bm{k}\cdot\bm{r}}}{r}= \frac{e^2}{m_e}\,\frac{1}{\omega_s^2-\omega_0^2+i\omega\gamma} \bm{E}_0\, e^{-i\omega_0t}\frac{e^{i\bm{k}\cdot\bm{r}}}{r}=\\
	=&\,\alpha(\omega_0)\,\bm{E}_0\,e^{-i\omega_0t}\,\frac{e^{i\bm{k}\cdot\bm{r}}}{r}
\end{align}
Now that we have the Hertz vector, we can calculate the electromagnetic field by using equations (\ref{maxwellHertzsolution}). We are only interested in the electric field, which will be expressed as:
\begin{align}
\notag	\bm{E}^{(sc)}\left(\bm{r},t\right)=&\bm\nabla\left(\bm\nabla\cdot\bm\Phi\left(\bm{r},t\right)\right)-\frac{1}{c^2}\frac{\partial^2\bm\Phi\left(\bm{r},t\right)}{\partial t^2}=\\
	=&\,\frac{e^2}{m_e}\,\frac{1}{\omega_s^2-\omega_0^2+i\omega\gamma}\left(\bm{\nabla}\,\bm{\nabla}-\frac{1}{c^2}\frac{\partial^2}{\partial t^2}\right)\left(\bm{E}_0\,e^{-i\omega_0t}\,\frac{e^{i\bm{k}\cdot\bm{r}}}{r}\right)=\\
	=&\,\frac{e^2}{m_e}\,\frac{1}{\omega_s^2-\omega_0^2+i\omega\gamma}\left(\bm{\nabla}\,\bm{\nabla}-\frac{1}{c^2}\frac{\partial^2}{\partial t^2}\right)\,\bm{E}^{(in)}\left(\bm{r},t\right)=\\
	=&\,\alpha(\omega_0,\omega_s,\gamma)\left(\bm{\nabla}\,\bm{\nabla}+k_0^2\right)\,\bm{E}^{(in)}\left(\bm{r},t\right)\label{complete_formula_with_prefactor_dependence}
\end{align}
in which we have called the incoming plane, monochromatic wave $\bm{E}^{(in)}$ and in the last step we used the fact that the partial derivative acts only on the time exponential.\\

Equation (\ref{complete_formula_with_prefactor_dependence}) defines a tensor, which allows us to express the scattered radiation as a function of the incoming radiation: because we obtained it under a set of assumptions that made us retain only the dipole term in the multipole expansion, we called it \textbf{dipole propagator}. Its explicit expression is:

\begin{equation}
	\hat{\bm{T}} = \left(\bm{\nabla}\,\bm{\nabla}+k_0^2\;\hat{\bm{I}}\,\right)
\end{equation}
in which we have called $\hat{\bm{I}}$ the identity tensor. We can also write the relation by components:
\begin{equation}
	\hat{T}_{ij}=\left(\frac{\partial}{\partial r_i}\frac{\partial}{\partial r_j}+k_0^2\,\delta_{ij}\right)
\end{equation}

Using the dipole propagator, we can explicitly write the scattered field both in vectorial form and by components:
\begin{align}
	\bm{E}^{(sc)}\left(\bm{r},t\right) =& \alpha(\bm{r})\,\hat{\bm T}\,\bm{E}^{(in)}\left(\bm{r},t\right)\label{scattered_as_func_of_inc}\\
	E_i^{(sc)}\left(\bm{r},t\right)=& \alpha(\bm{r})\,\sum_j\hat{T}_{ij}\,E_{j}^{(in)}\left(\bm{r},t\right)\label{scattered_as_func_of_inc_comp}
\end{align}


\section{The Hamiltonian of the optical spin glass}\label{theory_opticalspinglass}
The aim of this section is to show that the intensity of the wave front after the scattering sample can be written - in each point - with a dyadic, bilinear expression which has the form:
\begin{equation}\label{dyadic_result_beforeproof}
	I = \sum_{i,j}\,J_{ij}\sigma_i \sigma_j
\end{equation}
in which each $\sigma_i=\{0,1\}$, is the variable associated to the state of the $i-th$ spin, as seen in chapter \ref{Ch:spin_glass_theory}. \\\null\\
We assume that the incident wave is a monochromatic plane wave, travelling - for example - along the $z$ direction; its associated electric field can be written as:
\begin{equation}
\bm{E}^{(in)} = \bm{\epsilon}\,E_0\,e^{i\left(kz-\omega t\right)}
\end{equation}

we also assume that the emitting source is capable of generate a perfectly coherent wave, both in time and space.\\

We now want to screen part of the wavefront of the incoming beam. Experimentally, the most common method to do so is through a reflection on a digital micromirror device (see chapter \ref{Ch:experimental_setup} for the details). The mathematical modelling is way simpler: a variable $\sigma(x,y)$ is defined, which takes only values in $\{0,1\}$. This way, the points for which $\sigma(\bm{x})=0$ are effectively screened: there is no electrical field there to reach the scattering medium. For all other points, namely those for which $\sigma(\bm{x})=1$, are left unaffected by such operation. The screened wave will be expressed as:
\begin{equation}
\bm{E}^{(refl)}(x,y)=\bm{\epsilon}\,E_0\,e^{i\omega t}\,e^{ikz}\,\sigma\left(x,y\right)
\end{equation}

We now want to discretize the wavefront of the incoming wave. Namely, we fix a set of $\{x_1,x_2,...,x_N\}$, $\{y_1,y_2,...,y_N\}$ such that:
$$ \bm{r} \in \left[x_i^{min},x_i^{max}\right]\times\left[y_i^{min},y_i^{max}\right]\quad $$
The following identification for $\bm{E}_0\left(\bm{r}_i\right)$ applies:
\begin{equation}\label{discretization}
\bm{E}(\bm{r},t)=\bm{\epsilon}\,E_0\,e^{i\omega t}\,e^{ikz_i}\,\sigma\left(\bm{r}\right)\;\longrightarrow\;\bm{E}_i(t)=\bm{\epsilon}\,E_0\,e^{i\omega t}\,e^{ikz_i}\,\sigma_i
\end{equation}
in which now we want to prove that the variables $\sigma_i=\{0,1\}$ can be identified as the ones that appears in equation (\ref{dyadic_result_beforeproof}). This discretization is natural in the experimental framework, as the digital micromirror device is composed of pixels. However, once again we postpone this discussion to chapter \ref{Ch:experimental_setup}.  \newline

We need to enforce the \emph{autoconsistency} equations for the electric field in a scattering medium, namely the relation:
\begin{equation}\label{autoconsistency_naive}
\bm{E}_1=\bm{E}_0+\sum_i\,\{\mbox{scattered spherical waves } S_i\}
\end{equation}
which expresses the resulting electric field in terms of all the fields generated in the sample in the scattering process.
More formally, the condition (\ref{autoconsistency_naive}) can be written in terms of the \emph{dipole propagator} as seen in section \ref{sec_dipoleporp}.
We recall the expression of the dipole propagator:
\begin{equation}
\mymatrix{T}= \left(\bm{\nabla}\,\bm{\nabla}+k_0^2\;\hat{\bm{I}}\,\right)\end{equation}
We then want to use equation (\ref{scattered_as_func_of_inc_comp}) to write the scattered field as a function of the incoming field. We do so for each  pixel of the DMD:
\begin{equation}
\bm{E}\left(\bm{r}_i,t\right)=\bm{E}_0\left(\bm{r}_i,t\right)+\sum_j\mymatrix{T}_{ij}\;\alpha_j\,\bm{E}_0\left(\bm{r}_j,t\right)
\end{equation}
Now we look at the last equation at fixed time, so that the time dependence can be dropped. This can be done because of the assumption of perfect \emph{time coherence} of the source.
\begin{equation}
\bm{E}\left(\bm{r}_i\right)=\bm{E}_0\left(\bm{r}_i\right)+\sum_j\mymatrix{T}_{ij}\;\alpha_j\,\bm{E}_0\left(\bm{r}_0j\right)
\end{equation}
We once again apply the discretization of the electric filed, in the very same way we have seen it in equation (\ref{discretization}):
\begin{equation}\label{recurrence}
\bm{E}_i=\bm{E}_{0i}+\sum_j\mymatrix{T}_{ij}\;\alpha_j\,\bm{E}_j
\end{equation}
In this last equation $\bm{E}_i$ is the discretized scattered field in the $i-th$ cell of space, while the field with the double subscript $\bm{E}_{0i}$ represents an electric field in the $i-th$ cell of volume which has not been scattered yet.
Now we are ready to iteratively expand this equation, writing $\bm{E}_j$ as a function of $\bm{E}_{0j}$, taking its expression from the same equation. Some steps shown below:
\begin{align}\label{iterative_j}
\notag\bm{E}_i &= \bm{E}_{0i} +\sum_j\mymatrix{T}_{ij}\;\alpha_j\,\left[\bm{E}_{0j}+\sum_k\mymatrix{T}_{jk}\;\alpha_k\,\bm{E}_k\right]=\\
\notag&=\bm{E}_{0i}+\sum_j\mymatrix{T}_{ij}\;\alpha_j\,\bm{E}_{0j} +\sum_{j,k}\mymatrix{T}_{ij}\;\mymatrix{T}_{jk}\;\alpha_j\,\alpha_k\,\bm{E}_k=\\
\notag&=\sum_j\delta_{ij}\,\bm{E}_{0j}+\sum_j\mymatrix{T}_{ij}\;\alpha_j\,\bm{E}_{0j} +\sum_{j,k}\mymatrix{T}_{ij}\;\mymatrix{T}_{jk}\;\alpha_j\,\alpha_k\,\bm{E}_k=\\
\notag&=\sum_j\delta_{ij}\,\bm{E}_{0j}+\sum_j\mymatrix{T}_{ij}\;\alpha_j\,\bm{E}_{0j} +\sum_{k,j}\mymatrix{T}_{ik}\;\mymatrix{T}_{kj}\;\alpha_k\,\alpha_j\,\bm{E}_j=\\
\notag&=\sum_j\delta_{ij}\,\bm{E}_{0j}+\sum_j\mymatrix{T}_{ij}\;\alpha_j\,\bm{E}_{0j} +\sum_{k,j}\mymatrix{T}_{ik}\;\mymatrix{T}_{kj}\;\alpha_j\,\alpha_k\,\bm{E}_{0j}+...=\\
\bm{E}_i&=\sum_j \mymatrix{M}_{ij}\;\bm{E}_{0j}
\end{align}
in which we have defined $\mymatrix{M}$ as the series of the operators appearing in the expansion:
\begin{equation}
\mymatrix{M}_{ij}=\delta_{ij}+\mymatrix{T}_{ij}\;\alpha_j+\sum_k\mymatrix{T}_{ik}\;\mymatrix{T}_{kj}\;\alpha_j\,\alpha_k+...
\end{equation}
We have reached a point in which the expression of $\bm{E}_i$ is quite compact:
\begin{equation} \label{eq:OSG_E} 
\bm{E}_i=\sum_j\mymatrix{M}_{ij}\;\bm{E}_{0j} = \sum_j\mymatrix{M}_{ij}\;\hat{\epsilon}\,E_0\,e^{ikz}\sigma_j
\end{equation}
If we assume that the detected light is proportional to the light scattered from a given corresponding point of the dense scattering sample, it can be written as $\sim a_i\,\bm{E}_i$. Its square absolute value will be the \emph{energy function}:
\begin{align}
\notag I_i&=\frac{\Delta\Omega}{4\Phi}\,a^2_i\,\left|\bm{E}_i\right|^2=\frac{\Delta\Omega}{4\Phi}\,a^2_i\,\left(\bm{E}_i \cdot \bm{E}^*_i\right)^2=\\
\notag	&=\frac{\Delta\Omega}{4\Phi}\,a^2_i\,\left(\sum_{j}\mymatrix{M}_{ij}\,\hat{\epsilon}\,e^{ikz_j}\,\sigma_j\right)\cdot\left(\sum_{j'}\mymatrix{M}_{ij'}\,\hat{\epsilon}\,e^{ikz_j'}\,\sigma_{j'}\right)^*=\\
I_i&=\frac{\Delta\Omega}{4\Phi}a_i^2E_0^2\sum_{j,j'}\left(\mymatrix{M}_{ij}\,\hat\epsilon\right)\cdot\left(\mymatrix{M}_{ij'}\,\hat\epsilon\right)\,e^{ik(z_j-z_{j'})}\sigma_j\,\sigma_{j'}\label{Hamiltonian_fulldetail}
\end{align}
We now want to drop the index $i$, which is just the dependence of the written quantity on the specific point we look at on the scattering target. We choose a given point and therefore we get the formula we wanted to prove in equation (\ref{dyadic_result_beforeproof}):
\begin{equation}\label{Hamiltonian}
I=\frac{\Delta\Omega}{4\Phi}\alpha^2E_0^2\sum_{j,j'}J_{jj'}\sigma_j\sigma_{j'}=I_0 \sum_{j,j'}J_{jj'}\,\sigma_j\,\sigma_{j'}=I_0\,\bm{\sigma}\mymatrix{J}\bm\sigma
\end{equation}
In writing the latter equation, we have defined two characteristic quantities of the system, namely:
\begin{align}
I_0&=\frac{\Delta\Omega}{4\Phi}\alpha^2E_0^2\\
\notag J_{jj'}&=e^{ik(z_j-z_{j'})}\left(\mymatrix{M}_{0j}\,\hat\epsilon\right)\cdot\left(\mymatrix{M}_{0j'}\,\hat\epsilon\right)=\\
	&=e^{ik(z_j-z_{j'})}\sum_{\alpha,\beta,\gamma}\epsilon_\alpha\,\epsilon_\beta\,M_{ij}^{\alpha\beta}\,M_{ij'}^{\gamma\beta}\label{Jdef}
\end{align}
It is fundamental to note here that \emph{the matrix} $\mymatrix{J}$ \emph{depends on the specific point} at which we look. This happens implicitly through the dependence on the on the index $i$; however, the dependence is hidden because usually we do not want to change the point of observation during our experiment.
\newline On the contrary, the motion of the speckle pattern is to avoid at all costs, because two different points of the speckle pattern effectively correspond to two different spin glasses: namely, they are characterised by two different distributions of the coupling matrix $\mymatrix{J}$.
\chapter{Experimental design}\label{Ch:experimental_setup}
\epigraph{Il faut que l'observation de la nature soit assidue, que la réflexion soit profonde, et que l'expérience soit exacte.}{\textit{Denis Diderot, Pensées sur l'interprétation de la nature (1754)}}

The idea behind the experiment is to reproduce a spin glass system using the theoretical background of chapter \ref{Ch:theory_optics}. The system will use optical components in order to reproduce the light intensity according to the result of equation (\ref{Hamiltonian}): the standard spin glass Hamiltonian. Once such optical setup is in place, the plan is to use a standard Metropolis-Hastings algorithm \cite{hastings1970monte} to sample the phase space and extract the relevant information for the corresponding spin glass system, much as it is done with computer simulations.

\section{Experimental setup}
According to the theory in section \ref{theory_opticalspinglass}, to obtain a light intensity in the form of equation (\ref{Hamiltonian}), we need  three ingredients:
\begin{itemize}
	\item a \emph{source} of monochromatic light, coherent both in time and space
	\item a \emph{digital micromirror device} which spatially modulate the wavefront of the radiation emitted by the source
	\item a \emph{dense scattering medium}, scrambling light rays and yielding the interference of the modulated wavefront, thus producing the speckle pattern
\end{itemize}

The final speckle pattern is the result of all processes described in section \ref{theory_opticalspinglass} and its intensity in each point will satisfy equation (\ref{Hamiltonian}). It can therefore be imaged and its intensity can be used as the output signal of the system.

\subsection{Light source}
The obvious choice for the source of monochromatic, coherent light is a LASER. In particular, the \emph{Azur Light ALS GR532}, a HeNe laser, was used. The full specifications are available in table \ref{tab:laserspecs}. \\ 
\begin{table}[ht]
\begin{center}
	\begin{tabular}{|c|c|}
	\hline
	Specification	&	Value \\
	\hline
	\hline
	Wavelength	&	$532.00\pm0.25$ nm\\
	\hline
	Output power&	$2$ W\\
	\hline
	Output power tunability	& 1-100\\
	\hline
	Beam quality &	$M^2<1.1$\\
	\hline
	Beam diameter&	$1.0\pm0.2$\\
	\hline
	Beam divergence	&	$<\,0.5$ mrad\\
	\hline
	Spatial mode	&	TEM\textsubscript{00}\\
	\hline
	Spectral width	&	$<\,200$ kHz\\
	\hline
	Power stability (short term)	&	$\pm0.3\,\%$\\
	\hline
	Power stability (long term)&		$\pm0.5\,\%$\\
	\hline
	Noise	&	$<\,0.05$ \% rms\\
	\hline
	Frequency stability	&	$<\,0.1$ pm\\
	\hline
	Output polarization	&	$>\,300:1$\\
	\hline
	Pointing stability	&	$<\pm0.5$ $\mu\mbox{rad}/$°C\\
	\hline
	Supply requirements: voltage	&	90-240 V\\
	\hline
	Supply requirements: current frequency	&	50-60 Hz\\
	\hline
	Cooling	&	Air\\
	\hline
	\end{tabular}
	\caption{Specifications table for the \textit{Azur Light ALS GR532} laser}\label{tab:laserspecs}
\end{center}	
\end{table}
The main characteristic needed from the laser was the output stability in intensity, time coherence and spatial coherence. In fact, a change - even slight - of one of these parameters greatly affects the stability of the experimental setup and introduces an error in the output quantity $S$, as given in equation (\ref{Hamiltonian}). We now analyse separately how these three factors influence the output of the system: 
\begin{itemize}
	\item an instability in the \emph{intensity} emitted by the laser source affects the coefficient $E=\left|\bm{E}^{(in)}\right|$ in equation (\ref{scattered_as_func_of_inc_comp}), which then propagates through all the calculations, up to the expression of the intensity $I_0$ in equation (\ref{Hamiltonian}) and shifts the effective temperature of the system because changes the exponent of the Boltzmann factor
	\item a change in the wavelength $\lambda_0$ of the incoming radiation has the effect of changing the prefactor $\alpha(\omega_0)$ in equation (\ref{complete_formula_with_prefactor_dependence}. This produces a change in the way the light is scattered, therefore changing the path travelled by the photons on the surface of the scattering medium before getting collected towards the camera. As a result, the global effect is to rigidly shift the speckle pattern\footnote{This actually comes from an approximation for the autocorrelation function. A more detailed treatment of this phenomenon can be found in \cite{george1974space}}. This effect is particularly dangerous: the intensity in the target point, which we use as the output signal of the system, can vary considerably because of the high contrast between light and dark spots that characterises the speckle pattern
	\item an imperfection in the spatial coherence is to avoid because it introduces an error in the definition of the dipole propagator. In fact, we would have a factor of $e^{-i\bm{\Delta k}\cdot\bm{z}}$ when calculating some pairings in equation (\ref{Jdef}). This would provoke a phase shift for each term in the recursive expansion in equation (\ref{iterative_j}), thus invalidating the result
\end{itemize}

\subsection{Spatial modulator: the digital micromirror device}\label{sec:DMD}
A spatial modulator is any instrument capable of modifying the spatial wavefront of an incoming light beam. We use a DMD (short for \emph{digital micromirror device}) which is, essentially, a screen in which each pixel is a small square mirror that can tilt along an axis along the diagonal. DMDs have seen a quite wide usage in imaging experiments, such as fluorescence imaging\cite{articleDMD} and imaging through dense scattering systems \cite{Akbulut}.\\

The mirrors can be individually rotated of $\pm 12^\circ$ with respect to the direction perpendicular to the surface of the device (or $10^\circ$, which is the other industry standard), depending on its activation state. In the \emph{on} state, the light coming from the laser source is reflected into the correct optical path, towards the sample. In the \emph{off} position, instead, light is reflected at an angle of $24^\circ$ (twice the tilting angle) away from the system and it is dumped.\\

An important advantage of DMDs over other spatial modulation devices lies in its very low switching speed between the two states. An important value for estimating the switching speed is the \emph{settling time}, which is the time required for a pixel to become stable after changing its state. For a standard DMD the settling time is $18\,\mu s$, nearly three orders of magnitude below the switch speeds of the LCD-based SLMs. Such high commuting frequency is due to the original DMDs purpose: they were firstly designed for the film industry, and the commuting speed was intended to be used to obtain different shades of the colours: the higher the available commuting frequency, the more shades obtainable.\\

\begin{table}[ht]
\begin{center}
	\begin{tabular}{|c|c|}
	\hline
	Specification	&	Value \\
	\hline
	\hline
	DLP chip		&	Discovery\textsuperscript{TM} 4100\\
	\hline
	DMD Type		&	0.7" XGA 2xLVDS\\
	\hline
	Window Options	& Visible wavelenght\\
	\hline
	Micromirror Array &	$1024 \times 768$\\
	\hline
	Active Mirror Array Area&	$14.0 \times 10.5\,mm^2$\\
	\hline
	Controller Board Type	&	V4100\\
	\hline
	Control Board Dimensions	&	$71\times 68\,mm^2$\\
	\hline
	DMD Board Dimensions	&	$67\times 50\,mm^2$\\
	\hline
	Flexible Cable Length	&	$90\,mm$\\
	\hline
	RAM Capacity on Board&		32 Gbit\\
	\hline
	Binary Patterns on Board	&	43690\\
	\hline
	Hardware Trigger	&	master/slave\\
	\hline
	Controller Suite	&	ALP - 4.2\\
	\hline
	\multirow{2}{*}{Array Switching Rate}	&	22727 Hz @ 1bit B/W\\
	&	1091 Hz @ 6bit grayscale\\
	&	290 Hz @ 8bit Gray\\
	\hline
	PC Transfer Rate	&	800 fps\\
	\hline
	\end{tabular}
	\caption{Specifications table for the \textit{Vialux DLP7000} DMD}\label{tab:dmdspecs}
\end{center}	
\end{table}

\begin{figure}[h!t]
  \centering
  \begin{subfigure}[b]{0.45\linewidth}
    \centering\includegraphics[width=0.99\linewidth]{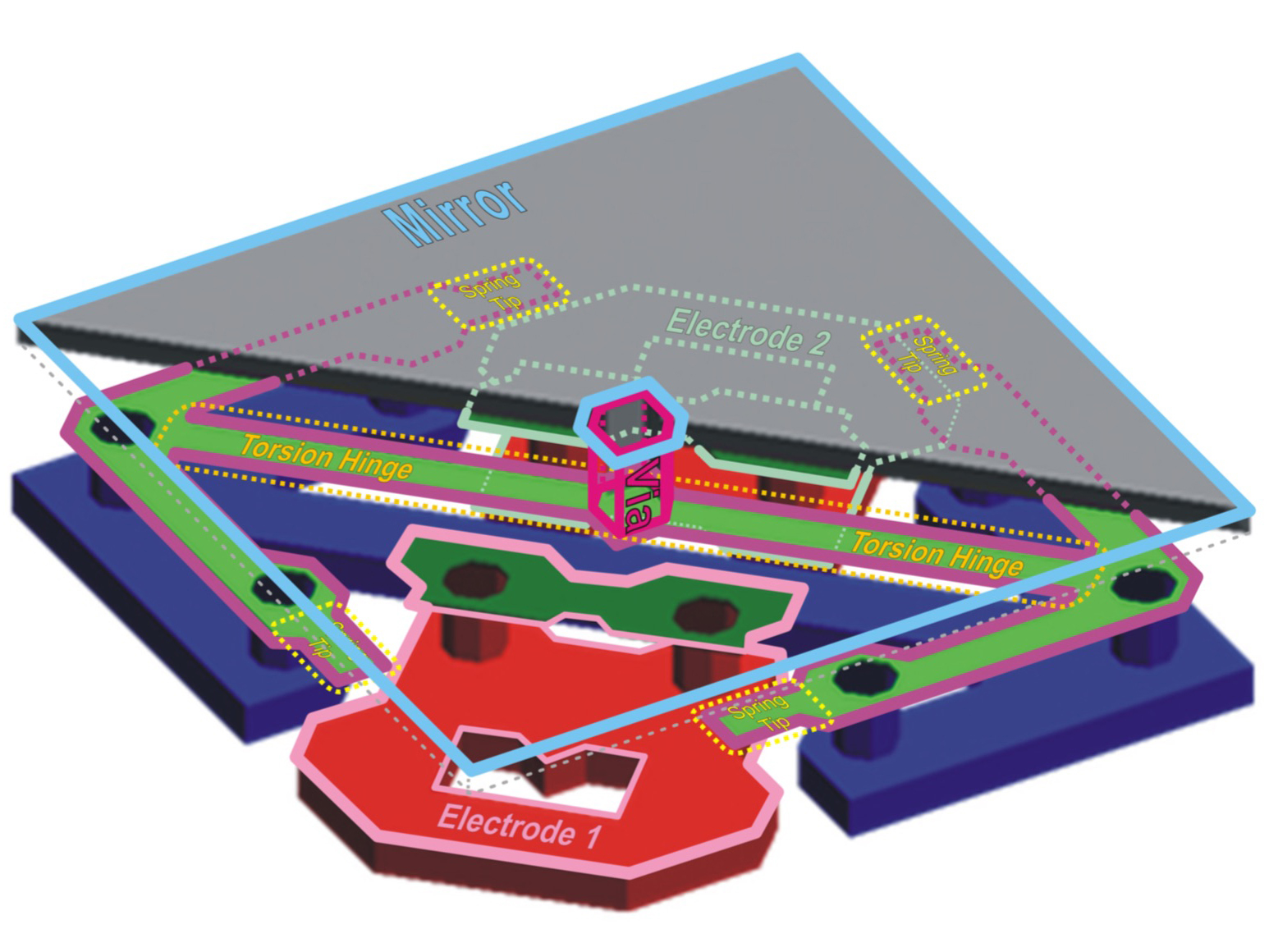}
    \caption{\label{dmd:fig1}The electronic design of each pixel. Figure from reference\cite{dmd_technical}.}
  \end{subfigure}\hspace{0.05\linewidth}
  \begin{subfigure}[b]{0.45\linewidth}
    \centering\includegraphics[width=0.99\linewidth]{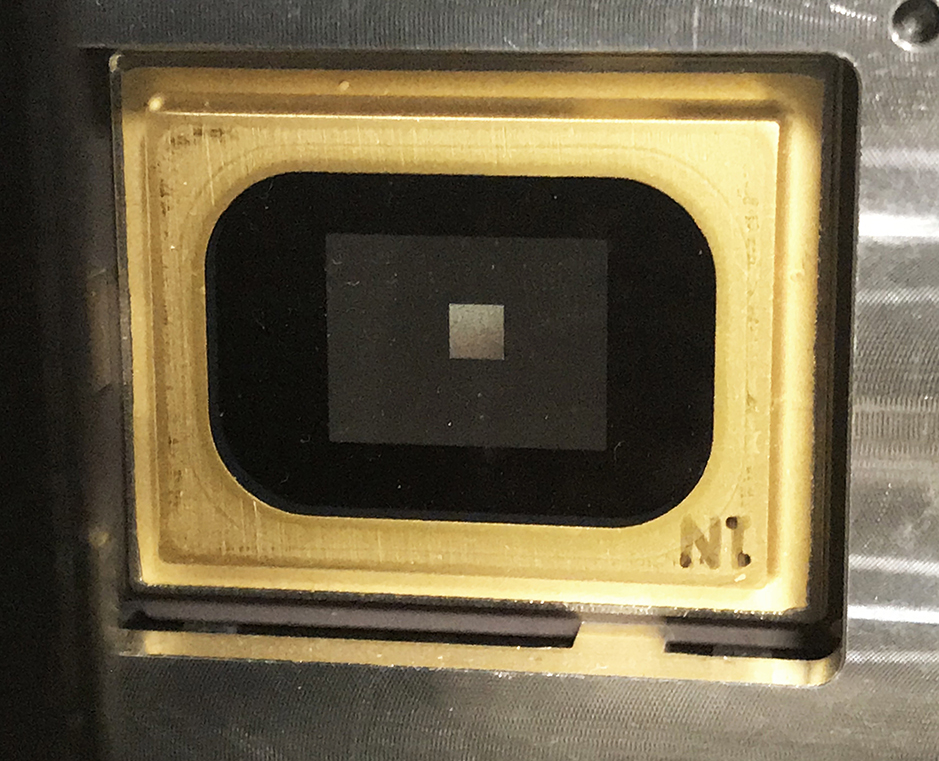}
    \caption{\label{dmd:fig2}A photo of the DMD, in which the activated area is visible.}
  \end{subfigure}
  \label{dmd_pic}
  \caption{Reference pictures for the DMD functioning}
\end{figure}

The device apparatus works both mechanically and electronically. The pixels are made by an aluminium micromirror attached to a torsional hinge placed below the mirror, along a diagonal of the pixel. The underside of the mirror is in contact with the spring tips: see figure \ref{dmd:fig1} for reference. The two electrodes are controlled electronically, and they change the mechanical position of the mirror by means of electrostatic attraction. In the idle state, an equal bias charge is applied to both electrodes simultaneously. This makes the mirror stay in its current position, instead of returning to the intermediate flat position. In fact the angle of the mirror located in the direction towards which the mirror is pointing is much closer to the underside than the opposite angle. As a result, despite the charges being equal, the force that holds the mirror in place is greater than the titling force applied by the opposite electrodes. During the commutation phase, the new state for each pixel is firstly loaded into a SRAM cell located beneath each pixel and connected to the electrodes. Once all the SRAM cells have been loaded, the bias voltage is removed from the electrodes, allowing the charges from the SRAM cell to prevail, moving the mirror. After the commutation, the mirror is once again held in position by reapplying the bias charge. The next required movement can be loaded into the memory cell, allowing the cycle to start over. A more detailed description of the technologies behind the DMD can be found in \cite{dmd_technical}.

\subsection{Scattering sample}
The scattering sample chose for the experiment is a titanium dioxide dust deposited on the surface of a glass slide. The deposit is obtained by dissolving the TiO\textsubscript{2} dust in ethanol. A solution with  concentration of 4 mol/L has been used: it was obtained by mixing $350\, mg$ of TiO\textsubscript{2} dust with $100\,mL$ of ethanol. Then $20\,mL$ of the solution is deposited on the glass slide and left to dry for two hours. When the evaporation of the ethanol is completed, we are left with the TiO\textsubscript{2} molecules arranged in a random fashion. We note that, since the optical setup works with reflected scattered light, the thickness of the sample is of little to no importance.\\

The sample with its glass slide is then glued on a holder which is screwed onto the cage system, to ensure maximum stability and prevent vibrations.

\begin{figure}[ht]
\begin{center}
	\includegraphics[width=10cm]{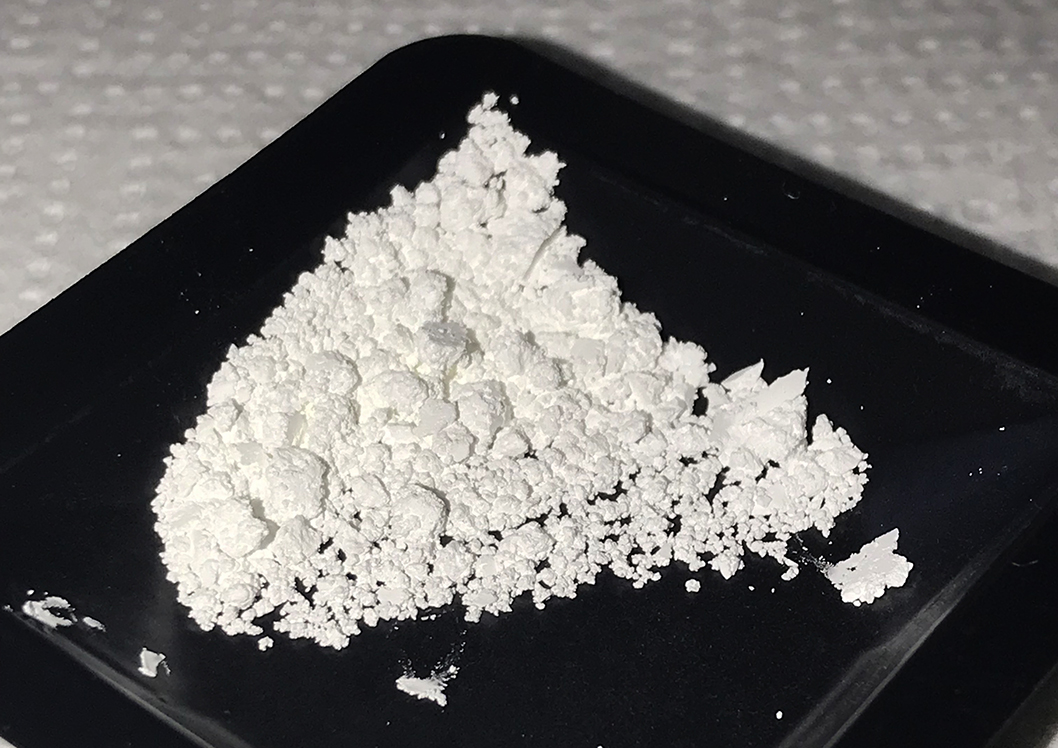}
	\caption{TiO\textsubscript{2} in crystalline form before being put in solution with ethanol to create the scattering sample.}\label{optical_setup}
\end{center}
\end{figure}

\subsection{Optical setup and detector}
The optical line encompasses the rest of the experimental apparatus: its aim is to collect the laser light reflected by all DMD pixels in the active state, collimate the light, bring it to the scattering sample, collect the reflected light and send it to the detector. The optical setup itself is quite simple, and its scheme is represented in figure \ref{optical_setup}.\\

\begin{figure}[ht]
\begin{center}
	\includegraphics[width=10cm]{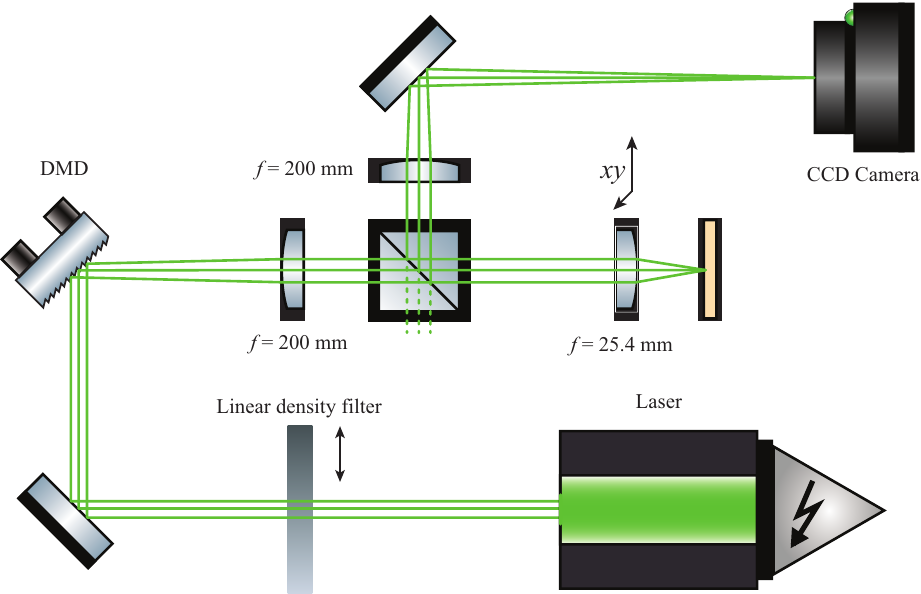}
	\caption{Experimental apparatus scheme}\label{fig:tio2}
\end{center}
\end{figure}

The main element is the beam-splitter, that takes care of separating the light going from the DMD to the sample from the light reflected by the sample, which has to go towards the detector. Three piano-convex lenses are used to create two focalising systems, the lens nearest to the sample being in common among the two systems. All elements in the optical system has been mounted with the \emph{Thorlabs Cage System\textsuperscript{\textcopyright}} to ensure maximum stability.\\

Both optical telescopes are composed by the lens in front of the sample, which has a focal length of $f_s=25.4\,mm$ and a lens of focal length $f=200.0\,mm$. This was means that no magnification occurs between the DMD and the sensor, while the light spot on the sample is demagnified by a factor of $f_s/f=7.84$. This allows a smaller light spot on the sample, increasing the overall stability and allowing all couples of mirrors to produce interacting light beams.\\

\begin{figure}[h!t]
  \centering
  \begin{subfigure}[b]{0.45\linewidth}
    \centering\includegraphics[width=0.99\linewidth]{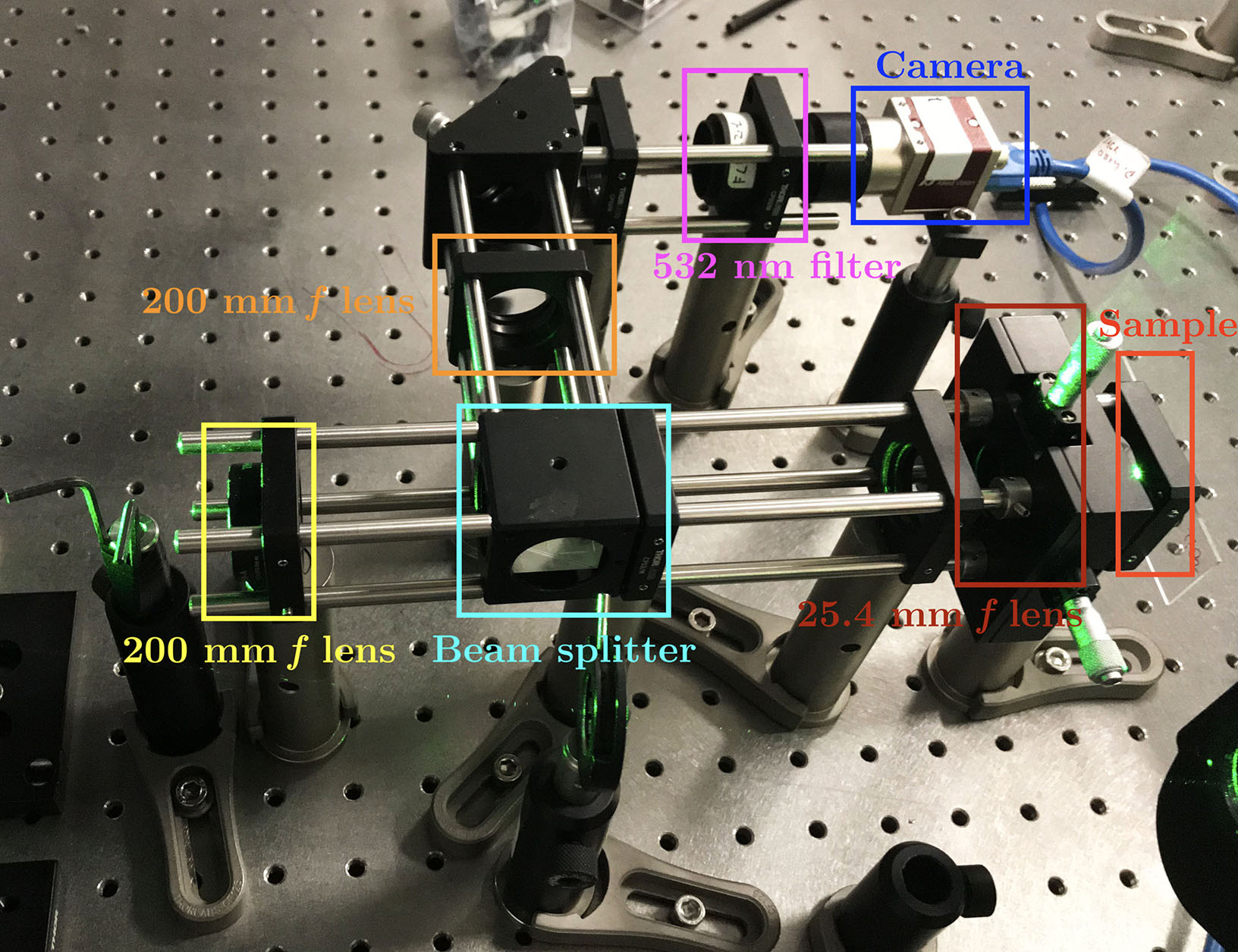}
    \caption{\label{fig:setup1}Experimental setup: optical cage.}
  \end{subfigure}\hspace{0.05\linewidth}
  \begin{subfigure}[b]{0.45\linewidth}
    \centering\includegraphics[width=0.99\linewidth]{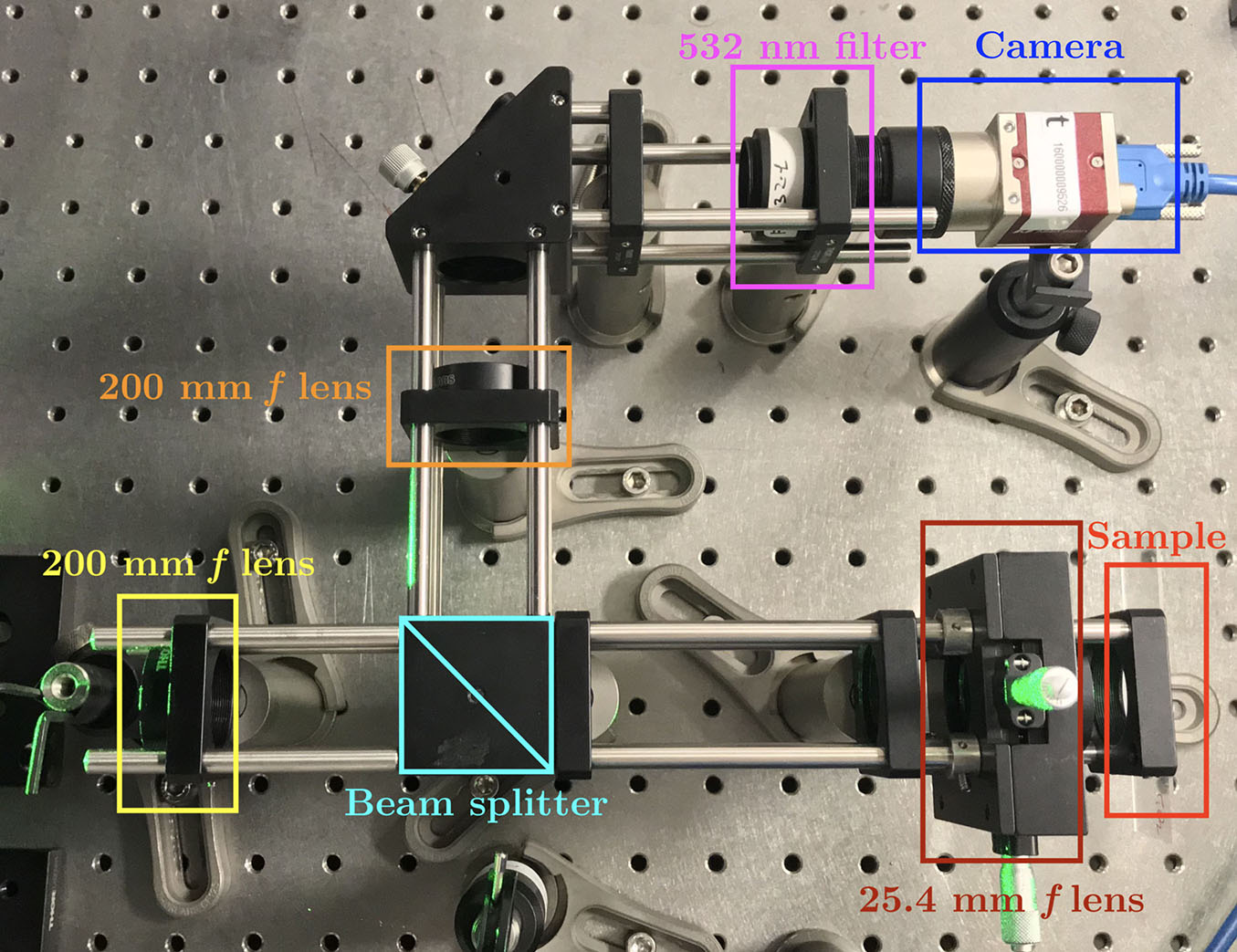}
    \caption{\label{fig:setup2}Experimental setup: optical cage.}
  \end{subfigure}\\\vspace{0.2cm}
  \begin{subfigure}[b]{0.45\linewidth}
    \centering\includegraphics[width=0.5858\linewidth]{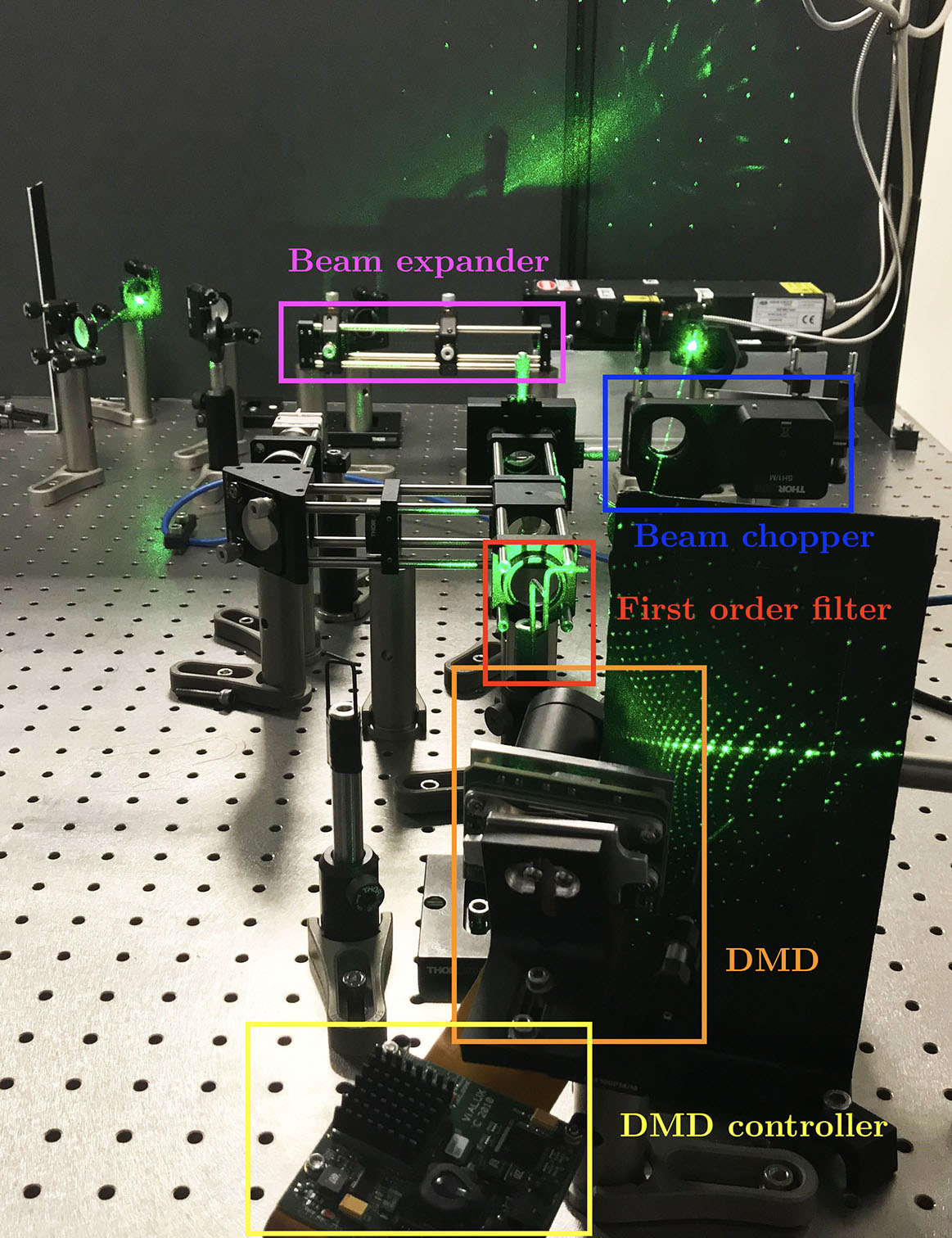}
    \caption{\label{fig:setup3}Optical setup from behind the DMD}
  \end{subfigure}\hspace{0.05\linewidth}
  \begin{subfigure}[b]{0.45\linewidth}
    \centering\includegraphics[width=0.99\linewidth]{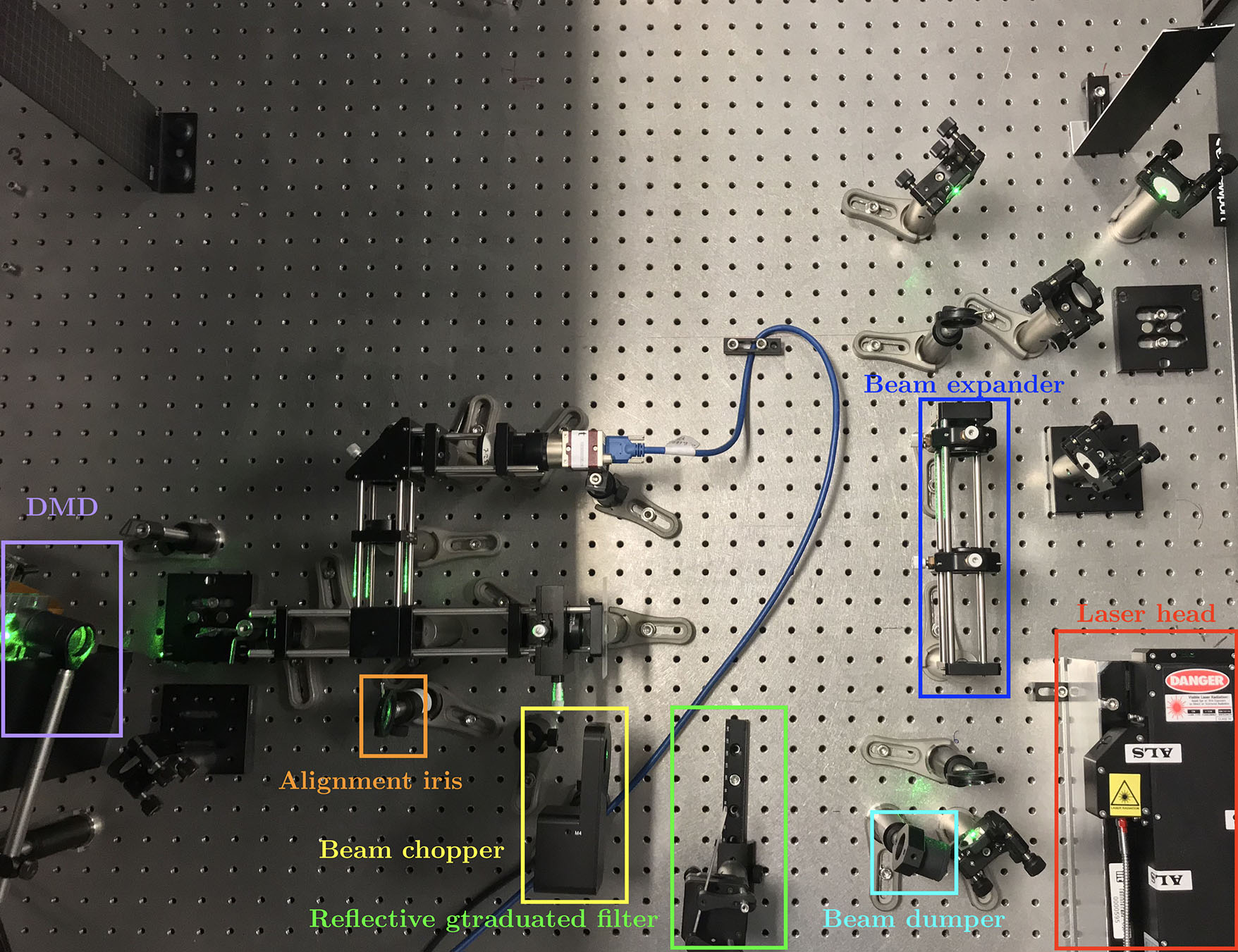}
    \caption{\label{fig:setup4}Complete optical setup}
  \end{subfigure}
  \caption{Different images of the optical setup: in (\subref{fig:setup1}) and (\subref{fig:setup2}) the optical line after the DMD is presented in detail, while in (\subref{fig:setup4}) the setup is presented as a whole.}
  \label{fig:opticalsetup}
\end{figure}

\begin{table}[h!t]
\begin{center}
	\begin{tabular}{|c|c|}
	\hline
	Specification	&	Value \\
	\hline
	\hline
	Interface		&	USB3 Vision\\
	\hline
	Resolution		&	$640\times 480$\\
	\hline
	Sensor	& ON Semi PYTHON 300\\
	\hline
	Sensor type &	CMOS\\
	\hline
	Sensor size &	Type 1/4\\
	\hline
	Pixel size	&	$4.8 \mu m × 4.8 \mu m$\\
	\hline
	ADC &	10 bit\\
	\hline
	Image buffer (RAM) 	&	128 MByte \\
	\hline
	Bit depth	&	8 bit, 10 bit\\
	\hline
	Monochrome pixel formats &		Mono8, Mono10p\\
	\hline
	Binary Patterns on Board	&	43690\\
	\hline
	Hardware Trigger	&	master/slave\\
	\hline
	Controller Suite	&	ALP - 4.2\\
	\hline
	\multirow{2}{*}{Array Switching Rate}	&	22727 Hz @ 1bit B/W\\
	&	1091 Hz @ 6bit grayscale\\
	&	290 Hz @ 8bit Gray\\
	\hline
	PC Transfer Rate	&	800 fps\\
	\hline
	\end{tabular}
	\caption{Specifications table for the \textit{Mako U-029B} camera}\label{tab:makospecs}
\end{center}	
\end{table}

\section{Algorithm implementation}\label{sec:algorithm}
The algorithmic part of the experiment is the implementation of the Metropolis-Hastings algorithm on the system. To better explain how the algorithm is applied to such system, we start by identifying the correspondence between the components of the setup and the elements which appears in the spin glass model, in particular the one which is describes in terms of the Amit representation: 
\begin{itemize}
	\item the role of the spins is played by the pixels of the DMD. A spin which takes the value $+1$ corresponds to an active pixel, that is, which reflects the light in the optical path toward the sample. Conversely, a spin taking the value $0$ corresponds to a pixel which reflects the light to the beam dumper
	\item the role of the coupling matrix $\mymatrix{J}$ is taken by the scattering sample and the interference it creates, which mixes the  radiation incoming from each pixel according to equation (\ref{Hamiltonian_fulldetail})
	\item the energy function $H$ corresponds to the intensity read by a pixel of the camera. Since we are in the case of an \emph{objective} speckle pattern, the whole image is formed on the plane of the camera sensor. It is important to remark the following: any pixel of the camera sensor is suitable to be chosen as the target intensity, but different pixel will correspond to different spin glass systems. This is a consequence of the fact - already pointed out in section \ref{theory_opticalspinglass} - that the coupling matrix is a function of the position of the speckle pattern at which we look.
\end{itemize}

Before continuing, it is important to note that this setup represent an \emph{infinite dimensional} spin glass system, in which each spin interacts with each other spin. Alternatively, we can say that in this system the coupling constant $J_{ij}$ is nonzero for any couple of spins of the system. The reader should not be deceived by the bidimensional nature of the spins on the surface of the DMD: because the image on the scattering sample is small and the radiation focalised, the interference superpose all electric fields, no matter their position on the DMD.\\

Once this identification is in place, the application of the Metropolis-Hastings Monte Carlo algorithm is quite straightforward. We give hereafter a pseudo-code description of the implemented algorithm, only focusing on the details regarding the application to the optical system. The general theory behind this class of algorithms is well established and can be found, for example, in section \ref{sec:metrhast} or in \cite{chib1995understanding,gilks1995markov}.\\

\begin{algorithm}[H]
 \vspace{0.1cm}
 \KwData{Initial DMD configuration $\mymatrix{S_0}$, effective temperature $\beta$, ROI region for the camera $\mymatrix{R}$}
 \KwResult{Next configuration in the Monte Carlo Markov chain}
 \vspace{0.3cm}
 $\mymatrix{S}$ $\leftarrow\mymatrix{S_0}$\;
 choose$\_$camera$\_$ROI($\mymatrix{R}$)\;
 send$\_$config$\_$to$\_$DMD($\mymatrix{S}$)\;
 $\mymatrix{C}$ $\leftarrow$ camera$\_$measurement( )\;
 exposure$\_$feedback($\mymatrix{C}$)\;
 
 \For {$i\leftarrow 1$ \KwTo MC$\_$steps}{
  I($\mymatrix{S}$) $\leftarrow$ energy$\_$from$\_$camera$\_$reading($\mymatrix{C}$)\;
  $\mymatrix{T}$ $\leftarrow$ MC$\_$trial($\mymatrix{S}$)\;
  send$\_$config$\_$to$\_$DMD($\mymatrix{T}$)\;
  $\mymatrix{C}$ $\leftarrow$ camera$\_$measurement( )\;
  exposure$\_$feedback($\mymatrix{C}$)\;
  I($\mymatrix{T}$) $\leftarrow$ energy$\_$from$\_$camera$\_$reading($\mymatrix{C}$)\;
  \vspace{0.1cm}
  \eIf{I($\mymatrix{T}$) $\leq$ I($\mymatrix{S}$)}{
  	$\mymatrix{S}$ $\leftarrow$ $\mymatrix{T}$\;
  }{
  	ran = random(0,1)\;
  	\eIf{ran $\leq$ $exp\left[{-\beta\left(I(\mymatrix{S})-I(\mymatrix{T})\right)}\right]$}{
  		send$\_$config$\_$to$\_$DMD($\mymatrix{S}$)\;
  	}{
  		$\mymatrix{S}$ $\leftarrow$ $\mymatrix{T}$\;
  	}
  }

}
\caption{Metropolis Monte Carlo algorithm on the optical system}
\end{algorithm}
\vspace{0.5cm}

\section{Experimental problems and limitations}\label{sec::exp_problems}
The algorithm in itself is quite straightforward once each piece of the setup has been correctly associated to the correct element of the spin glass theory. Nevertheless, due to some intrinsic limitations of real systems, some care is to be taken during the implementation of the algorithm.

\subsection{Negative defined energy Exposure time feedback}\label{sec:negen}
The value of the energy of a spin glass is given by the dyadic expression we have seen in chapter \ref{Ch:spin_glass_theory}, namely:
\begin{equation}
	H = -\sum_{i,j}J_{ij}\,\sigma_i\sigma_j
\end{equation}
As we seen in chapter \ref{Ch:theory_optics}, the energy function (the Hamiltonian $H$) is substituted by the intensity function $I$, as in equation (\ref{Hamiltonian_fulldetail}). The Metropolis-Hastings algorithm produces the Boltzmann distribution as equilibrium distribution:
\begin{equation}
	P_{MH}\propto e^{-\frac{H}{k_BT}}
\end{equation}
which means the states corresponding to a \emph{lower} energy are favourite. If we apply this straightforwardly to our system, we would have the system exploring mainly the portion of phase space at lower intensity. This is not optimal from an experimental point of view, because a low intensity means that the noise on the result can potentially become very high.

To bypass this problem, we slightly change the Metropolis-Hastings algorithm as follows. Normally, once the intensity (or its corresponding energy function) has been measured both for the last state of the system $I_{last}$ and after the trial move has been performed $I_{trial}$, a variable $\Delta I =I_{trial}-I_{last}$ is defined and the trial move is accepted with the standard Metropolis choice:
\begin{enumerate}
	\item if $\Delta I<0$ the trial move is always accepted
	\item if $\Delta I>0$ the move is accepted with probability
\begin{equation}
	P_{acc} = e^{-\frac{\Delta I}{k_BT}}
\end{equation}
	To do so, a random number $\zeta$ is generated, uniformly in $[0,1]$, and the move is accepted if $\zeta<P_{acc}$
\end{enumerate}
The implemented algorithm invert the last two inequalities, effectively changing the sign in the definition of $\delta I$. This favours the high energies configurations and solves the problems of the noise at very low energy.

\subsection{Exposure time feedback}
The modification of the algorithm suggested in the last section does indeed solve the problem with low energy states, but favouring the high intensity configuration can potentially saturate the camera sensor. The camera has an 8 bit output, meaning that only an intensity between $0$ and $255$ can be obtained as output. Everything above the $255$ will not be recorded, as the sensor saturates. Thus we use the exposure time to keep the recorded intensity into the correct range. A feedback loop is used to decrease the exposure time if the read intensity exceeds a target value, and to decrease it if the read intensity falls below a given threshold. To maintain the compatibility between measures performed at different exposure times, each time it is corrected by a factor $\zeta$, the camera output is multiplied by a factor $\/\zeta$.

The following is a pseudo-code example of how the feedback loop works:\\

\begin{algorithm}[H]
 \vspace{0.1cm}
 \KwData{The camera output $\mymatrix{C}$ with $C_{ij}\in[0,255]$, the original exposure time $t_c$, the original exposure time $t_0$}
 \KwResult{Decide whether to change the current exposure time for the camera, and correct the measure reading accordingly}
\While  {$max(C_{ij})\geq 220$ \textbf{or} $max(C_{ij})\geq 80$}{
 \If{$max(C_{ij})\geq 220$ }{
  $t_c\leftarrow \frac{t_c}{2}$ decrease the exposure time\;
  $\alpha \leftarrow  \frac{t_c}{t_0}$ update the multiplication constant\;
 }
 \If{$max(C_{ij})\leq 80$ }{
  $t_c\leftarrow 2\times t_c$ increase the exposure time\;
  $\alpha \leftarrow  \frac{t_c}{t_0}$ update the multiplication constant\;
 }$\mymatrix{C}$ $\leftarrow$ camera$\_$measurement( )\;
 }
 $\mymatrix{C} \leftarrow \alpha\,\mymatrix{C}$\;
 I $\leftarrow $ sum$\_$elements($\mymatrix{C}$)\;
 \caption{Exposure time control algorithm}
\end{algorithm}
\vspace{0.5cm}
This allows the system to stay in a non saturated regime and to obtain a dynamic range of approximately $10^6$ counts. This figure is obtained through a $10^4$ order of magnitude in the exposure time and another $10^2$ from the camera counts. Of course, this come at the cost of slowing down the simulation when low intensities, and therefore high exposure times, are in play.\\

A method to ease this problem, along with the noise problem of the last section, is to use a camera with a better signal to noise ratio and a wider dynamic range. For example, a $16-bit$ camera has a maximum count number of 65535, allowing two more orders of magnitude in the global dynamic range, but also allowing the system to run one hundred times faster in low intensities situations. Even better, a photodiode could be used in combination with an acquisition card, to maximise both the dynamic range and the signal-to noise ratio: see chapter \ref{Ch:conclusions} for perspectives in setup development.

\subsection{Multi-spin flipping} 
Another problem with the detector is the \emph{signal to noise ratio}. For the 8 bits camera used, this is around one in a hundredth. In fact, the shot noise can create one or two counts when the CCD is completely dark, compared to a maximum count of 255. For this experimental limitation, we need to change at least one hundredth of the pixel at each move. In fact, the exposure time is set in such a way that the global intensity ranges from 0 to 255. If we change less than one hundredth of the DMD pixels, we change the intensity on the detector by a value in the order of some units, hence the change is dominated by noise.\\

Fortunately, the Metropolis-Hasting algorithm in robust with respect to a change in the number of spins that are changed in the trial move (see section \ref{sec:msf} for a detailed description of the Metropolis-Hasting algorithm invariance with respect to the number of spins involved in the trial move).
\chapter{Preliminary measurements and checks}\label{Ch:preliminary}

\section{Stability analisys}
It is important to emphasise that the formula (\ref{Hamiltonian}) is valid no matter which point of the speckle pattern is considered to measure the intensity. However - this is crucial - the coupling coefficients will be different for each of them.
In other words, the intensity at different points of the speckle pattern can always be written in a bilinear form as the Hamiltonian of a spin-glass-like system, but the particular expression of the coefficients $J_{ij}$ will differ.\\

As a consequence, it is fundamental to ensure a perfect stability of the optical setup: even a slight drift of the system can dramatically undermine the outcome of the measure, because it changes the effective Hamiltonian of the system under analysis. If this happens, the Monte Carlo algorithm is no longer able to give the correct equilibrium distribution. Indeed, such a distribution is no longer well-defined.\\

The easiest way to sample the stability of the system is to measure the system with the dmd in a fixed position. We choose a given starting configuration for the DMD, which is not change thereafter. The camera acquires a small area of the speckle pattern, the \emph{region of interest} (ROI), which is broader than the target area later used to run the Monte Carlo simulation. In such a way, a successful stability test ensure that the ROI is unchanged by experimental bias, and therefore target area in the Monte Carlo simulation is unchanged as well.\\

We repeat the same measure many times, with a time-step $t_0$ between each measure, and we calculate the correlation of the most recent measure with the original one. The resulting quantity is the time correlation function $C_C$ of the camera sensor, and therefore of the speckle pattern:
\begin{equation}
	C_C(t=i\,t_0) = \sum_{i=1}^{N_{pixel}}\left<p_i(t=0)\;p_i(t=i\,t_0)\right>
\end{equation}

The subscript $C$ is used to differentiate this correlation function form the other one used later in the this work, which will be fundamental in the next chapter. The fundamental difference is that this correlation function is calculated on the \emph{camera pixels}, while the other one will be defined on the DMD mirrors.\\

We have not defined yet the precise form of the correlation. Nevertheless, we want the setup to produce a result for $C_C$ as close as possible to $1$ at any time:
\begin{equation}
	\left\{
	\begin{array}{ll}
	C_C(t)\equiv 1 & \mbox{for a perfectly stable system}\\
	\lim_{t\rightarrow\infty} C_C(t) = 0 & 	\mbox{for an unstable system}
	\end{array}
	\right.
\end{equation}

We now want to define the specific formula of the correlation we study.

\subsection{Pearson correlation coefficients}
The Pearson correlation coefficients of two random variables is a measure of their linear dependence. If each variable has $N$ scalar observations, then we define the Pearson correlation coefficient as:
\begin{equation}
	\rho\left(A,B\right)=\frac{1}{N-1}\sum_{i=1}^N\left(\overline{\frac{A_i-\mu_i}{\sigma_A}}\right)\left(\frac{B_i-\mu_B}{\sigma_B}\right)
\end{equation}
where $\mu_A$ and $\sigma_A$ are the mean and standard deviation of $A$, respectively, and $\mu_B$ and $\sigma_B$ are the mean and standard deviation of $B$.\\

An alternative definition of the correlation coefficient cab be given in term of the covariance:
\begin{equation}
	\rho\left(A,B\right)=\frac{\mbox{cov}\left(A,B\right)}{\sigma_A\sigma_B}
\end{equation}

The correlation coefficient matrix of two random variables is the matrix of correlation coefficients for each pairwise variable combination:
\begin{equation}
	R=\left(
	\begin{array}{cc}
	\rho(A,A)	&	\rho(A,B)\\
	\rho(B,A)	&	\rho(B,B)
	\end{array}
	\right)
\end{equation}

Since $A$ and $B$ are always directly correlated to themselves, the diagonal entries are just 1, that is:
\begin{equation}\label{corrcoef}
	R=\left(
	\begin{array}{cc}
	1	&	\rho(A,B)\\
	\rho(B,A)	& 1
	\end{array}
	\right)
\end{equation}

The correlation coefficients are particularly convenient in MATLAB, due to the way the camera processes the data. At each measurement, a matrix is produced: each element of the matrix corresponds to a pixel of the region of interest (ROI) on the camera sensor. We therefore need to correlate a set of matrices, all with the same numbers of rows and columns (the ROI does not change during the stability test, nor during any other performed measurement).\\

MATLAB provides the {\verb corrcoef } function, located in the standard library. Given two matrices $A$ and $B$ such function returns the correlation coefficient matrix as expressed in equation (\ref{corrcoef}). A non diagonal element is then plotted as a function of time (note that the specific element is irrelevant, the correlation matrix being symmetric).\\

The results of the described procedure are plotted in figures \ref{corr1}. The plot includes many iteration of the experiment: the initial measurement is replicated many times (20) and each subsequent frame in the stability test is correlated with each one of them. The results are then grouped to retain only the relevant decorrelation information and dump noise due to the camera.\\

\begin{figure}[H]
\begin{center}
	\includegraphics[width=\linewidth]{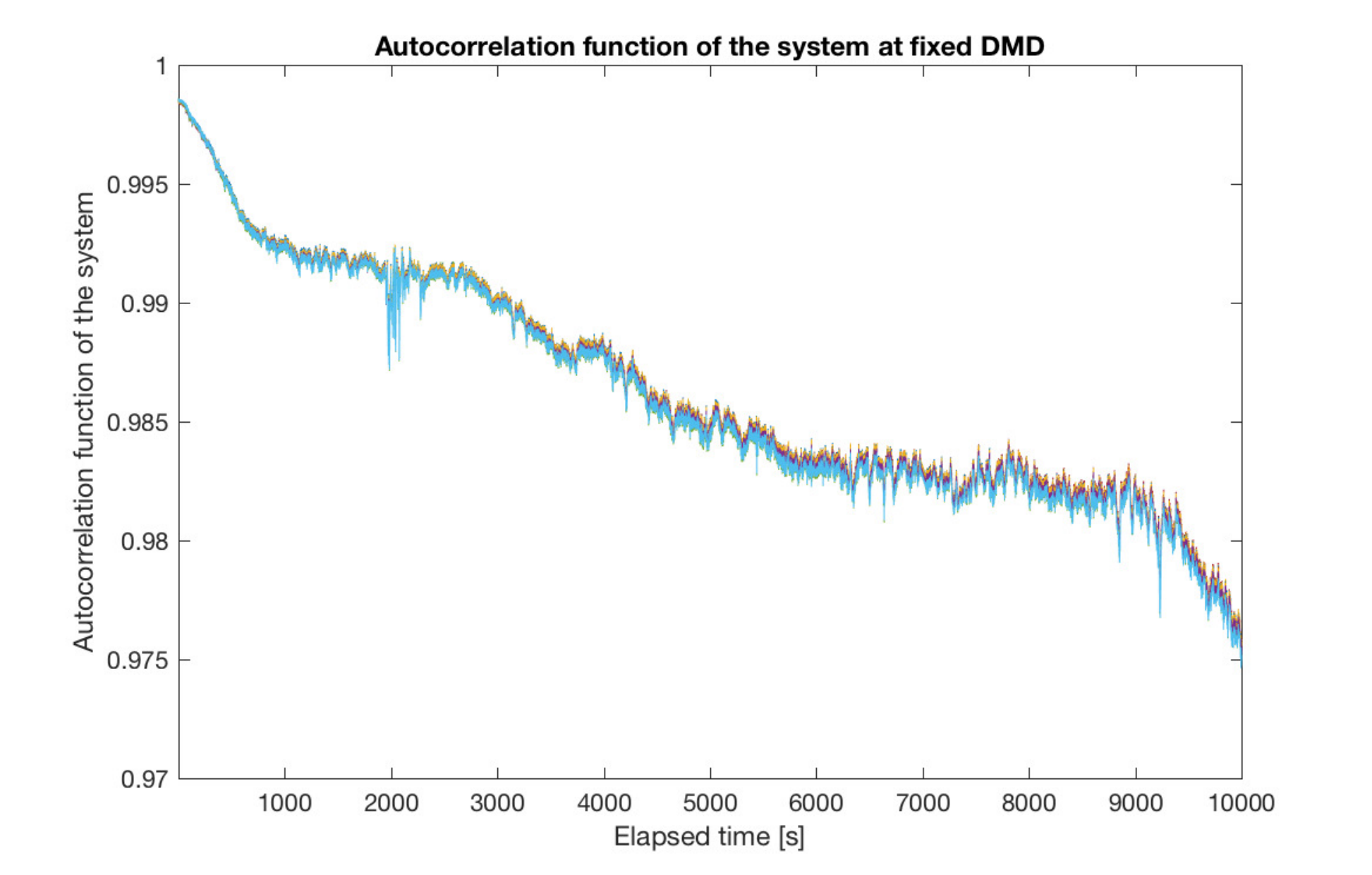}
	\caption{Autocorrelation function of the camera measurement at fixed DMD. The stability test has been performed for 10000 seconds (approximately 3 hours).}\label{corr1}
\end{center}
\end{figure}

We find a decorrelation of the system even when the DMD is not changing its configuration. This is obvious and expected, as we are dealing with an experimental setup: changes in the laser frequency or intensity, air temperature in the room or thermal agitation of molecules inside the sample are a small group of variables that may interfere with the measurement. However, much care has been taken to build a stable system, and the results are clearly visible in figure \ref{corr1}. The correlation coefficient of the time evolution when the setup is fixed do not fall below $98\%$ for at least two hours from the beginning of the measurement. \\

The Pearson coefficient is quadratic in the signal, hence having a $98\%$ value in the stability run corresponds to having a $1\%$ error on the output signal. This reasoning is a little hand-waiving, but ultimately correct. We also note that the Pearson coefficient never reaches one, not even at time zero. This first drop represent the immediate decorrelation due to camera noise, which is the only one likely to happen in just one frame: between the sample frame and the first used in the measure. This is the upper limit we can not beat unless a better detector is used.\\

One could object that the stability of the system may decrease if we move the DMD during the measure. To test this scenario, a second stability test has been performed. Instead of leaving the DMD fixed and taking a camera image at every time-step, the DMD is switched to a random configuration after the image has been taken by the camera. The DMD is switched back to its original position before the next camera acquisition starts, in order to have comparable quantities. The result of this stability test is reported in figure \ref{corr3}.\\

As expected, this modification introduces movement into the system and the decorrelation is more pronounced, but still perfectly under control. Recalling that the average measure time for the experiment of interest is around two hours, we are still well above $95\%$. This means, for what has been said above, that the signal is stable up to an error of around $3\%$ over two hours.\\

\begin{figure}[H]
\begin{center}
	\includegraphics[width=\linewidth]{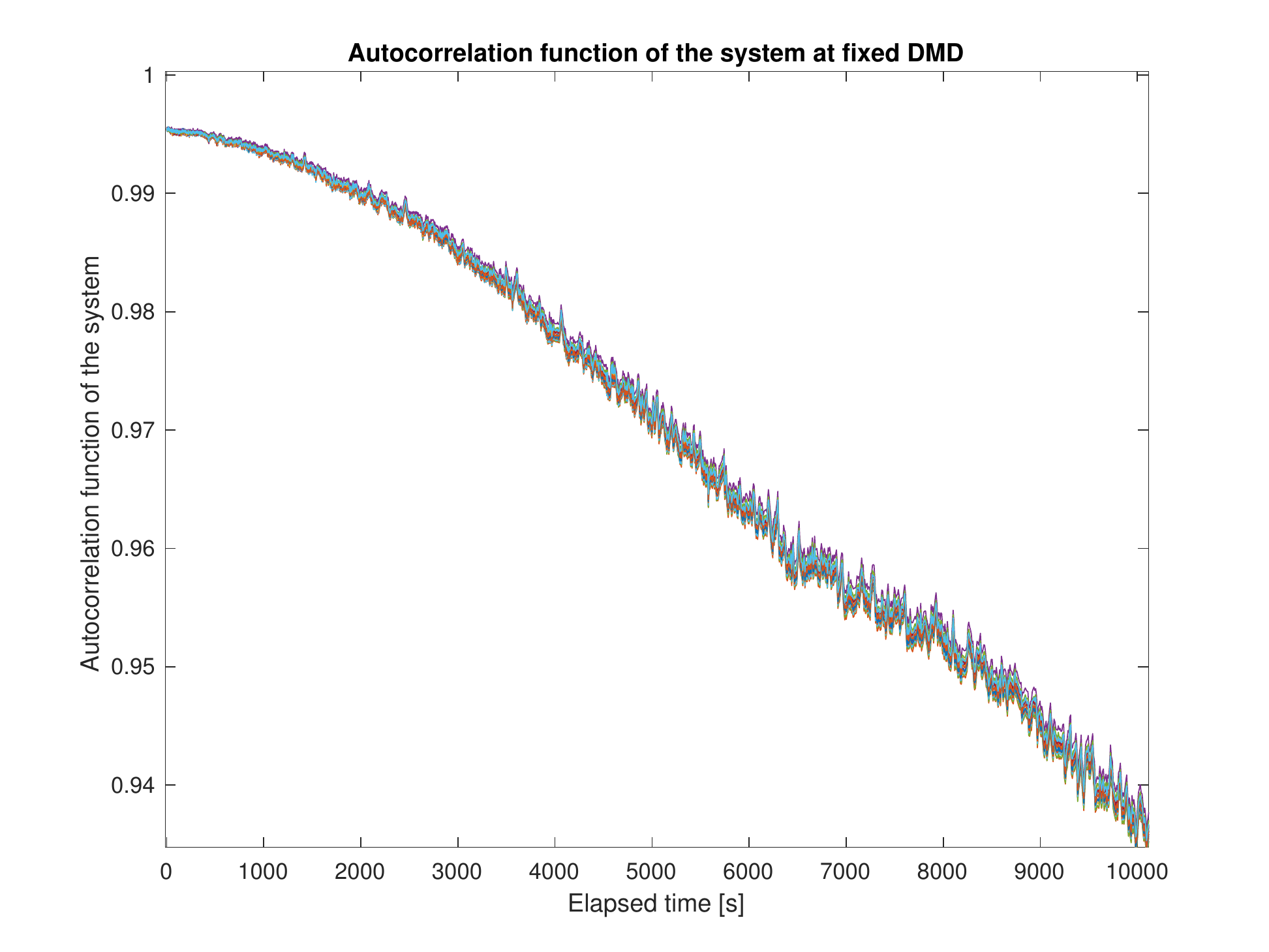}
	\caption{Autocorrelation function of the camera measurement with the DMD in the same position at each measure. This time, however, the DMD is switched to a random configuration and back to the original one between each measure. The stability test has been performed for 10000 seconds (approximately 3 hours).}\label{corr3}
\end{center}
\end{figure}

\section{Measure of couplings}
As showed in Chapter \ref{Ch:theory_optics}, the light intensity after a scattering sample can be written as in equation (\ref{Hamiltonian}), which is reported here for convenience:
\begin{equation}
	I=I_0 \sum_{i,j}J_{ij}\,\sigma_i\,\sigma_j
\end{equation}

The first measure needed to characterise the system is the strength of the couplings constants $J_{ij}$. This measure will provides useful information about the system, the most important being:
\begin{itemize}
	\item the reference spin glass model;
	\item the ferromagnetic or antiferromagnetic nature;
	\item the order of magnitude of the critical temperature $T_C$.
\end{itemize}

Indeed, we want to compare with the Sherrington-Kirkpatrick model, which entails a perfectly gaussian distribution of the coupling coefficients, namely:
\begin{equation}
	P\left(J_{ij}\right) = \frac{1}{\sqrt{2\pi}J}\,e^{-\frac{{\left(J_{ij}-J_0\right)}^2}{2J^2} }
\end{equation}

The distribution has zero means and standard deviation $J$. It is a well know result \cite{thouless1977solution} that the critical temperature $T_C$ of the Sherrington-Kirkpatrick model is the same as the standard deviation of the coupling distribution:
\begin{equation}\label{eq:criticaltemp}
	T_C = J
\end{equation}

For the purpose of this work, this quantity is of great interest. In fact, it allows to chose in which range of the parameter $\beta$ we need to run the simulations. We note that the result of equation (\ref{eq:criticaltemp}) is true only in the case of an exactly gaussian distribution of the coupling coefficients. Since we have no \emph{a priori} knowledge of the $J$ distribution, nor the ability to tune their value (at least at this experimental stage: see chapter \ref{Ch:conclusions} for developments on this topic), we will use the critical temperature found in equation (\ref{eq:criticaltemp}) as a reference value to tune the Monte Carlo parameter $\beta$ rather than a result to compare to. Simulations are run in a wide range around $1/J$, from approximately one order of magnitude below to one order of magnitude above, to ensure that the critical point is sampled.

\subsection{Experimental procedure for the measure}
We ideally want to measure the coupling constant between each couple of spins. This is not practically possible since the number of couplings we need to measures is of order $O\left(N^2\right)$. We therefore sample the distribution as we randomly choose some spin couples on which the measure is performed.\\

Once two spins are chosen, say $\sigma_1$ and $\sigma_2$, how is the measure carried on? The idea is to send to the DMD a globally \emph{off} state, namely a state in which no micromirror sends light into the correct optical path: all the incoming light is disperse and the camera measures no photons (or, rather, only rare counts due to noise).\\

Then, the two micromirrors corresponding to the chosen spins are set to the active position and the speckle pattern appears on the camera sensor. This, however, is not a direct measurement of the coupling constant: in fact, we measure the following contribution:

\begin{align}
	I_{12} &= I_0 \sum_{i,j}J_{ij}\,\sigma_i\,\sigma_j \left[\left(\delta_{i1}+\delta_{i2}\right)\left(\delta_{j1}+\delta_{j2}\right)\right]=\\
	& = I_0\,\left(J_{11} + J_{12} + J_{21} + J_{22}\right)
\end{align}
and, because the $J_{ij}$ is the intensity contribution from having activated the $i$-th and the $j$-th pixel, $\mymatrix{J}$ is symmetric. We can therefore write:
\begin{equation}
	I_{12} = J_{11} + 2\,J_{12}+J_{22}
\end{equation}
We are only interested in the $J_{12}$ term, so we need to invert the last formula, obtaining:
\begin{equation}
	J_{12}=\frac{1}{2}\left(I_{12} -J_1-J_2\right)
\end{equation}

In which $J_1$ and $J_2$ are the intensity contributions of having only one active spin. Hence, a cycle of three camera measures is needed for each coupling constant. The algorithm used to perform the measure is really simple and is presented here in pseudo-code.\\

\begin{algorithm}[H]
 \vspace{0.1cm}
 \KwData{The spins of which we want to measure the coupling constant: $\left(\sigma_i,\sigma_j\right)$}
 \KwResult{Measure the coupling constant $J_{ij}$}
\vspace{0.1cm}
send$\_$config$\_$to$\_$DMD(0)\;
DMD$\_$flip$\_$spin($\sigma_i$)\;
$\mymatrix{C}$ $\leftarrow$ camera$\_$measurement( )\;
$I_i$ $\leftarrow $ sum$\_$elements($\mymatrix{C}$)\;
\vspace{0.4cm}
DMD$\_$flip$\_$spin($\sigma_i$)\;
DMD$\_$flip$\_$spin($\sigma_j$)\;
$\mymatrix{C}$ $\leftarrow$ camera$\_$measurement( )\;
$I_j$ $\leftarrow $ sum$\_$elements($\mymatrix{C}$)\;
\vspace{0.4cm}
DMD$\_$flip$\_$spin($\sigma_i$)\;
$\mymatrix{C}$ $\leftarrow$ camera$\_$measurement( )\;
$I_{ij}$ $\leftarrow$ sum$\_$elements($\mymatrix{C}$)\;
\vspace{0.4cm}
$J_{ij}$ $\leftarrow$ $\frac{1}{2}\left(I_{ij} -I_i-I_j\right)$
\vspace{0.2cm}
 \caption{Algorithm for measuring the coupling constant}
\end{algorithm}
\vspace{0.7cm}

The results of the measurements are shown in figure \ref{Jdistribution} and in table \ref{tab:Jmoments}. We obtain for the coupling constant a nearly-gaussian distribution. The distribution has zero mean and is basically a small deviation form a gaussian distribution, in the sense that the standard deviation is much greater than both the skewness and the kurtosis.\\

We therefore will be able to compare the spin glass results with the one available for the Sherrington-Kirkpatrick model, since this was the only assumption of the model not straightforwardly coming from the experimental setup and still to be checked.

\begin{figure}[H]
\begin{center}
	\includegraphics[width=\linewidth]{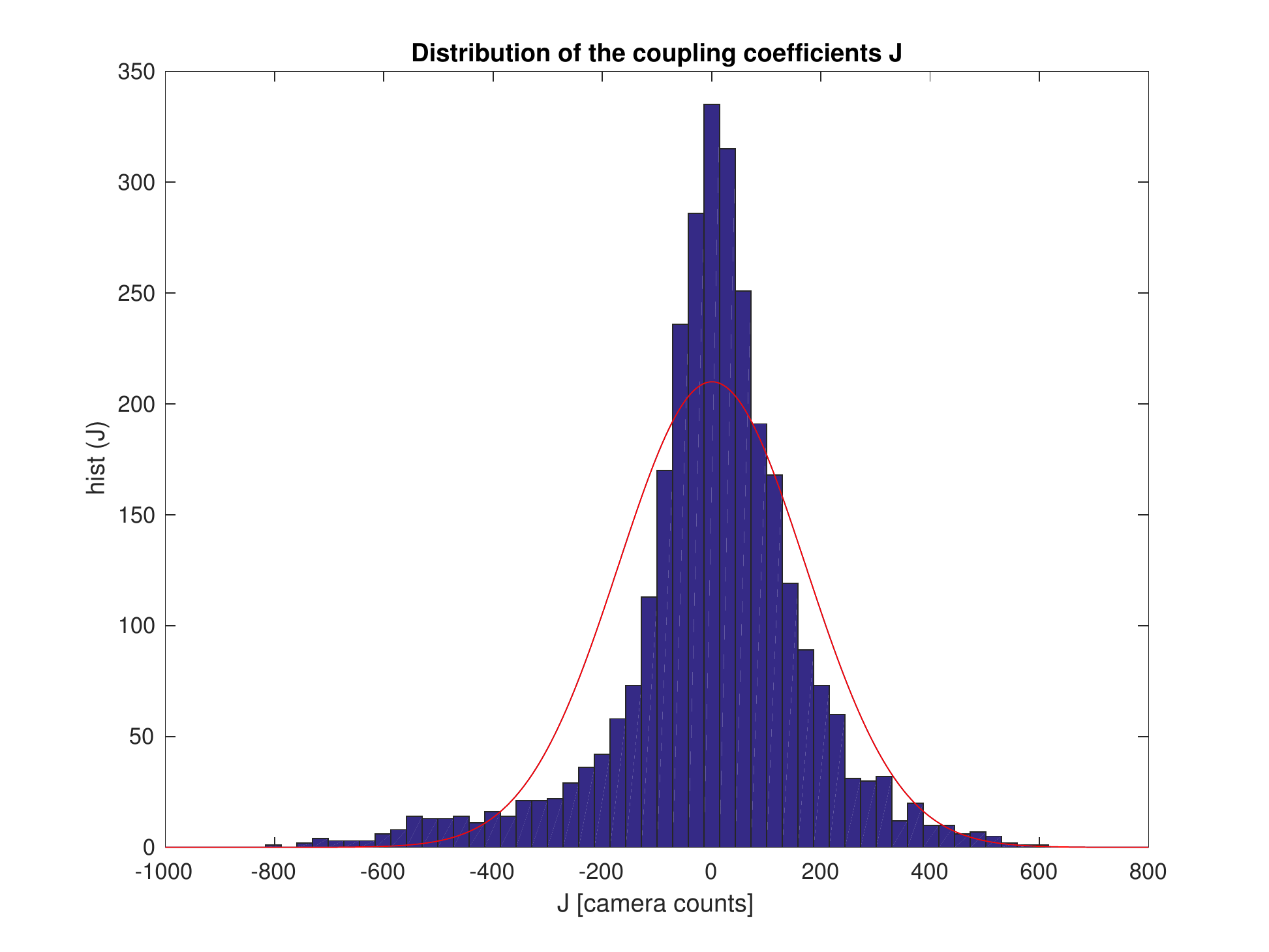}
	\caption{Coupling constants distribution form a random sampling of their intensity. The gaussian distribution with the same average, standard deviation and normalisation has been plotted over, as a reference for the corresponding SK model.}\label{Jdistribution}
\end{center}
\end{figure}

\begin{table}[H]
	\begin{center}
		\renewcommand\arraystretch{1.2}
		\begin{tabular}{|c|c|}
		\hline
		Moment	&	Value\\
		\hline
		\hline
		Mean	& $\left<J\right>=0.779\pm0.012$ \\
		\hline
		Standard deviation &	$\sigma_J=171\pm9$\\
		\hline
		Skewness	& $\gamma_1(J)=-0.747\pm0.010$\\
		\hline			
		Kurtosis & $\mbox{Kurt}(J) = 5.71\pm0.06$\\
		\hline
		\end{tabular}
		\renewcommand\arraystretch{1}
	\end{center}
	\caption{The moments of the distribution of the coupling constant. Results are given with three significant digits and errors are assigned accordingly.}\label{tab:Jmoments}
\end{table}

\chapter{Data analysis}\label{Ch:data_analysis}
\epigraph{On fait la science avec des faits, comme on fait une maison avec des pierre; mais une accumulation de faits n'est pas plus une science qu'un tas de pierres n'est une maison.}{\textit{Henri Poincaré, La Science et l'Hypothèse (1902)}}

\section{Autocorrelation measure}

This is the principal measurement performed on the optical spin glass system. The idea is to save the full state of the system at each step to calculate the autocorrelation function of the spins, namely:
\begin{equation}\label{eq:exp_autocorr}
	C(t) = \frac{1}{N}\sum_{i=1}^N\left<\sigma_i(0)\sigma_i(t)\right>
\end{equation}
This quantity is very useful from an experimental point of view: because the average is taken over all spins of the system, we have a good statistics even with a reasonably small number of Monte Carlo steps. The presented results are relative to $50000$ steps Monte Carlo runs with a system size of $21\times21$\footnote{This is arguably an unusual number of spins for a simulations, and was chosen due to the method used to address the spin variable by the DMD controller. Basically, a centre mirror chosen and a square frame is activated around it. Having a $21\times21$ system corresponds to having a $10$ spins frame around the central mirror.} spins, so just above 100 steps per spin. However, since the total number of spins is $21\times21=441$, we can potentially get a number of terms to average in the order of millions: in fact, we get the complete expression of (\ref{eq:exp_autocorr}) with time average:
\begin{equation}\label{eq:exp_autocorr_t}
	C(t) = \frac{1}{N}\frac{1}{t_{max}}\sum_{i=1}^N\sum_{\tau=0}^{t_{max}-1}\left<\sigma_i(\tau)\sigma_i(\tau+t)\right>
\end{equation}

Actually, two subsequent configurations are strongly correlated and therefore little to no information is added by averaging equation (\ref{eq:exp_autocorr}) with a configuration with just one step shift. Hence, the average is performed only for values of the shift $\tau$ bigger than the size of the system, thus averaging only some thousands of terms to obtain $C(t)$ from equation (\ref{eq:exp_autocorr_t}).\\

We firstly tried to fit the resulting function with it a combination of stretched exponentials of the form:
\begin{equation}
	C(t) = q_{EA}+(1-q_{EA})\,\left(e^{-\frac{t}{\tau}}\right)^\beta
\end{equation}
It was soon realised that such complicated non-linear fit required such precise starting guesses for the parameter to make extremely cumbersome the automation of the fitting script. In addition, the chosen fitting function yielded results for the parameter $\beta$ which were always very close to one (never was it observed a difference greater than 10\%). It was therefore decided to use a simpler function, namely the very same combination without the $\beta$ parameter:
\begin{equation}
	C(t) = q_{EA}+(1-q_{EA})\,e^{-\frac{t}{\tau}}
\end{equation}
this formula provided perfectly equivalent results for the $q_{EA}$ parameter, and only slightly different values for $\tau$, with an observed difference never greater than 5\%. On the other hand, the fit algorithm greatly improved the stability with the latter formula, so that the fit procedure could be successfully automated. We therefore decided to drop the exponential stretch and increase by 10\% the error on the fitted $\tau$ values, while retaining the fitted result for the $q_{EA}$ parameter.\\

The analysis is performed for several runs and the dependence on the temperature parameter is studied. In figure \ref{fig:acorrfuncts_T} some autocorrelation functions are plotted together with their fitted curve. They are ordered from the bottom-left towards the top-right in decreasing temperature. The data used for this figure are the system configurations reached in the Monte Carlo simulation, after having cut off the the states reached during the thermalisation steps, plus the first 10\% of data of the already thermalised system. This procedure was chosen to ensure  the Monte Carlo simulation was running on a perfectly equilibrated system. We note that not all the sampled temperatures are reported, mainly for readability issues.

\begin{figure}[H]
	\begin{center}
		\includegraphics[width=0.9\linewidth]{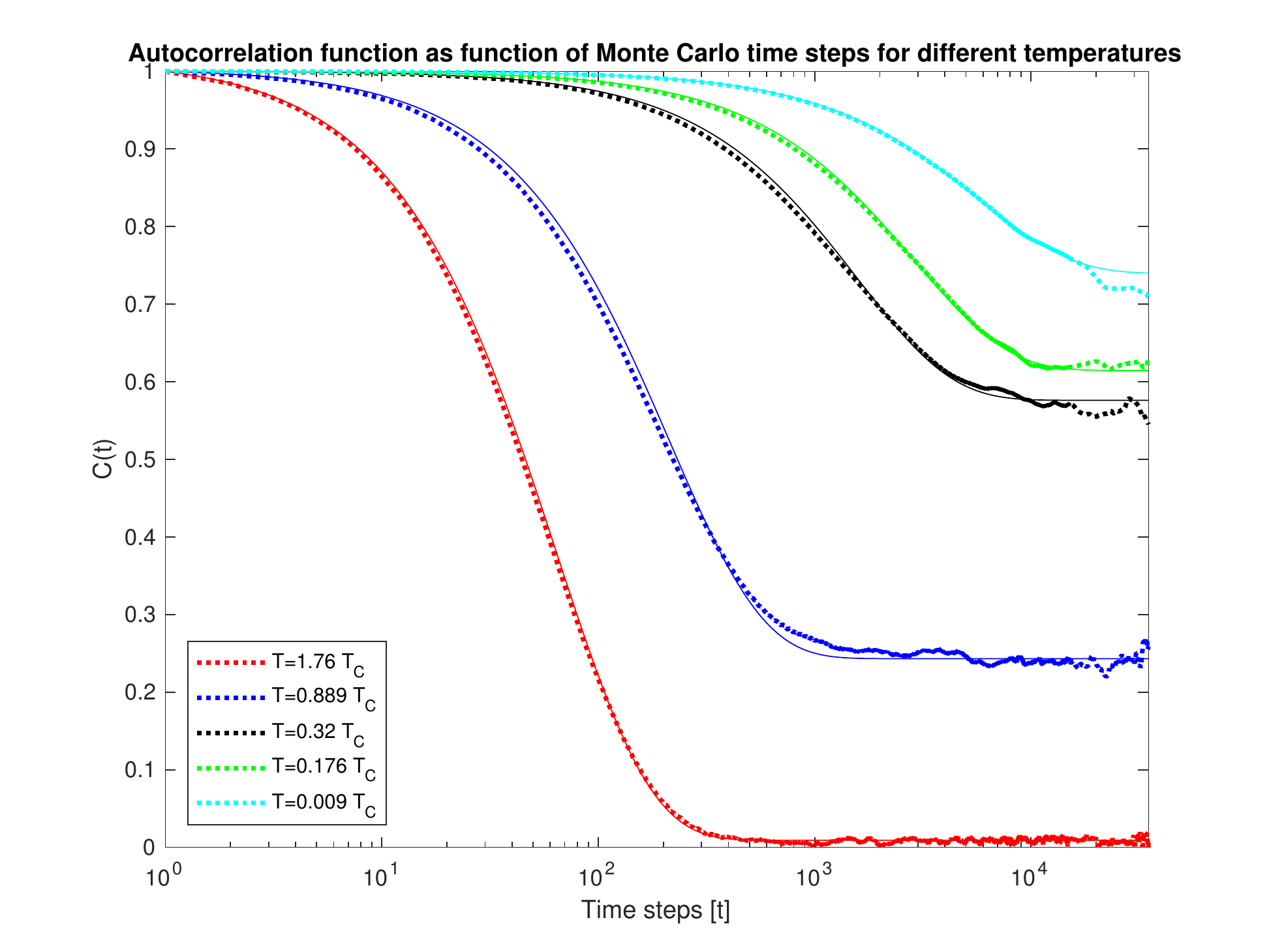}
		\caption{Autocorrelation functions at different effective temperature $\beta$ calculated from the data of the Monte Carlo algorithm. Data have been obtained from the simulation on the \emph{optical spin glass}.}\label{fig:acorrfuncts_T}
	\end{center}
\end{figure}

We can see that the autocorrelation function of the system never stays fixed at unity, not even at very low temperature, a situation in which a thermalised system should remain in a fixed state and therefore have an autocorrelation function flat at $C(t)\equiv1$.  This can be easily understood by closely  analysing the state of the thermalised system. Suppose we are running a simulation at $T\ll T_C$, a typical situation in which little to no decorrelation should be found. The thermalised system is in a local (deep) maximum of the measured intensity $I_{max}$\footnote{Usually, this point is considered to be a minimum, but in section \ref{sec:negen} we changed the sign in the definition of the energy: each minimum point is therefore a maximum, and vice-versa.}. The Monte Carlo algorithm has a negligible probability of accepting a move that decreases the measured intensity, but it can happen that the trial move bring the algorithm to a state with corresponding energy $I_T\lessapprox I_{max}$. If this happen, and if the intensity difference between the two measure is comparable to the noise $\sigma_I$ on the intensity measurement itself:
\begin{equation}
\delta I = \left|I_T- I_{max}\right| \approx \sigma_I
\end{equation}
it can happen that the measured intensity is higher than the one on the maximum (either because the noise increased the non-maximum intensity, or dumped the maximum one). This causes the system to perform a random walk in the phase space in a neighbourhood of the local maximum, and this phenomenon produces the observed decorrelation. The effect of the measured intensity can be observed in figure \ref{int:fig1} and \ref{int:fig2} in which the intensity change during the Monte Carlo simulation is plotted against the Monte Carlo time, and the instability close to the maximum is zoomed.

\begin{figure}[!ht]
  \centering
  \begin{subfigure}[b]{0.99\linewidth}
    \centering\includegraphics[width=0.75\linewidth]{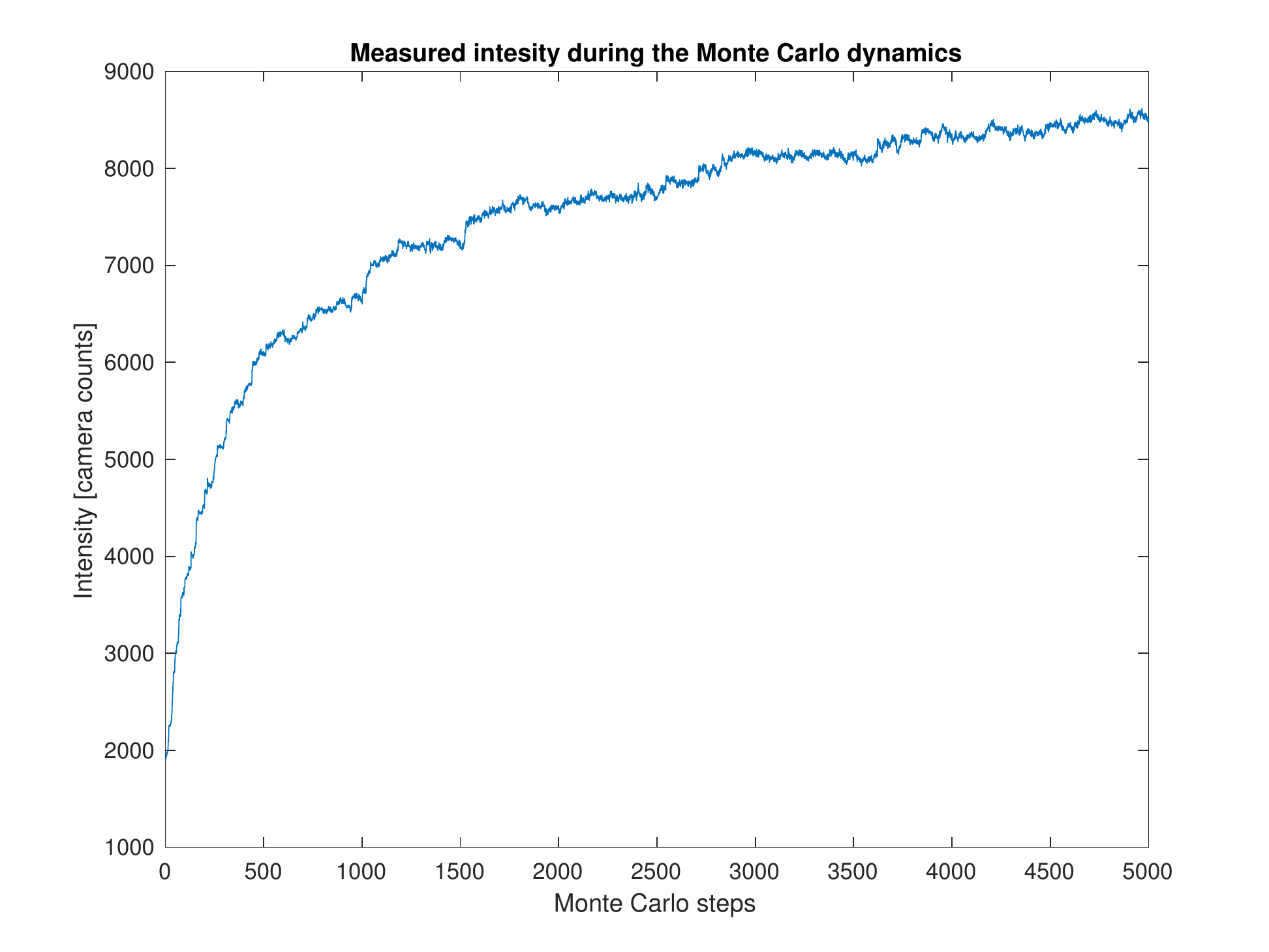}
    \caption{\label{int:fig1}Measured intensity during the Monte Carlo simulation, recorded at every time step. This simulation was run at $T=3.6\,10^{-5}\,T_C$, hence in a regime in which no moves that decrease the total intensity should be accepted.}
  \end{subfigure}\hspace{0.05\linewidth}\\\vspace{0.2cm}
  \begin{subfigure}[b]{0.99\linewidth}
    \centering\includegraphics[width=0.75\linewidth]{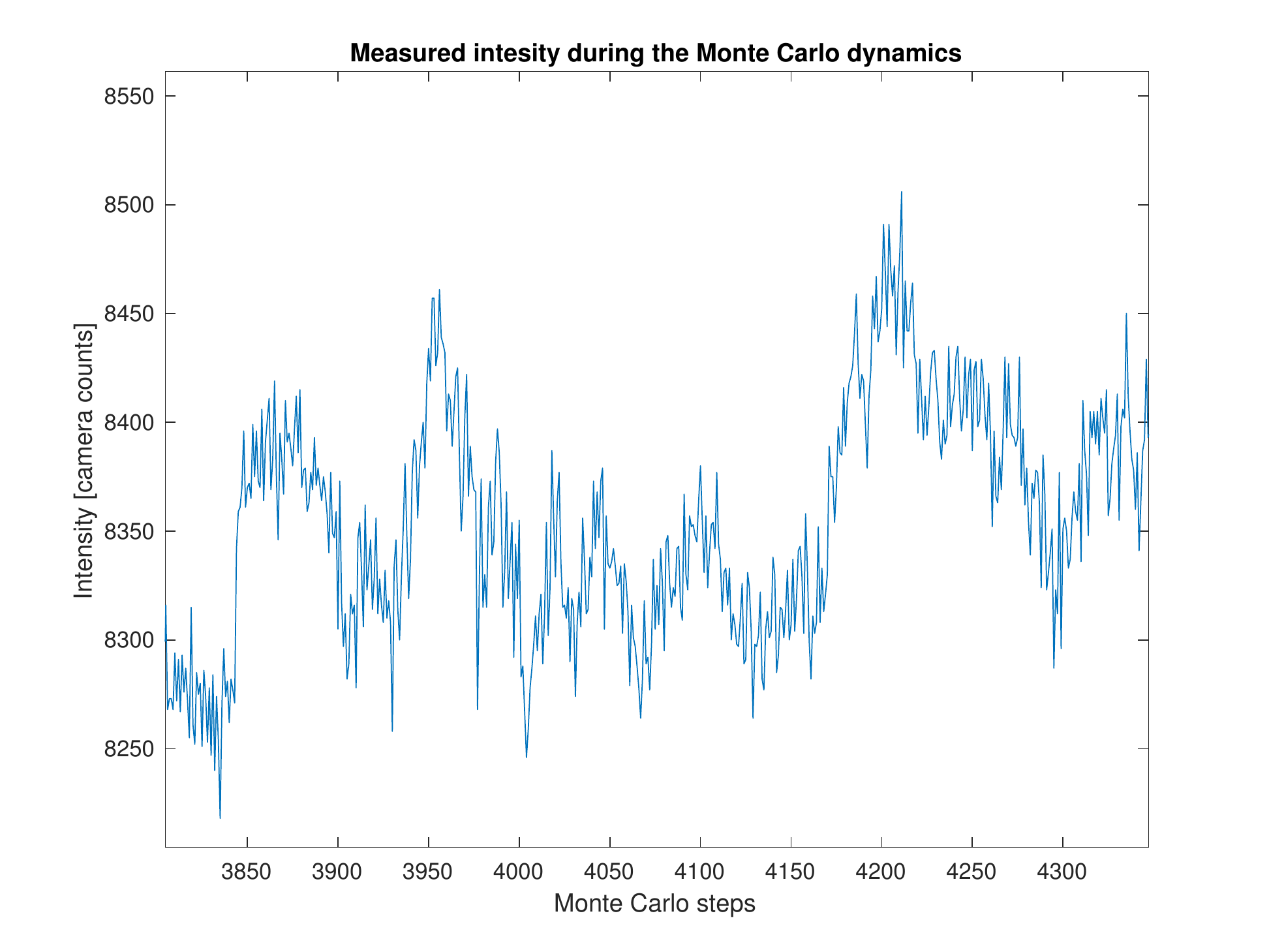}
    \caption{\label{int:fig2}Zoom of figure \ref{int:fig1}, in which is clear the phenomenon of instability near the maximum, and the consequent random walk near the maximum in phase space.}
  \end{subfigure}
  \caption{}
\end{figure}

In figure \ref{fig:qea_t} and \ref{fig:tau_t} the two fit parameter are shown as function of the temperature and the inverse temperature, respectively. The instability analysis we just mentioned above is clear in the trend of the Edward-Anderson parameter $q_{EA}$, which do not reach 1 at zero temperature as expected. Instead, we are capped at a value of approximately
\begin{equation}
q_{EA}(T=0)=0.64
\end{equation}

This is in not a problem fo the Edward-Anderson parameter, whose main feature is the discontinuity at the critical temperature, but may be a problem for the autocorrelation time $\tau$, which indeed deviates from a power law trend at low temperatures.

\begin{figure}[H]
	\begin{center}
		\includegraphics[width=0.9\linewidth]{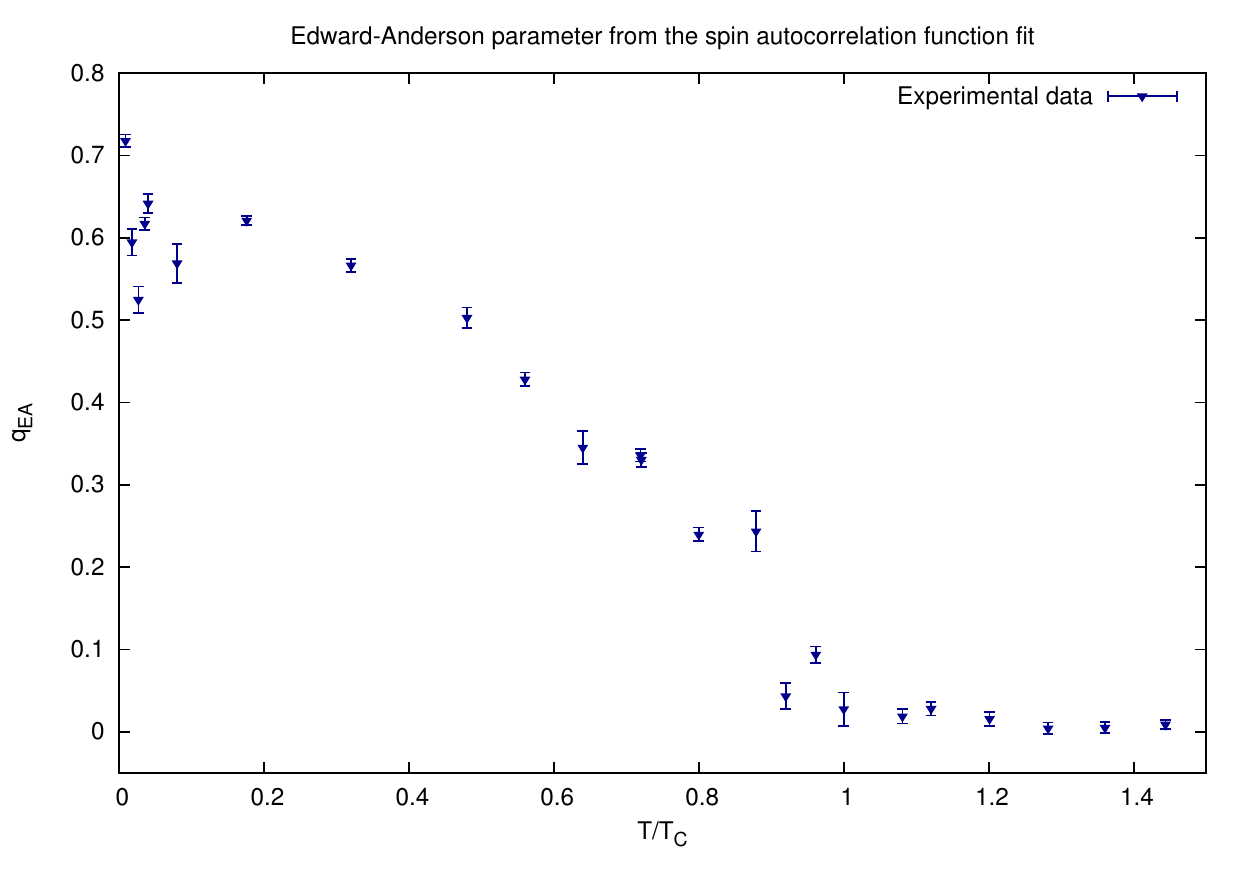}
		\caption{Edward-Anderson parameter fitted from the long time behaviour of the spin autocorrelation function, as a function of the reduced temperature}\label{fig:qea_t}
	\end{center}
\end{figure}

\begin{figure}[H]
	\begin{center}
		\includegraphics[width=0.9\linewidth]{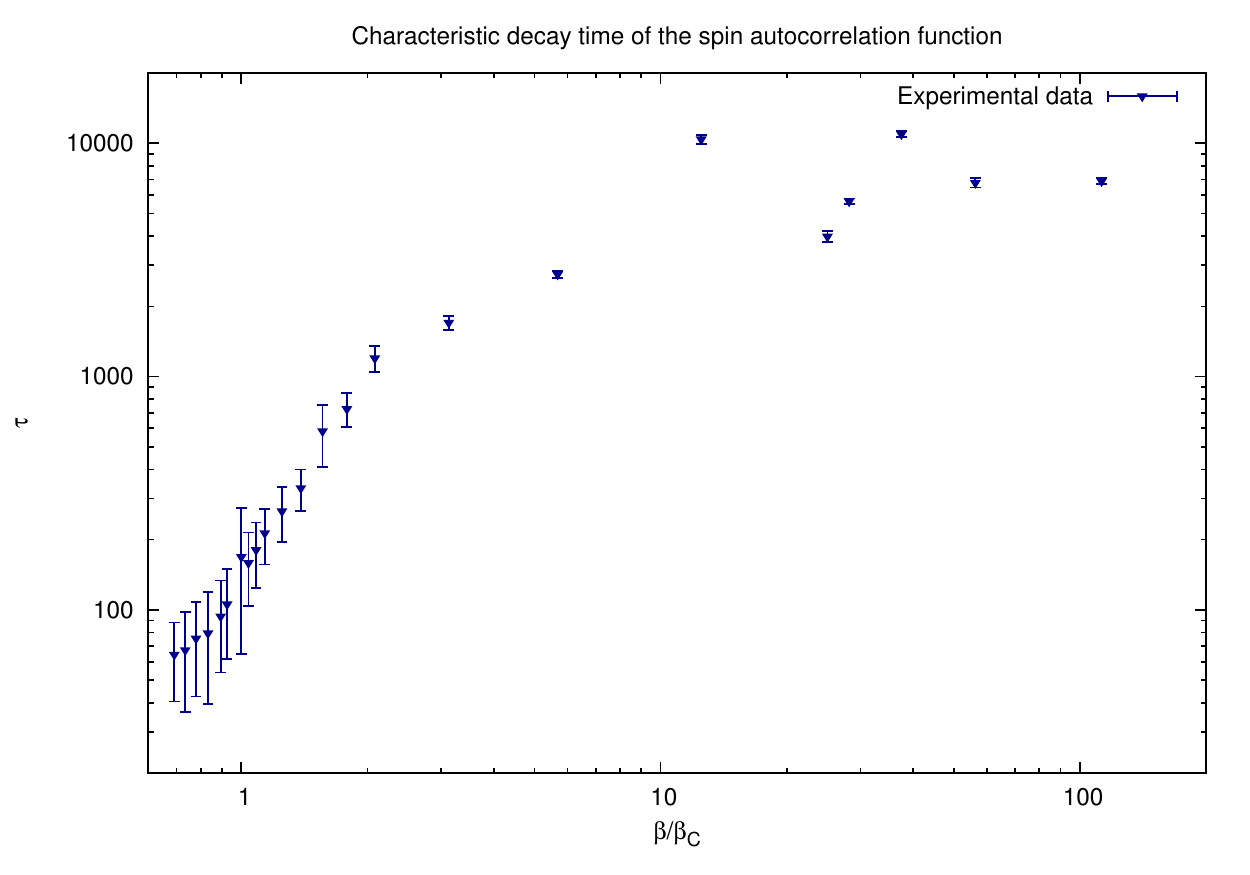}
		\caption{Characteristic autocorrelation time fitted from the long time behaviour of the spin autocorrelation function, as a function of the reduced temperature.}\label{fig:tau_t}
	\end{center}
\end{figure}

To investigate whether this effective temperature shift due to the experimental instability affects the characteristic autocorrelation time $\tau$, we try to get rid of the \emph{noise temperature} $T_C$. This is done by extracting $T_C$ from a linear interpolation of the Edward-Anderson parameter. Namely, we aim to rescale the temperature in such a way that:
\begin{itemize}
	\item the critical temperature remains fixed at $T_C=1$ in reduced units
	\item the minimum sampled temperature by the experiment corresponds to the interpolated temperature $T_i$
\end{itemize}

Because we want to interpolate over a line with a $-1$ angular coefficient, we immediately get:
\begin{equation}
	T_i = 1-q_{EA}(T=0) = 0.36
\end{equation}
and the corresponding transformation is:
\begin{equation}
	T \longrightarrow \frac{T+T_i}{1+T_i}
\end{equation}

The same quantities are plotted again in the new rescaled temperature:
\begin{figure}[H]
	\begin{center}
		\includegraphics[width=0.8\linewidth]{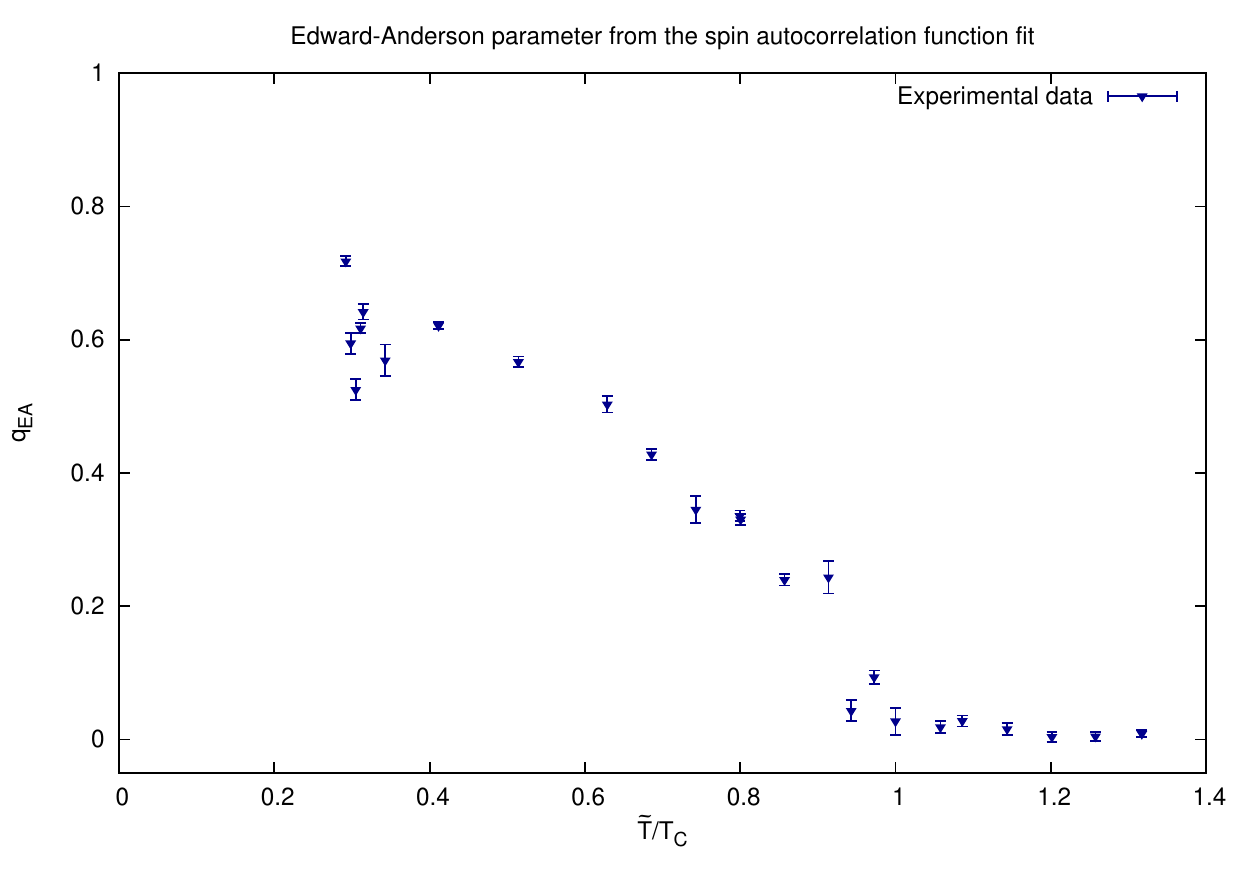}
		\caption{Rescaled Edward-Anderson parameter to compensate for the experimental temperature shift. It is clearly visible the predicted discontinuity at the critical temperature, typical for a order parameter.}\label{fig:qea_SHIFTED}
	\end{center}
\end{figure}

\begin{figure}[H]
	\begin{center}
		\includegraphics[width=0.8\linewidth]{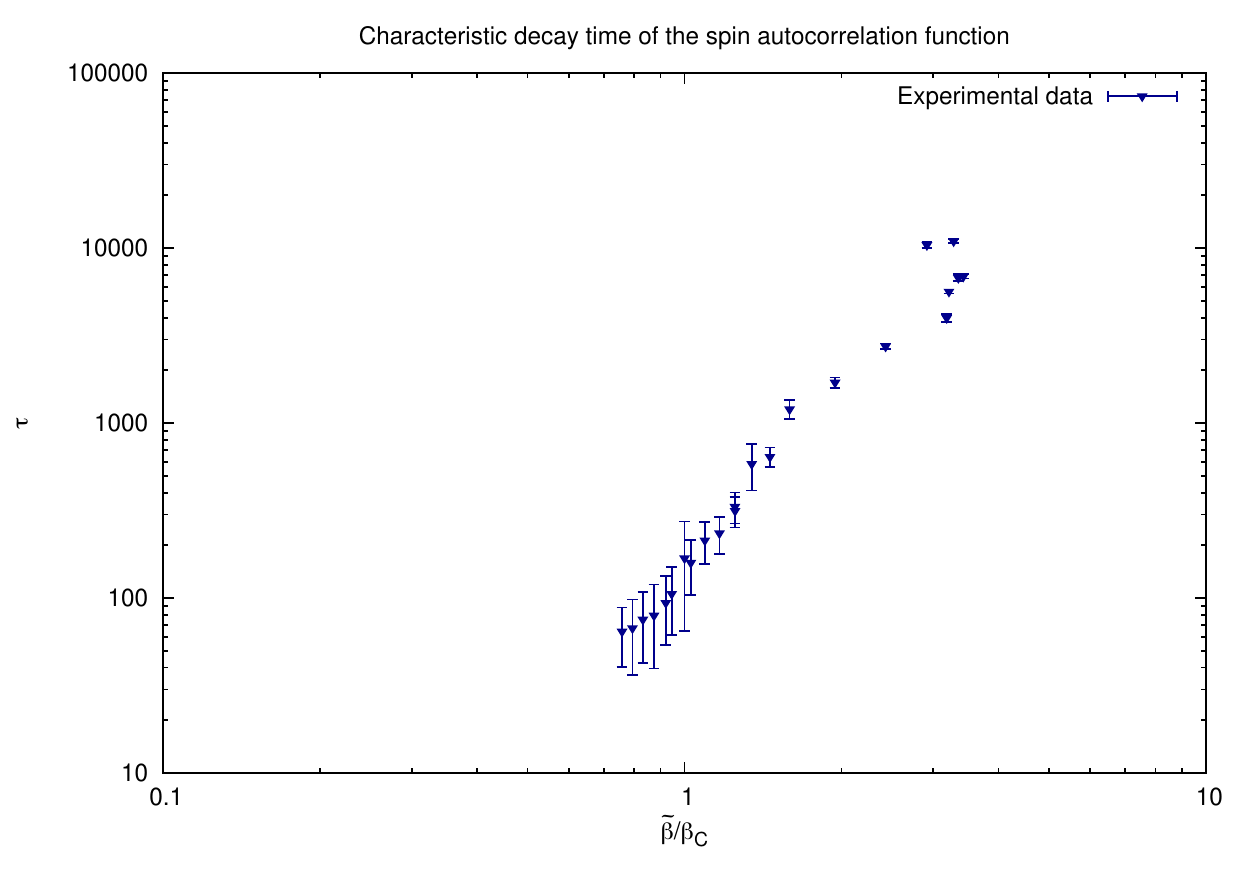}
		\caption{Rescaled characteristic autocorrelation time to compensate for the experimental temperature shift. The rescaling strongly rectifies the result, yielding a power law trend for the $\tau$ parameter.}\label{fig:tau_SHIFTED}
	\end{center}
\end{figure}

We see that the rescaled temperature has indeed a significant rectifying effect on the trend of the characteristic autocorrelation time $\tau$.

\chapter{Conclusions and perspectives}\label{Ch:conclusions}
We have proposed an optical system to simulate a spin glass. Firstly, in chapter~\ref{Ch:theory_optics}, we theoretically proved that the intensity function of such setup has the same mathematical formulation as the spin glass Hamiltonian. We then tried to prove the equivalence experimentally through the measure of the order parameter associated to the optical system. The autocorrelation functions produced a result for the Edward-Anderson parameter of the optical spin glass which is remarkably close to the results found via computer simulations.\\

In particular, the result is compared to the one found by David Sherrington and Scott Kirkpatrick for the system they proposed, a result which is given in absence of external magnetic field \cite{PhysRevLett.35.1792}. We note that we do not expect a perfect superposition because of the results of the coupling distribution. Indeed we recall that the optical spin glass is not a perfect Sherrington-Kirkpatrick spin glass, but differs for two details:
\begin{itemize}
	\item we have no control over the coupling, which are embedded in the disorder of the scattering sample. It's impossible to have a perfectly gaussian distribution of the $J$'s: the actual moments of their distribution has been estimated in chapter \ref{Ch:preliminary} and the results presented in table \ref{tab:Jmoments};
	\item we are using the Amit representation, namely our spins can assume the values $\sigma=\{-1,+1\}$, and this produced an effective magnetic field (see section \ref{sec:twoising} for the details).
\end{itemize} 

Despite these limitation, we see in figure \ref{qeaparameter} an almost perfect superposition of the two results. We take this as the final check that \emph{the proposed optical system indeed behaves as a spin glass-like system}.

\begin{figure}[!ht]
  \centering
  \begin{subfigure}[b]{0.45\linewidth}
    \centering\includegraphics[width=0.99\linewidth]{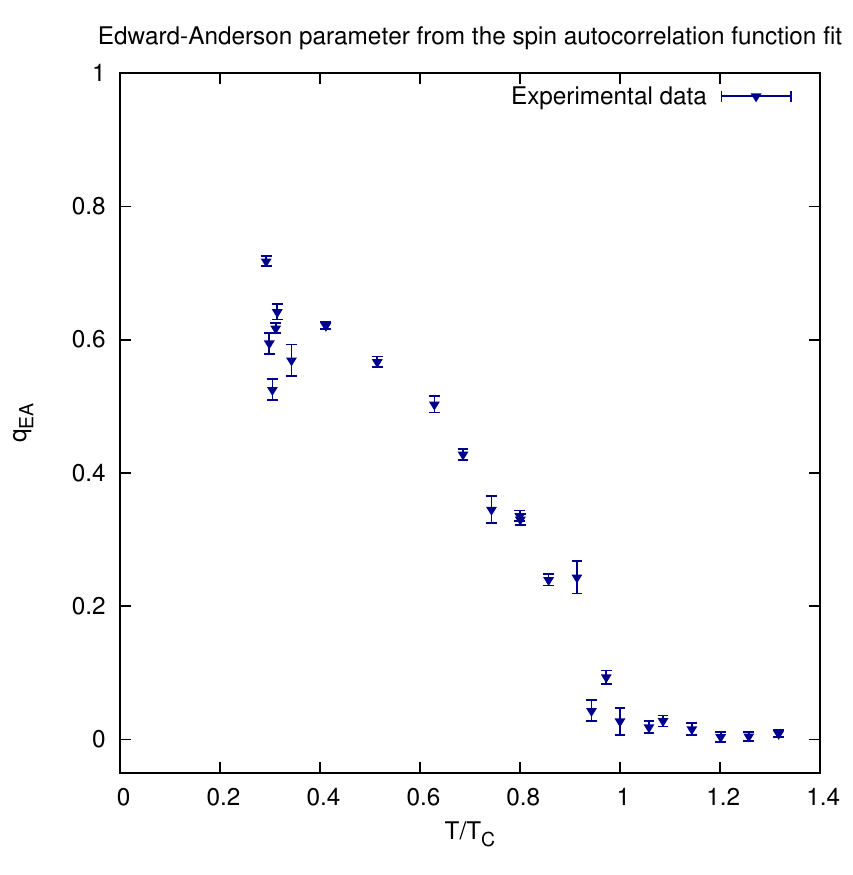}
    \caption{\label{fig:qeame}Edward-Anderson parameter for the Sherrington-Kirkpatrick optical spin glass\\}
  \end{subfigure}\hspace{0.05\linewidth}
  \begin{subfigure}[b]{0.45\linewidth}
    \centering\includegraphics[width=0.99\linewidth]{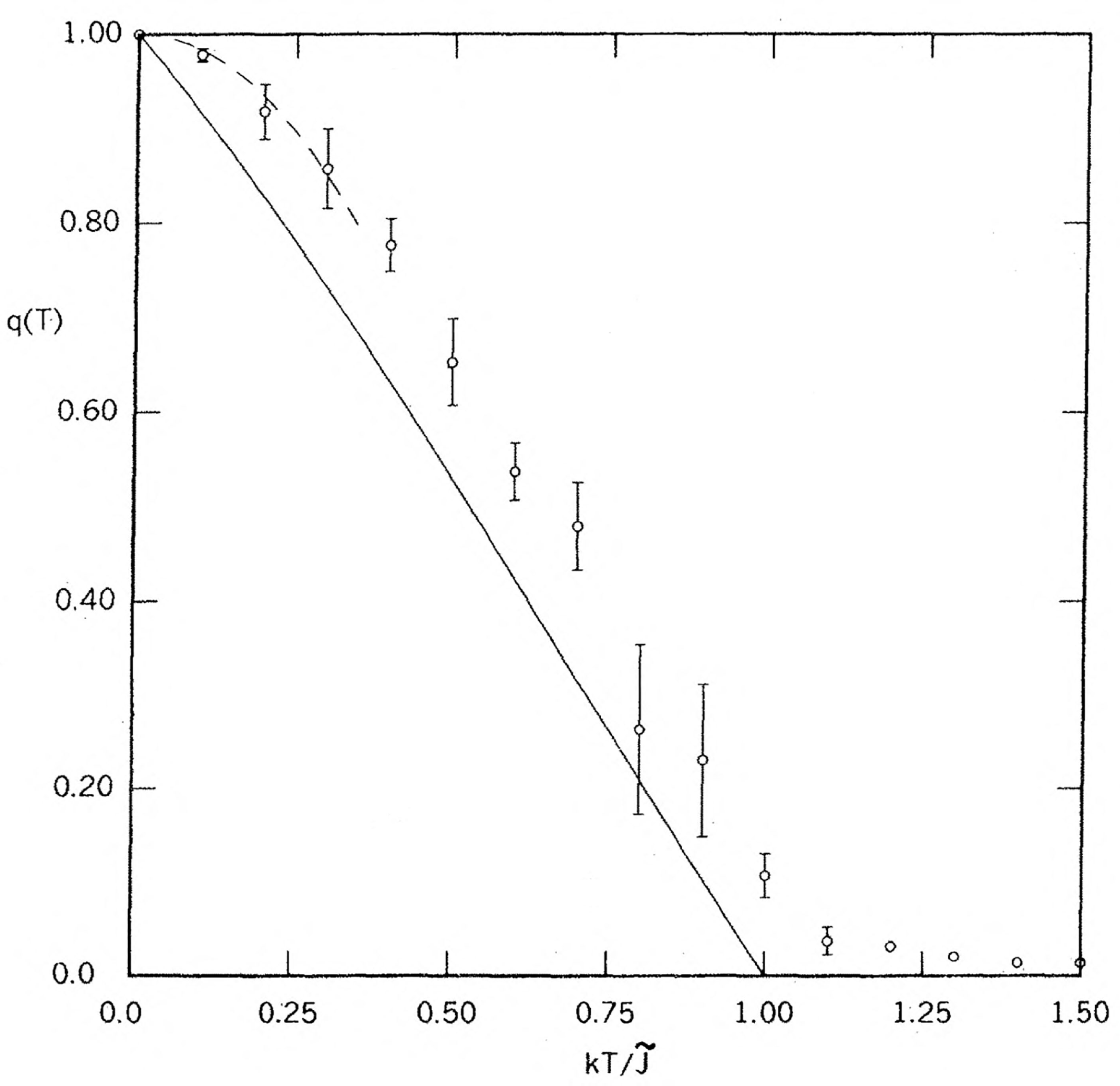}
    \caption{\label{fig:qeask}Edward Anderson parameter found by Sherrington and Kirkpatrick for their own system \cite{PhysRevB.17.4384}}
  \end{subfigure}
  \caption{Comparision between the original Sherrington and Kirkpatrick result for the Edward-Anderson parameter obtained for the system they proposed and the corresponding result for the optical system.}
  \label{qeaparameter}
\end{figure}

\section{Future experimental updates}
The experiment fall in the field of the so called \emph{all optical computations}, a branch of photonics in which the light is used to perform calculations that would require much time to be done on a traditional computer. The optical computation performed in the analysed system is the intensity measure, which substitute the complete calculation of the energy at each configuration in the Monte Carlo simulation. As mentioned in section \ref{sec:NP}, such numerical calculation scales as $N^2$, in which $N$ is the total number of spins in the system.\\

We now want to compare the time $\tau^{(0)}$ needed to perform a single step of the Monte Carlo algorithm both on a computer and on the optical spin glass system. For a computer we have:
\begin{equation}
	\tau^{(0)}_C = \alpha_C + \beta_C\,N^2
\end{equation}
in which $\alpha_C$ is the flipping time for a spin - a bit in the RAM of the computer - and $\beta_C$ is the time needed to read from memory and perform the calculation of each term in the summation of equation (\ref{eq:SKham})\footnote{This is true if we change an arbitrary number of spins at each step like in the Wolff or Swendsen-Wang algorithm. Usually the scaling is better that the one presented here, but is no better than $$\tau^{(0)}_C = \alpha_C + \beta_C\,(N-1)$$ because even if we change just one spin we need to calculate the new contribution for the other $N-1$ spins}. We neglect the time needed to perform the exponential operation in the calculation of the Boltzmann factor in equation (\ref{eq:Metropchoice}) and for moving data from the RAM to the processor.  On the other hand, the single step of the optical spin glass involves a time:
\begin{equation}
	\tau^{(0)}_O = \alpha_O + \beta_O
\end{equation}
hence presents no scaling at all: no matter how many spins we have it takes exactly the same time to perform a Monte Carlo step, because one intensity measure takes always exactly one exposure time of the camera.\\

The drawbacks of the optical spin glass are significative. Despite the absence of scaling, the coefficients in the two relation presented above are much greater that the one on the computer:
\begin{equation}
	\alpha_O \gg \alpha_C \qquad \beta_O \gg \beta_C
\end{equation}
the reasons for these are quite simple: in the $\alpha$ coefficient is embedded the spin flip time. While a computer only need to change the logical value stored in a random access memory, the optical spin glass needs to move a physical micromirror, which comes with an associated mass, angular momentum, stabilisation time. The $\beta$ coefficient, the measure time parameter, is much bigger for the optical setup as well: while for the computer the operation only involves a multiplication which the CPU performs in a few clock cycles, the measure on the optical spin glass needs a complete exposure of the camera sensor.\\

For all these reasons, it is clear that \emph{the optical spin glass can be competitive only for very large values of N}. The idea is therefore to enhance the experimental setup with two objectives: make the system as fast and stable as possible, while adding the capability of handling a great number of spins. The first step is to expand the laser beam to illuminate a bigger area of the DMD and a greater number of micromirrors. This can be done with relative simplicity via a Keplerian telescope lens system or its equivalent Galilean arrangement. Some beam expander are available on the market as one pre-assembled piece. This ensures the scalability to a great number of spins, the limit being the full resolution of the DMD array. However, since the DMDs come from the cinema industry (see section \ref{sec:DMD}), they commonly comes in standard resolutions up to HD, having over 2 millions micromirrors in a $1080\times 1920$ resolution.\\

While the optical spin glass has no scaling on the single Monte Carlo step, we still have an exact one-to-one correspondence between the number of moves to complete a full Monte Carlo sweep of the system. To do it as quickly as possible, a photodiode should be used as sensor. The photodiode performs much better than the camera used until now. In particular, based on the specification given in table \ref{tab:photodspecs}, it has a dynamic range:
\begin{equation}
	DR = \frac{\mbox{maximum input}}{\mbox{NEP}}=\frac{1\,\mbox{mW}}{0.09\mbox{pW}/\sqrt{\mbox{Hz}}}=3.5\,10^8
\end{equation}
at a sampling rate of 1 kHz, compared just to 255 of the camera (which was, however, limited by the 8-bit ADC). This will allow to increase the number of the spins $N$, without having to flip too many spins at each move.\\

\begin{table}[h!t]
\begin{center}
	\begin{tabular}{|c|c|}
	\hline
	Specification	&	Value \\
	\hline
	\hline
	Detector		&	UV-enhanced silicone APD\\
	\hline
	Wavelength range		&	$\lambda = 200nm$ to $\lambda = 1000nm$\\
	\hline
	Detector Active Area	& 0.5 mm (diameter)\\
	\hline
	Maximum conversion gain &	$12.5\times10^6$ V/W\\
	\hline
	Output bandwidth &	DC to $10$ MHz\\
	\hline
	CW saturation power	&	$0.32 \mu W$ @ M=50 to $3.20 \mu W$ @ M=5\\
	\hline
	Maximum input power &	$1$ mW\\
	\hline
	M factor adjustment 	&	$M = 5$ to $50$ \\
	\hline
	Minimum NEP	&	$0.09$ pW/$\sqrt{\mbox{Hz}}$\\
	\hline
	Integrated noise &	0.28 nW\\
	\hline
	Electrical output &	BNC\\
	\hline
	Impedance	&	50 $\Omega$\\
	\hline
	\end{tabular}
	\caption{Specifications table for the photodiode THORLABS \emph{APD410A2}}\label{tab:photodspecs}
\end{center}	
\end{table}

To ensure the necessary stability for such a complex system, a thermal insulation is needed: it will be provided by moving the setup into a box with insulating walls made of polystyrene. This is much easier than setting up a tube around the laser beam and provides similar result. There is no protection against air movements, but the overall stability should improve greatly anyway.\\

These are the three planned enhancements to the system to be able to simulate the very same Sherrington-Kirkpatrick system with a great number of spins, thus in situation in which the described system can be competitive with, or even outperform, the traditional approach of computer simulations.\\

\section{Other applications} 
All-optical computations are an extremely hot topic of research in these days especially because they can easily be applied in a number of fields: from quantum mechanics \cite{smith2018radix} to electronics \cite{bharti2018design} to neural networks \cite{lin2018all}.\\

The idea is to conclude the present work by hinting some possible developments of the proposed optical system to simulate other similar systems. 

\subsection{$p$-spin model}
The first generalisation we want to see is the $p-spin$ model spin glasses \cite{barrat1997p}. Such systems differs from the standard Sherrington-Kirkpatrick model for the many-body approximation order they include in the Hamiltonian. The $p-spin$ Hamiltonian has the form:
\begin{equation}
	H = -\sum_{1\leq i_1\leq i_2\leq\dots\leq i_p\leq N}J_{i_1i_2\dots i_p}\;\sigma_{i_1}\sigma_{i_2}\dots\sigma_{i_p}
\end{equation}
with $p\geq 3$ (for $p=2$ we get the Sherrington-Kirkpatrick model) and gaussian couplings $J_{i_1i_2\dots i_p}$ with zero associated mean and associated variance ${p!}/{\left(2\,N^{p-1}\right)}$. The unscattered beam must be dumped, for example by a set of two orthogonal polarisers, since the scattered intensity is much smaller with respect to the original intensity emitted by the laser source.\\

It is possible to create the $4-spin$ model optically by changing the sample to a non linear scattering material with a non trivial second-order nonlinear susceptibility. Such a sample would sustain a second harmonic generation (SHG) process \cite{franken1961generation}, in which two photons with the same frequency generate a new photon with twice the energy of the initial photons. This means that the resulting wavelength is half the one of the original radiation. One can therefore add a $\frac{\lambda}{2}$ filter in order to retain only the light intensity created by the frequency doubling process.\\

Each photons created in this way carries the information of two spins (the two mirrors from which the two forming photons are coming) and the calculation of the intensity doubles the number of spins involved, just as it happens in the Sherrington-Kirkpatrick model: compare equations (\ref{eq:OSG_E}) and (\ref{Hamiltonian_fulldetail}). Therefore, such a setup yields the following Hamiltonian:
\begin{equation}
	H = -\sum_{1\leq i_1\leq i_2\leq i_3 \leq i_4\leq N}J_{i_1i_2i_3i_4}\;\sigma_{i_1}\sigma_{i_2}\sigma_{i_3}\sigma_{i_4}
\end{equation}

which is exactly the $4$-spin Hamiltonian. Hence, with the \emph{trick} of second harmonic generation it's possible to create a $p$-spin model with $p=4$, but it's not easy to generalise the $p$-spin model further for arbitrary values of $p$. In figure \ref{persp:figb} the optical scheme for the optical $p$-spin setup is presented. 

\subsection{Hopfield model and neural networks}
For the applications to neural networks, we want to create a dynamical evolution in the form:
\begin{equation}\label{eq:hopfield_dynamics}
	\sigma(t+1) = \theta\left[\sum_{i} J_i \sigma_i\right]
\end{equation}
which represents the dynamical system associated to a Hopfield network. To do so, we let interact both the scattered field $E_1$ and the unscattered field $E_0$, coming directly form the laser source:
\begin{equation}
	E_R=\varepsilon E_0 + E_1
\end{equation}
Since the scattered field is usually much smaller that the other one, we set up a couple of almost orthogonal polarisers in order to cut off much of the intensity of $E_0$: we call $\varepsilon E_0$ the unscattered field which is let through by the second polariser. We the calculate the total intensity received by the detector:
\begin{equation}\label{eq:hopield_int}
	{\left|E_R\right|}^2 = \varepsilon^2\,E_0^2 + E_1^2 +\varepsilon\, E_0\, E_1
\end{equation}
we now express the intensity ratio as:
\begin{equation}
	\frac{E_1}{E_0}\approx k\varepsilon
\end{equation}
we want to enforce the following condition:
\begin{equation}\label{eq:hopfield_wcond}
	E_1^2 \ll \varepsilon \,E_0 \, E_1 \lessapprox \varepsilon^2 \, E_0^2
\end{equation}
in which $\epsilon\ll k\ll 1$: 
\\

We know that the intensity reaching the camera has an expression in the following form:
\begin{equation}
	H^{(k)}=\sum_{i,j}\,J_{ij}^{(k)}\,\sigma_i\,\sigma_j
\end{equation}
in which the $^{(k)}$ stresses that the intensity depends on the chosen observation point $k$. Since this will represent just the $E_1$ contribution of equation (\ref{eq:hopield_int}), we will have a total intensity on the camera given by:
\begin{equation}
	H^{(k)}=\varepsilon^2\,E_0^2 + \sum_{i,j}\,J_{ij}^{(k)} \sigma_i E_0^{(j)}
\end{equation}
in which we have used the conditions in equation (\ref{eq:hopfield_wcond}) to drop the term bilinear in the $\sigma$ variables. Since all the incoming unscattered fields $E_0^{(j)}$ are plane waves, we can drop the sum over the $j$ index by replacing the specific $J_{ij}$ with the averaged quantity:
\begin{equation}\label{eq:droppped_j}
	\overline{J_i} = \sum_{j} J_{ij}\,E_0^{(j)}
\end{equation}

If we plug this last equations into the previous one we get: 
\begin{equation}
	H^{(k)}=H_0 + \sum_{i} \overline{J_i}^{(k)}\,\sigma_i
\end{equation}
which has the linear combination we want to generate the Hopfield dynamics as expressed in equation (\ref{eq:hopfield_dynamics}).\\

To implement the correct dynamic, we need to connect the pixels of the camera with the DMD mirrors with a one-to-one correspondence. We want to remark that the only requirement of this correspondence it to be bijective: there is no specific spatial requirements. In particular, the relative spatial position of a mirror and a pixel has no meaning whatsoever in the model.\\

Once this correspondence has been established, say $k=k(i)$, we can control the $i$-th mirror with the intensity read by the $k$-th pixel. Wewant to have:
\begin{equation}
	\sigma_i(t+1) = \theta\left[H^{(k(i))}-H_0\right]
\end{equation}
in which the theta function has to be done either by the control electronics or digitally by a simple algorithm.The optical setup to achieve the presented Hopfield dynamics is shown in figure \ref{persp:figc}; the electronics which performs the non-linear operation is not shown.

\begin{figure}[ht]
  \centering
  \begin{subfigure}[b]{0.95\linewidth}
    \centering\includegraphics[width=0.99\linewidth]{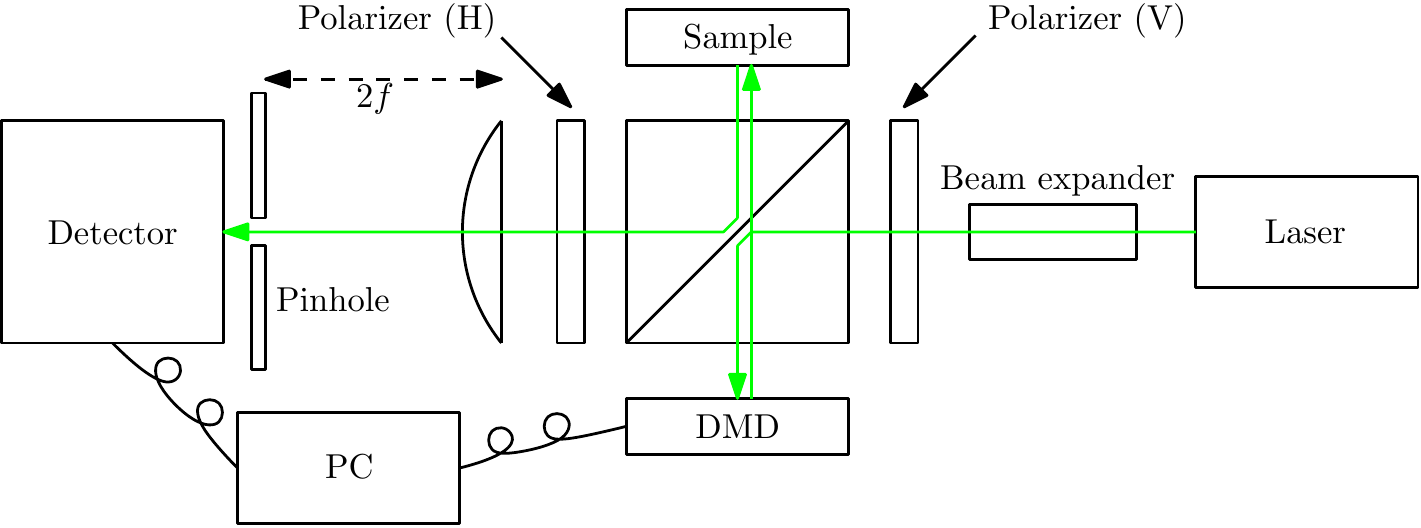}
    \caption{\label{persp:figa}Optical scheme for the simulation of a Sherrington-Kirkpatrick spin glass system}
  \end{subfigure}\hspace{0.05\linewidth}
  \\\vspace{1cm}
  \begin{subfigure}[b]{0.95\linewidth}
    \centering\includegraphics[width=0.99\linewidth]{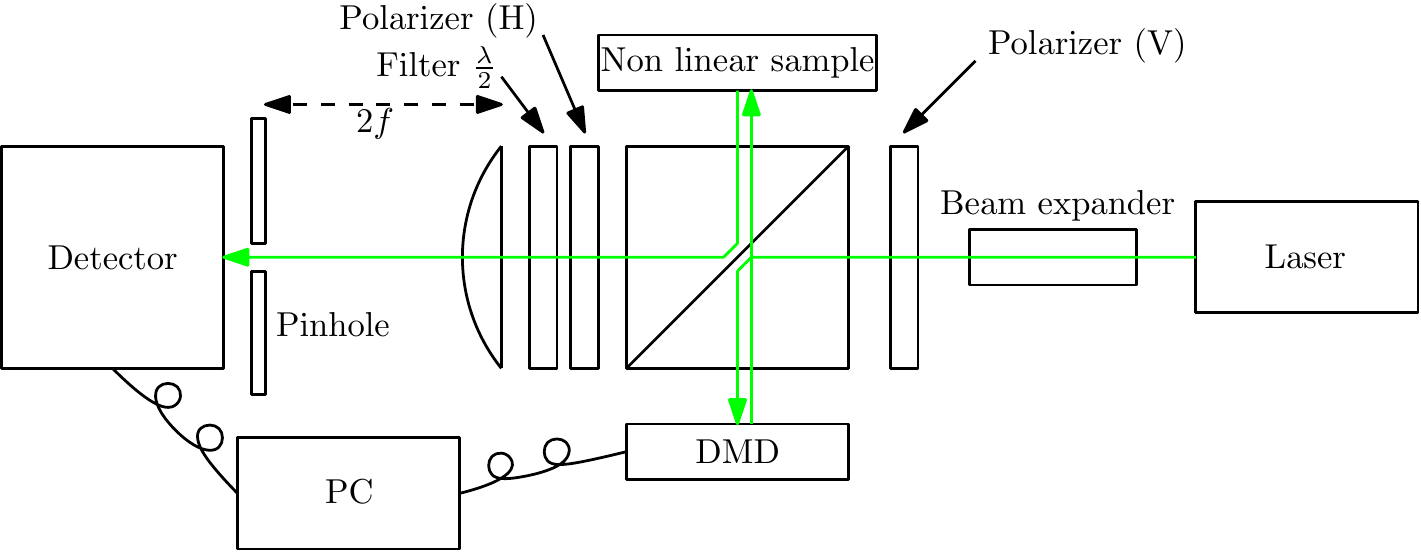}
    \caption{\label{persp:figb}Optical scheme for the simulation of a p-spin spin glass system, with $p=4$}
  \end{subfigure}\hspace{0.05\linewidth}
  \\\vspace{1cm}
  \begin{subfigure}[b]{0.95\linewidth}
    \centering\includegraphics[width=0.99\linewidth]{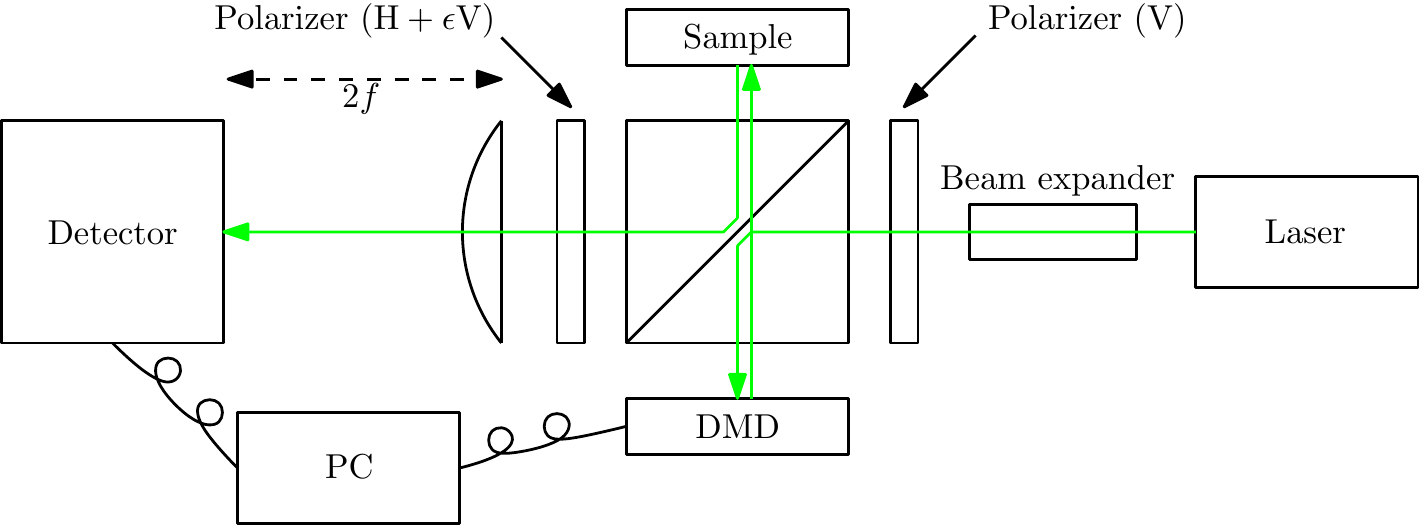}
    \caption{\label{persp:figc}Optical scheme for the simulation of a Hopfield neural network}
  \end{subfigure}\hspace{0.05\linewidth}
  \\\vspace{1cm}
  \caption{The three optical setup needed to simulate the Sherrington-Kirkpatrick spin glass model (figure \ref{persp:figa}), the p-spin spin glass model (figure \ref{persp:figb}) and the Hopfield neural network (figure \ref{persp:figc}). The similarity of the setups - and therefore the versatility of the proposed method - is self-evident.}
  \label{Future_experiments} 
\end{figure} 

To implement a learning algorithm to the proposed system, we will need to control the values of the Hopfield couplings $\overline{J_i}$. This can be changed via a \emph{spatial light modulator} (SLM) of liquid crystals, which can change the phase of the incoming fields $E_O^{(j)}$. We recall that the couplings $\overline{J_i}$ are indeed the result of an average over all possible incoming fields, namely:
\begin{equation}\label{eq:slmeffect}
	\overline{J_i} = \sum_{j}J_{ij} \, E_0^{(j)}
\end{equation}
normally, when this is inserted into the Hopfield dynamics
\begin{equation}
	H^{(k)}= H_0 +\sum_{i,j}J{ij}^{(k)}\,\sigma_i\, E_0^{(j)}
\end{equation}
the index $j$ is dropped from the average because the incoming fields $E_0^{(j)}$ are all plane waves, as in equation \ref{eq:droppped_j}), which we recall here for convenience:
\begin{equation}
	H^{(k)}= H_0 +\sum_{i}\overline{J_{i}^{(k)}}\,\sigma_i
\end{equation}
this last step in no longer possible if a SLM changes the relative phases of the $E_0^{(j)}$, because this changes the coupling constants in equation (\ref{eq:slmeffect}).\\

Thus we showed a convenient way of changing the coupling constants of the optical setup: this changes can be set up to be controlled by a suitable machine learning algorithm to produce a learning-capable optical system. The complete optical configuration to allow this learning implementation is shown in figure \ref{fig:complete_setup_hopfield_learning}.

\begin{figure}
	\begin{center}
		\includegraphics{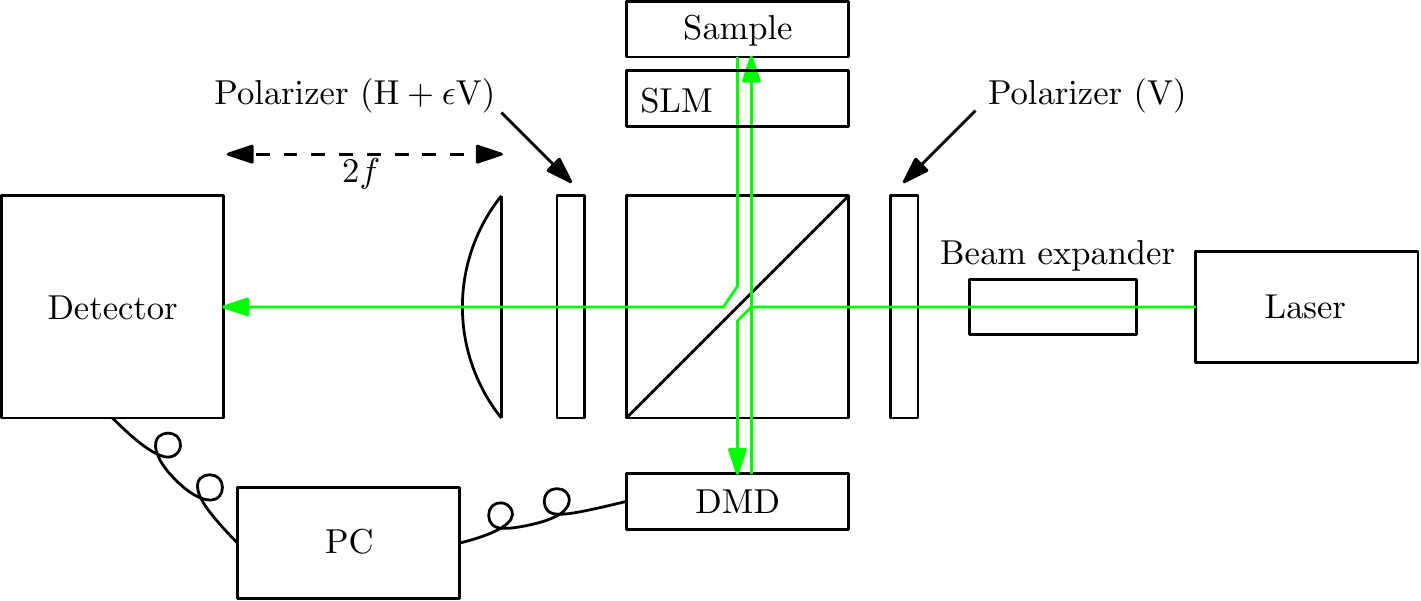}
		\caption{Optical scheme for the simulation of a Hopfield neural network with a control on the coupling constant to allow the implementation of a learning algorithm.}\label{fig:complete_setup_hopfield_learning}
	\end{center}
\end{figure}
  
}

\backmatter
{\setstretch{1.2}
\addcontentsline{toc}{chapter}{Bibliography}
\bibliography{mybib.bib}{}
\bibliographystyle{unsrt}
\clearpage
}

\end{document}